\documentclass{aa}

\usepackage[]{graphicx}
\usepackage[]{exscale}
\usepackage[]{amsmath, amsfonts, amssymb}

\def\aa#1#2#3{#1, {A\&A, }{\bf#2}, #3}
\def\aj#1#2#3{#1, {AJ, }{\bf#2}, #3}
\def\apj#1#2#3{#1, {ApJ, }{\bf#2}, #3}
\def\aarev#1#2#3{#1,   {Astron. Astrophys. Rev., }{\bf#2}, #3}

\def\mn#1#2#3{#1, {MNRAS, }{\bf#2}, #3}
\def\MTAC#1#2#3{#1, {Math. Tables Aids Comp., }{\bf#2}, #3}
\def\nnfm#1#2#3{#1, {Notes Numer Fluid Mech, }{\bf#2}, #3}
\def\pasj#1#2#3{#1, {Publ. Astron. Soc. Japan, }{\bf#2}, #3}
\def\ptps#1#2#3{#1, {Prog. Theor. Phys. Supp., }{\bf#2}, #3}
\def\pp#1#2{#1, {Protostar and Planets V, eds Mannings, V., Boss, A.P., Russell, S. S.}, #2}
\def\pptwo#1#2{#1, {Protostar and Planets I$\!$I, eds Black, D. C., Mathews, M. S.}, #2}
\def\phyrev#1#2#3{#1,    {Phys. Rev., }{\bf#2}, #3}
\def\rgp#1#2#3{#1, {Reviews of Geophys., }{\bf#2}, #3}

\begin{document}

\title{Grain charging in protoplanetary discs}

\author{Martin~Ilgner}


\institute{Max-Planck-Institut f\"ur Astronomie, K\"onigstuhl 17, D - 69117 Heidelberg, Germany \\
\email{ilgner@mpia-hd.mpg.de}}

\date{Received 30. September 2011 / Accepted 26. November 2011}

\titlerunning{Grain charging in protoplanetary discs}

\authorrunning{M.~Ilgner}

\abstract
{Recent work identified a growth barrier for dust coagulation that originates in the electric 
repulsion between colliding particles. Depending on its charge state, dust material may have 
the potential to control key processes towards planet formation such as MHD 
(magnetohydrodynamic) turbulence and grain growth which are coupled in a two-way process.}
{We quantify the grain charging at different stages of disc evolution 
and differentiate between two very extreme cases: compact spherical grains and 
aggregates with fractal dimension $D_f = 2$.}
{Applying a simple chemical network that accounts for collisional charging of grains, 
we provide a semi-analytical solution. This allowed us to calculate the equilibrium population 
of grain charges and the ionisation fraction efficiently. The grain charging was evaluated for 
different dynamical environments ranging from static to non-stationary disc configurations.}
{The results show that the adsorption/desorption of neutral gas-phase heavy metals, such as 
magnesium, effects the charging state of grains. The greater the difference between the 
thermal velocities of the metal and the dominant molecular ion, the greater the change in 
the mean grain charge. Agglomerates have more negative excess charge on average than 
compact spherical particles of the same mass. The rise in the mean grain 
charge is proportional to $N^{1/6}$ in the ion-dust limit. We find that grain charging 
in a non-stationary disc environment is expected to lead to similar results.}
{The results indicate that the dust growth and settling in regions where the dust growth is 
limited by the so-called ''electro-static barrier'' do not prevent the dust material from remaining 
the dominant charge carrier.}


\maketitle

 
\section{Introduction}\label{sec1}
Small dust particles are regarded as the key ingredient to control the planet formation 
process in protoplanetary discs. At early stages of planet formation, submicron-sized 
compact particles are thought to grow towards larger but fluffy agglomerates. At mm to cm 
sizes these agglomerates are supposed to be compacted, whereby they loose their fractal 
structure. Because of the rise in the mass-to-surface ratio associated with the compaction, 
agglomerates tend to be more concentrated locally. However, there are a number of potential 
processes that may control grain growth, one of which is disc turbulence. The magnetorotational 
instability - MRI - (Balbus \& Hawley 1991; Hawley \& Balbus 1991) has been shown to be robust 
in generating turbulence in Keplerian discs. MHD turbulence provides 
the internal stress required for mass accretion, the rate of which is constrained by observations. 
Observations of young stars indicate that discs usually show signatures of active gas accretion 
onto the central star with mass flow rates of about $10^{-8\pm1} \ \rm M_{\odot} yr^{-1}$ (e.g., 
Sicilia-Aguilar et al. 2004). Recent observations from discs around young stars also set constraints 
on the turbulent linewidth and provide support for subsonic turbulence in discs (Hughes et al. 2011). 
However, numerical studies (e.g. Fleming \& Stone 2003) on MRI-driven MHD turbulence have identified 
locations within the planet forming regions, the so-called ''dead zones'', where the coupling 
between the gas and the magnetic field is not sufficient to maintain MHD turbulence. Fleming \& Stone 
also showed that a low Reynolds stress can be maintained in the dead zone, such that low levels of 
accretion are sustained there. That is why dead zones have been considered to advance the planet 
formation process. Dead zones provide a safe environment for grain growth and for planetesimals 
(Gressel et al. 2011). However, the presence of small grains in the weakly ionized dead zones leads to 
significant changes in the abundances of the charge carriers in the gas phase and therefore affects the 
MRI. In other words, MHD turbulence and grain growth are coupled in a two-way process.\\
\indent
Grain charging may also effect the dust growth: Okuzumi (2009) pointed out that the so-called 
''electro-static barrier'' may inhibit dust coagulation in planet forming regions. In a subsequent study, 
Okuzumi et al. (2011a) extended the analysis including the dust size distribution. They found 
that under certain conditions the dust undergoes bimodal growth. While the small aggregates sweep up 
free electrons, the large aggregates continue to grow. However, the growing stalls if the electrostatic 
repulsion becomes strong and the collisions are driven by Brownian motion. The results of Okuzumi et 
al. (2011b) indicate that under minimum mass solar nebula conditions the dust growth halts at aggregate 
sizes beyond $[4 \cdot 10^{-5}, 3\cdot 10^{-2}] \ \rm cm$ depending on the radial position. Conventional 
chemical models studying the dust-grain chemistry in protoplanetary discs account for up to $|Z| = 3$ grain 
charges (e.g. Sano et al. 2000). More recently, the range of grain charges was extended considerably for 
various purposes. Focusing on surface layers of T-Tauri and transitional discs, Perez-Becker  \& Chiang 
(2011) examined the charge distribution on grains covering grain charges 
$[Z_{\rm min}, Z_{\rm max}] = [-200, 200]$. Okuzumi (2009) calculated the charge distribution on dust 
aggregates. He demonstrated that aggregates made of $N=10^{10}$ monomers can carry excess charges 
of about $Z_{\rm min} \sim - 10^5$. To account for higher grain charges we therefore improved our 
simple chemical network introduced in Ilgner \& Nelson (2006a) allowing $[Z_{\rm min}, Z_{\rm max}] $ to 
become free of choice and appropriate to the conditions applied.\\
\indent
In this paper we examine the grain charging for various stages of disc evolution ranging from 
static to non-stationary disc environment. Linking the grain charging with the gas-dust dynamics 
provides a natural environment to mimic the variation of the dust-to-gas ratio 
$\Sigma_{\rm d}/\Sigma_{\rm g}$. The grain charging is assumed to originate in collisional 
charging processes between grains, electrons, and gas-phase ions. We include X-ray ionisation 
from the central star as the primary source for ionisation and consider, to some extent, the 
contributions from cosmic ray particles. We furthermore assume an equilibrium distribution of dust 
material in vertical direction balancing turbulent stirring and sedimentation. In particular, we apply 
the model of Takeuchi \& Lin (2005) to simulate the dynamics of the gas-dust disc. One of our primary 
goals is to compare the grain charging obtained for compact spheres and dust agglomerates and the 
mean grain charge in particular.\\
\indent
We have investigated the effect that thermal adsorption/desorption of metals has on grain charging 
using two different chemical networks. In comparison with the values obtained by switching off 
the thermal adsorption of metals, the modified Oppenheimer-Dalgarno model produces lower, 
the Umebayashi-Nakano model higher values for the mean grain charge. This originates in 
differences in the thermal velocities of the dominant gas-phase ion: 
$v_{\rm HCO^+} < v_{\rm Mg^+} < v_{\rm H_2^+}$. In line with expectations, we find that 
dust agglomerates have a higher charge-to-mass ratio carrying more charges than the 
corresponding compact spherical particles. Considering stationary disc configurations,  
we observe that grain charging on dust agglomerates is always associated with the so-called 
"ion-dust regime",  where the charge balance is mainly maintained by negatively charged grains 
and positively charged gas-phase ions. (The ion-electron regime is characterised by the balance 
between gas-phase ions and electrons.) In particular, we show that the fractional abundances of 
the charge carriers are independent of the number $N$ of constituent monomers. Concerning 
compact spherical dust particles, we identified regions $z/h_{\rm g} > 0$ associated with the ion-electron 
regime. Another conclusion of our work is that  - for the parameter range considered (i.e., compact radii 
$a_{\ast} \le 10^{-3} \ \rm cm$ and agglomerates with characteristic radii $a_{\rm c} \le 10^{-2} \ \rm cm$ 
with $N \le 10^6$, respectively\footnote{The characteristic and compact 
radius is defined in Eqs. (\ref{eq16:sec2}) and (\ref{eq18:sec2}), respectively.}) - grain charging in a 
non-stationary disc environment is expected to lead to similar results. We also present semi-analytical 
solutions for both agglomerates and compact spheres that allow us to determine the equilibrium distribution 
of  $<\!\!Z\!\!>$, $\sqrt{<\!\!\Delta Z^2\!\!>}$, and the fractional abundances of the charge carriers a priori.\\
\indent
The paper is organised as follows. In Sect.2  we discuss the disc model we apply. In Sect. 3 
we introduce the chemical network that we used to compare our results with the reaction network 
Okuzumi (2009) applied. Key problems concerning the ionisation rates are discussed 
in Sect. 4 while the numerical methods applied are described in Sect. 5. In Sect. 6 we present 
the results of our model, and finally in Sect. 7 we summarise the key findings of our study.

\section{Disc model}\label{sec2}
The geometry of the disc model is described in cylindrical coordinates $(R,\varphi,z)$, where 
$R, \varphi$, and $z$ denote the radial distance to the point of interest, the azimuthal angle, 
and the height. Here, we simply considered a two-dimensional (2D) geometry by assuming axial 
symmetry, i.e., $\partial / \partial \varphi = 0$.\\
\indent
Treating the disc as a two-component fluid of dust particles and gas, we assumed that the 
circumstellar gas disc is turbulent and accreting onto the central star. We also assumed 
that the gas is primarily heated by stellar radiation rather than viscous dissipation. We 
supposed that the mean gas motion is characterised by a subsonic flow in $R$, supersonic 
flow in $\varphi$, while the gas density is characterised by hydrostatic equilibrium in $z$-direction.\\
\indent
The underlying disc model we applied is based on the assumption of a two-component fluid 
that allows the dust particles and the gas to follow different velocity fields ${\vec{v}}_{\rm d}$ 
and ${\vec{v}}_{\rm g}$, respectively. In particular, we considered dust particles for which the 
two-component fluid is best described in the so-called ''free molecular approximation'', e.g., 
$a_0 \ll 1.44 \ R_{\rm au}^{11/4}$ cm under minimum mass solar nebula condition. ($a_0$ 
denotes the dust radius while $R_{\rm au}$ is the orbital radius in units of AU.) Since the 
mean free path of the gas molecules exceeds the dust particle sizes, the collisions between 
the gas and the dust particles are much more frequent than the collisions between gas particles. 
We moreover considered disc stages for which the gas density $\varrho_{\rm g}$ dominates over the 
density $\varrho_{\rm d}$ of the dust. The gas is therefore assumed to be decoupled from the 
dust disc. However, the gas drag may alter the velocity $v_{\rm d}$ of the dust particles. Assuming 
$v_{\rm d} \ll c_{\rm s}$, the drag coefficient $\beta$ is characterised by (Epstein, 1924)
\begin{equation}
\beta = \frac{4}{3} \varrho_{\rm g} v_{\rm th} \sigma \ ,
\label{eq01:sec2}
\end{equation}
where $v_{\rm th}$ denotes the thermal velocity. We can additionally specify the friction 
or stopping time $t_{\rm S} =  m/\beta$, which will turn out to be an important diagnostic 
quantity. $m$ and $\sigma$ refer to the mass of the dust particle and its projected surface 
area, respectively. The friction time basically characterises the time scale needed to couple 
the dust dynamics with the gas, which is why $t_{\rm S}$ is also called stopping time. Instead of  
$t_{\rm S}$, we used the dimensionless stopping time  $T_{\rm S}$ defined as the ratio 
between $t_{\rm S}$ and the dynamical time scale $t_{\rm \varphi}$.\\
\indent
We studied the grain charging for various stages of disc evolution ranging 
from static to non-stationary disc environment. A description of the gas-dust dynamics 
applied is given in Subsec. \ref{sec2:sub1} while Subsec. \ref{sec2:sub2} summarises 
the local disc structure.

\subsection{Dynamics of the gas-dust disc}\label{sec2:sub1}
The extreme case of disc dynamics is associated with the so-called ''minimum mass 
solar nebula'' (MMSN) model, which was proposed by Hayashi (1981).  He adopted the 
following radial profiles of the dust and gas surface densities 
$\Sigma_{\rm d}$ and $\Sigma_{\rm g}$ for $R < 36 \ \rm AU$ 
\begin{eqnarray}
\Sigma_{\rm g}    & = & 1.7 \cdot 10^3 \  R_{\rm au}^{-3/2} {\rm g cm^{-2}} \label{eq02:sec2} \\
\Sigma_{\rm d}   & = & 
\begin{cases}
 7.1 \ R_{\rm au}^{-3/2} {\rm g cm^{-2}}  & \text{if } 0.35 < R_{\rm au} \le 2.7 \\
 30  \ R_{\rm au}^{-3/2} {\rm g cm^{-2}}  & \text{if } 2.70 < R_{\rm au} < 36 \ .
\end{cases}  
\label{eq03:sec2}
\end{eqnarray}
We recall that Hayashi derived the power law above assuming no substantial transport 
of dust material. Although recent simulations of the radial migration of planets indicate that 
the radial profile of $\Sigma_{\rm g}$ (as well as the  gas temperature profile) differs 
significantly from the MMSN model (cf Desch 2007), we opted for Hayashi's MMSN model. 
That is because the assumed power index for the MMSN model nicely corresponds to a 
static solution of the disc model of Takeuchi \& Lin (2002), see below for more information. 
The grain charging obtained for that particular environment is discussed in Sec. \ref{sec3}.\\
\indent
Takeuchi \& Lin (2005) extended their disc model to study the dynamical evolution of the 
gas-dust disc. The corresponding set of continuity equations is
\begin{eqnarray}
\frac{\partial  \Sigma_{\rm g} }{\partial t} + 
\frac{1}{R} \frac{\partial}{\partial R} \left( R \Sigma_{\rm g} \overline{v}_{R, \rm g} \right)
& = & 0 \label{eq04:sec2} \\
 \frac{\partial  \Sigma_{\rm d} }{\partial t} + 
\frac{1}{R} \frac{\partial}{\partial R} \left( RF_{\rm diff} +  
R \Sigma_{\rm d} \overline{v}_{R, \rm d} \right) & = & 0, \label{eq05:sec2}
\end{eqnarray}
where $\overline{v}_{R, \rm g}$ and $\overline{v}_{R, \rm d}$ are the vertically 
integrated velocities of the  gas and the dust particles. The (ordinary) diffusion is described 
by the phenomenological equation, which is well-known as Fick's first law for a binary 
system. The corresponding mass flow is approximated by
\begin{equation}
F_{\rm diff}  =  - \Sigma_{\rm g} \frac{\nu}{\rm Sc} \frac{\partial}{\partial R} \left( 
\frac{\Sigma_{\rm d}}{\Sigma_{\rm g}} \right), \label{eq06:sec2}
\end{equation}
with the Schmidt number ${\rm Sc} = \nu / \cal{D}$ specifying the ratio of the rate of 
angular momentum transport to the rate of turbulent transport of grain particles. 
We used the $\alpha$ prescription for the viscous stress, such that the kinematic viscosity 
is $\nu = \alpha c_{\rm s} h_{\rm g}$.\\
\indent
The interpretation of the integrated velocities  $\overline{v}_{R, \rm g}$ and 
$\overline{v}_{R, \rm d}$ is crucial for our understanding of the species' transport in 
our particular  model. $\overline{v}_{R, \rm g}$ and $\overline{v}_{R, \rm d}$ refer rather to 
the velocities associated with the transport of the surface densities $\Sigma_{\rm g}$ and 
$\Sigma_{\rm d}$ than to the velocities the individual species are transported with. For 
the gas disc,  the velocity  $\overline{v}_{R, \rm g}$ can be obtained by combining the 
equation of angular momentum and the continuity equation
\begin{equation}
\overline{v}_{R, \rm g} = - \frac{3}{R^{1/2} \Sigma_{\rm g}} \frac{d}{dR} 
\left( R^{1/2} \nu \Sigma_{\rm g}\right).
\label{eq07:sec2}
\end{equation}
Inserting equation (\ref{eq07:sec2}) into the continuity equation for the gas surface 
density, we obtain the well-known diffusion equation reviewed, e.g., in Pringle (1981).\\
\indent
Assuming power laws for the pressure scale height $h_{\rm g}$, the sound speed 
$c_{\rm s}$, and the local gas density $\varrho_{\rm g}$ (see Subsec. \ref{sec2:sub2} below), 
Takeuchi \& Lin (2002) have presented steady-state solutions of Eqs. (\ref{eq04:sec2}) - 
(\ref{eq05:sec2}). Evaluating the mass fluxes 
$R \Sigma_{\rm g} \overline{v}_{R, \rm g}$ and $R \Sigma_{\rm d} \overline{v}_{R, \rm d}$, 
they obtained $\Sigma_{\rm g} \propto R^{-3/2}$ for $v_{R, \rm g} / c_{\rm s} = 0$ and 
\begin{eqnarray}
\Sigma_{\rm g} & = & \Sigma_0 R_{\bf \rm au}^{-1} \label{eq08:sec2} \\
\Sigma_{\rm d} & = & \dot{M}_{\rm d} / \left(2 \pi R \overline{v}_{R, \rm d}\right) \label{eq09:sec2}
\end{eqnarray}
for $v_{R, \rm g} / c_{\rm s} \ne 0$ where the mass flux $\dot{M}_{\rm d} $ of dust material 
enters as a free parameter. $\overline{v}_{R, \rm d}$ 
\begin{equation}
\overline{v}_{R, \rm d} = \frac{1}{\Sigma_{\rm d}} \int \limits_{-\infty}^{\infty} \varrho_{\rm d} v_{R, \rm d} dz
\label{eq10:sec2}
\end{equation}
can an be calculated semi-analytically (see Takeuchi \& Lin, 2002) assuming 
\begin{equation}
v_{R, \rm d} (R,z) = \frac{T_{\rm S}^{-1} v_{R, \rm g} - \eta v_{\rm K, mid}}{T_{\rm S} + T_{\rm S}^{-1}}.
\label{eq11:sec2}
\end{equation}
Here, $\eta$ and $v_{\rm K, mid}$ refer to the ratio of the pressure gradient to the gravity
and Keplerian velocity at disc midplane, respectively.\\
\indent
However, apart from steady-state conditions it is difficult to find  a reliable approximation 
\begin{figure*}[ht]
\includegraphics[width = .5\textwidth]{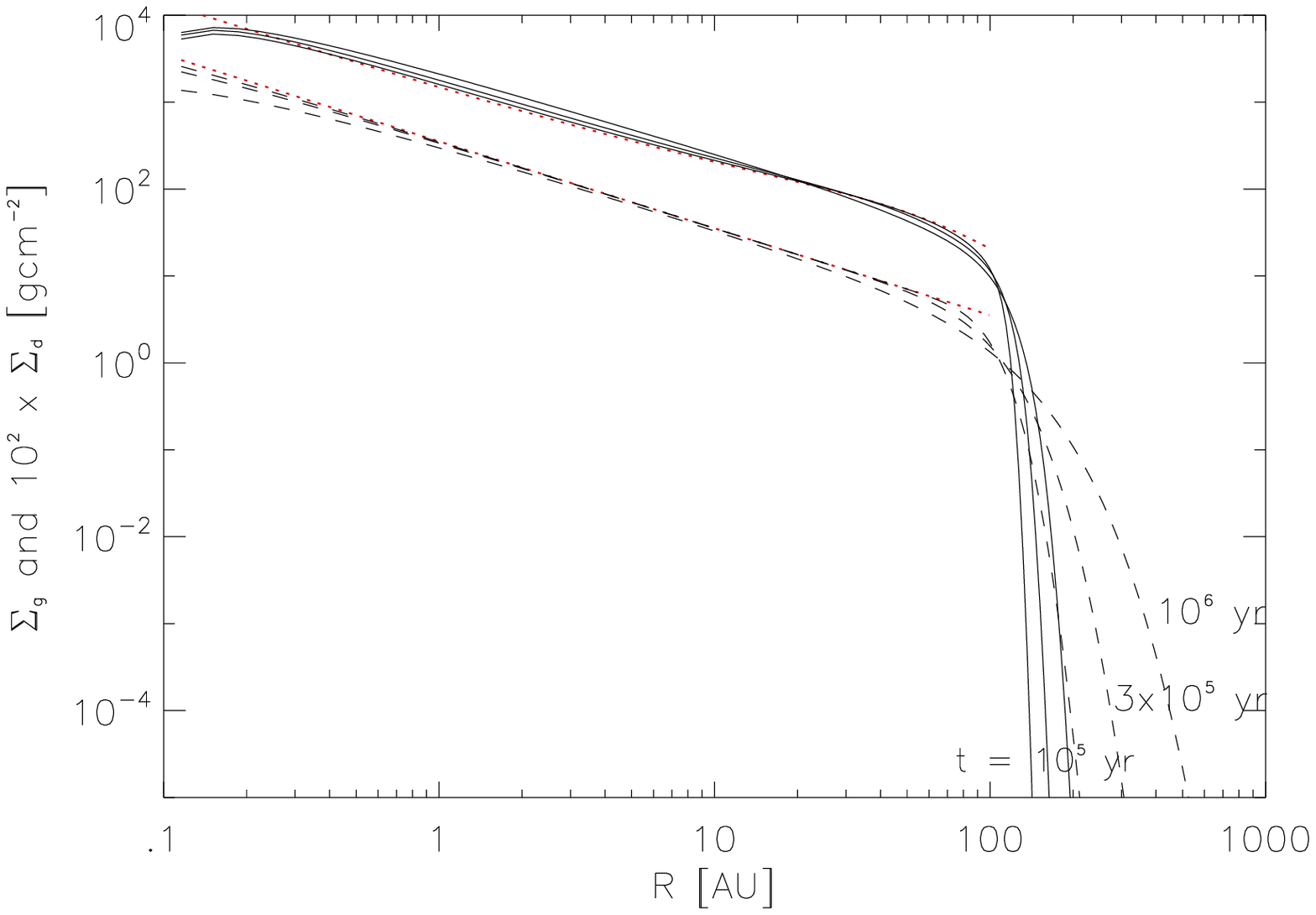}
\includegraphics[width = .5\textwidth]{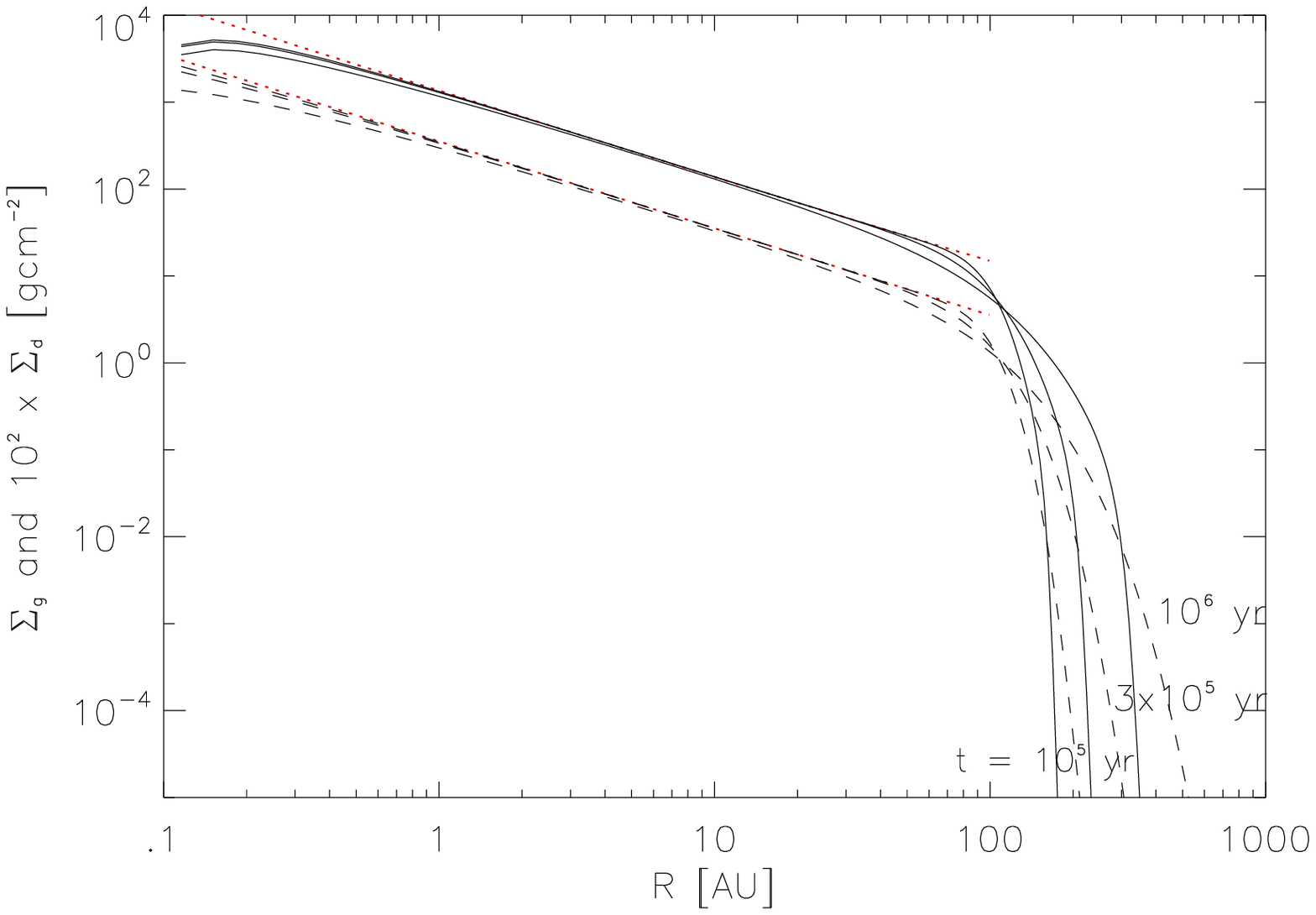}
\caption{Radial profile of the surface densities of the gas and the dust at different 
time snaps $t = 10^5, 3 \cdot 10^5$ and $10^6 \ \rm yr$. The solid lines correspond 
to the dust surface density while the gas surface density is shown using the dashed lines. 
The (red-coloured) dotted line refers to $\Sigma_{\rm g} \sim R^{-1}$ and 
$\Sigma_{\rm d} = \dot{M}_{\rm d} /(2\pi R \! <\!v_{R, \rm d}\!>\!) $ at $t/t_{\rm K} = 0$.   
$|\dot{M}_{\rm d}| = 10^{-10} \rm M_\odot / yr$, $\alpha = 10^{-3}$, and 
$\varrho_{\rm p} = 1.0 \ \rm gcm^{-3}$. The profiles shown in the left panel are obtained 
under the assumption of spherical grains with $a_0 = 10^{-3} \ \rm cm$. The corresponding 
profiles obtained for agglomerates with $D_{\rm BCCA} = 2$ and $N = 10^{6}$ monomers 
of $a_0 = 10^{-5} \ \rm cm$ are shown in the right panel.}
\label{fig1:sec2}
\end{figure*}
for $\overline{v}_{R, \rm d}$. We followed the instruction of Takeuchi \& Lin (2005) to calculate 
$\overline{v}_{R, \rm d}$ but we are aware of  the caveats inherent in their approach. This 
problem arises because a general analytical expression for the corresponding radial component 
$v_{R, \rm g}$ of the gas velocity field does not exist. In order to tackle this problem we simply 
approximate $\overline{v}_{R, \rm d}$ by 
\begin{equation}
\overline{v}_{R, \rm d}  \sim \frac{T_{\rm S}^{-1} \overline{v}_{R, \rm g} - \eta v_{\rm K, mid}}
{T_{\rm S}^{} + T_{\rm S}^{-1}}
\label{eq12:sec2}
\end{equation}
evaluating $T_{\rm S}$ and $\eta$ at disc midplane:
\begin{eqnarray}
\eta & = & 
- \left( \frac{h_{\rm g}}{R} \right)^2 \left( \frac{\partial \ln \Sigma_{\rm g}}{\partial \ln R} + \frac{q - 3}{2} \right) \label{eq13:sec2} \\[.3em]
T_{\rm S} & = & \frac{3}{8} \frac{\pi}{\Sigma_{\rm g}} \frac{m}{\sigma} \ ,
\label{eq14:sec2}
\end{eqnarray}
where $q$ denotes the power index  associated with the gas pressure scale height 
$h_{\rm g}$ (see next subsection).\\[.5em]
As we will demonstrate below, the dynamics of the gas-dust disc is controlled by the stopping 
time $T_{\rm S}$. Because of $T_{\rm S} \propto m/\sigma$ the stopping time relies on the 
particular topology of the dust particles considered. We studied two extreme cases of grain 
topology: compact spherical grains and irregular, very porous 
dust agglomerates. We begin addressing that particular question for compact spherical grains 
before discussing the dust agglomerates and its effect on the gas-dust dynamics.\\[.5em]
\noindent
\underline{Case A}: \emph{Compact spherical dust grains}\\[.3em]
Considering compact spherical grain particles of material bulk density $\varrho_{\rm p}$ and 
a radius $a_0$, the stopping time at disc midplane $z/h_{\rm g} = 0$ is given by
$$
T_{\rm S}  = \frac{\pi \varrho_{\rm p} a_0}{2 \Sigma_{\rm g}}.
$$
Takeuchi \& Lin (2005) studied in detail the effect of compact spherical grains on the 
dynamics of the gas-dust disc. We took their fiducial model, which nicely demonstrates 
the inward migration of larger (spherical) dust particles to test our implementation 
of the numerical schemes applied. We also adopted boundary and initial conditions 
suitable for our particular model.\footnote{We investigated these effects; the results 
obtained are discussed in the appendix.} To summarise, we set the inner boundary at  
$R_{\rm inner} = 0.1 \ \rm AU$ and replaced the zero torque boundary conditions with 
predefined analytical values for $\Sigma_{\rm g}(t)$. In particular, we applied the 
self-similarity solution 
\begin{equation}
\Sigma_{\rm g} = \Sigma_0 X^{-1} \exp\{ -aX/\tau\} \ .
\label{eq15:sec2}
\end{equation}
Lynden-Bell \& Pringle (1974) obtained under the assumption $\nu_{\rm t} \propto R$ with 
dimensionless variables $X = R/R_0$ and $\tau = t/t_0 + 1$. In the limit of  $X \ll 1$
the expression reduces to $\Sigma_{\rm g} \propto X^{-1}$, which represents the steady-state 
solution $\Sigma_{\rm g} \propto R^{-1}$ discussed above. The most appropriate value for 
the constant $a$ was applied to approximate $\Sigma_{\rm g}(t)$ in the outer disc regions. 
We replaced the initial assumption $\Sigma_{\rm d}/ \Sigma_{\rm g} = \rm const$ by 
$|\dot{M}_{\rm d}| = \rm const$. For the sake of convenience we applied the value of Takeuchi 
\& Lin (2002) $\dot{M}_{\rm d} = 10^{-10} M_{\odot} \rm yr^{-1}$, which is one order of magnitude 
smaller than the corresponding mass flux of the gas $\dot{M}_{\rm g} = 3 \pi \nu \Sigma_{\rm g} 
\sim 2.6 \times 10^{-9} M_{\odot} \rm yr^{-1}$.\\
\indent
Because of $T_{\rm S} \propto a_0$, the dynamics of the dust disc depends on the size of 
the spherical grain particle considered. This may also have important consequences for the 
dust growth. The simplest approach to mimic the dust growth is 
to assume that the entire available dust mass is represented by monodisperse compact 
spherical particles of a given size $a_\ast$. By changing $a_\ast$ one may trace different 
stages of dust growth. However, dust growth is supposed to be processed by coagulation 
leading to irregular dust agglomerates. The effect of dust agglomerates on the dust dynamics
is discussed below.\\[.5em]
\noindent
\underline{Case B}: \emph{Irregular porous dust agglomerates}\\[.3em]
The grain particles are now regarded as dust agglomerates. In particular, we considered 
agglomerates associated with the so-called ''ballistic cluster-cluster aggregation'' (BCCA) model. 
The BCCA model describes the growth of aggregates\footnote{We simplified the terminology 
used in this paper. Because fractal agglomerates with fractal dimension $D \sim 2$ can be modelled 
applying the BCCA, agglomerates with $D \sim 2$ are referred to as ''BCCA agglomerates''.} through 
collisions of equal-sized aggregates as cartooned in Fig. \ref{fig3:sec2}. We are aware of constraints 
associated with the BCCA growth process (see discussion in Okuzumi 2009). However, for the purpose 
of our paper we simply assumed that BCCA agglomerates are formed independently of the disc 
environment applied. That is because the BCCA agglomerates are 
used as test particles to trace the grain charging for different disc environments.\\
\indent
Simulations of dust growth have shown that for the BCCA model the characteristic 
radius\footnote{The characteristic radius $a_{\rm c}$ serves to measure the size of the 
agglomerate. In particular, its weighting 
is different from  the rms-definition of the gyration radius $r_{\rm g}$, cf. Okuzumi et al. (2009).} 
$a_{\rm c}$  obeys a power law (e.g., Okuzumi et al. 2009) while the mean value for the 
projected surface area is well approximated by a linear function of $N$ (cf Minato et al. 2006):
\begin{eqnarray}
a_{\rm c} & = & a_0 N^{1/D_{\rm BCCA}} \label{eq16:sec2} \\
\overline{\sigma}_{\rm BCCA}  & = & (0.351 N + 0.566 N^{0.862}) \pi a_0^2 \ {\rm for} \ N \ge 16.
\label{eq17:sec2}
\end{eqnarray}
$N$ denotes the number of the constituent monomers forming the agglomerate ($a_0$ 
is the radius of each monomer). $D_{\rm BCCA}$ is the fractal dimension of BCCA 
associated with $a_{\rm c}$ and $D_{\rm BCCA} \approx 1.9$ (Meakin 1991). We  considered 
$D_{\rm BCCA}  = 2$. The projected surface area is 
\begin{figure}[ht]
\includegraphics[width = 9cm]{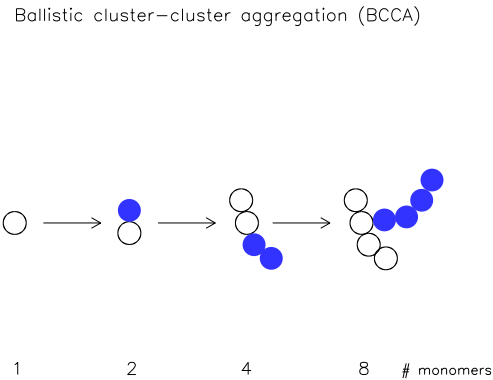}
\caption[]{Scheme representing the dust growth characterised as ballistic cluster-cluster 
aggregation. The agglomerates collide with a clone of each as marked with 
filled circles. The figure is adopted from Okuzumi et al. (2009).}
\label{fig3:sec2}
\end{figure}
$\sigma_{\rm BCCA} \approx \pi a_{\rm c}^2 = N \pi a_0^2$, which is in line with 
the linear approximation of Eq. (\ref{eq17:sec2}). The expression of the stopping 
time (\ref{eq14:sec2}) then is 
$$
T_{\rm S}  \approx  \frac{\pi}{2} \frac{\varrho_p a_0}{\Sigma_{\rm g}}.
$$
We note that for that particular model the stopping time is independent of the number of 
monomers considered. We conclude that the agglomerates of different sizes (i.e., different $N$) 
experience the same dynamical evolution contrasting the properties derived for 
compact spherical grains.\\
\noindent
We can relate the single  agglomerate made of $N$ monomers of size $a_0$ to a 
compact sphere of the same mass by 
\begin{equation}
a_\ast = a_0 N^{1/3} \label{eq18:sec2} \ ,
\end{equation}
which turns out to become an important diagnostic quantity for our calculations. According to Eq. 
(\ref{eq18:sec2}), a compact sphere of $a_{\ast} = 10^{-3} \ \rm cm$ corresponds to an 
agglomerate of $N = 10^6$ monomers with $a_0 = 10^{-5} \ \rm cm$. The surface density 
profiles obtained for this dust agglomerate are shown in the right panel of Fig. \ref{fig1:sec2}. 
We observe that the dust is tightly coupled to the gas dynamics that expands because of 
turbulent mixing towards larger orbital radii $R$. The advective inward drift of the dust (which 
is the dominant transport process for compact spherical grains $a_0 \ge 10^{-2} \ \rm cm$) 
cannot compete with the diffusive mixing along $R$.

\subsection{Local disc structure}\label{sec2:sub2}
We applied the conventional ansatz of the thin disc approximation $h_{\rm g}/R \ll 1$. 
We assumed that the gas disc is isothermal in $z$-direction. Then, we know that on 
a time scale $t_z = h_{\rm g}/c_{\rm s} = \Omega_{\rm K}^{-1}$ the hydrostatic 
equilibrium is  established while changes in the gas surface density 
$\Sigma_{\rm g}$ are determined by the viscous time scale $t_{\nu} = R^2/\nu$ 
with $t_z \ll \tau_{\nu}$. Hence the local structure of the 
gas disc is assumed to set up instantaneously during the viscous evolution.\\
\indent
Following Takeuchi \& Lin (2002), we assumed that the gas pressure scale height $h_{\rm g}$ 
and the isothermal sound speed $c_{\rm s}$ are simply defined by power laws associated 
with the scaling exponent $q = - 1/2$:
\begin{eqnarray}
h_{\rm g} & = & h_{0, \rm g} R_{\rm AU}^{(q + 3)/2}  \label{eq19:sec2} \\
c_{\rm s} & = & c_0 R_{\rm AU}^{q/2}  \label{eq20:sec2}  \\
& & \hspace*{-1cm} \text {so that the gas density is defined by} \nonumber \\ 
\varrho_{\rm g} & = & \frac{1}{\sqrt{2\pi}} \frac {\Sigma_{\rm g}}{h_{\rm g}} 
\exp\left\{ - \frac{1}{2} \frac{z^2}{h_{\rm g}^2} \right\} \ .
 \label{eq21:sec2}
\end{eqnarray}
\noindent
Regarding the dust disc, we are primarily interested in conditions for which the mass flows 
associated with the settling and (vertical) turbulent mixing of the dust particles approach 
or attain a state of equilibrium. Taking the fast settling limit 
$T_{\rm s}/\alpha \gg (h_{\rm g}/R)^2$ into account, the dust density profile becomes
\begin{eqnarray}
\varrho_{\rm d} & = & \varrho_{0, \rm d} 
\exp \left\{- \frac{1}{2} \frac{z^2}{h_{\rm g}^2} 
- {\rm Sc} \frac{ T_{\rm s, mid}}{\alpha} \left( \exp \frac{z^2}{2h_{\rm g}^2} - 1 \right) \right\} 
 \label{eq22:sec2} \\
\Sigma_{\rm d} & = & \int \limits_{-\infty}^{\infty} \varrho_{\rm d} dz,
 \label{eq23:sec2}
\end{eqnarray}
cf Takeuchi \& Lin (2002). Approximating the settling time scale by 
$t_{\rm sed} = (T_{\rm S} \Omega_{\rm K})^{-1}$,  we confirm in the fast settling limit 
the validity of
\begin{equation}
t_z \ll t_{\rm sed} \ll t_{\nu} = \frac{1}{\alpha} \frac{1}{ (h_{\rm g}/R)^2} \frac{1}{\Omega_{\rm K}}.
 \label{eq24:sec2}
\end{equation}

\noindent
We considered that the gas is transparent to the emitted visible radiation of the protostar 
and neglected the contributions to the gas temperature from viscous dissipation. The 
temperature $T_{\rm g}$ of the gas is then given by
\begin{equation}
T_{\rm g} = T_0 \left( \frac{L{\ }}{L_\odot}\right)^{\frac{1}{4}} R_{\rm AU}^{-\frac{1}{2}}
 \label{eq25:sec2}
\end{equation}
with $T_0 = 280 \ \rm K$ (Hayashi 1981). Regarding the temperature of the dust particles, 
we drastically simplified the physics by setting $T_{\rm d} = T_{\rm g}$, which is in line with 
previous studies (Sano et al. 2000, Bai \& Goodman 2009). However, we are aware of the effect 
stellar radiation has on dust particles at high altitude (i.e., the so-called ''superheated layer'') 
and its consequences for the interior gas temperature as demonstrated by Chiang \& Goldreich 
(1997).  Chiang \& Goldreich report that the superheated dust layer may change with $R$; 
for their particular model of the minimum mass solar nebula, they showed that the location 
of the superheated layer changes from  $z/h_{\rm g} = 5$ at $R = 3 \ \rm AU$ to  
$z/h_{\rm g} = 4$ at $R = 100 \ \rm AU$. For altitudes $z/h_{\rm g} \le 3$  we considered, that the 
dust particles may be located some distance below the superheated layer.

\section{Chemical model}\label{sec3}
We briefly recall the constituents of a chemical model: its components (or species), 
the chemical reactions that cause time-dependent changes in the abundances of species, and 
the underlying kinetic equation for each component.\\
\indent
Regarding the class of chemical species applied, we considered gas-phase species, 
grain species,  and mantle species. The latter are species adsorbed onto the grain  
surface and associated with the counterparts of the neutral gas-phase species. Using this 
nomenclature, we differentiate between reactions that involve mantle species and 
those that do not. As we did in Ilgner \& Nelson (2006a), we refer to the former as Òmantle 
chemistryÓ and the latter as Ògrain chemistryÓ. We considered two different types for the 
topology of grain particles:\\[.3em]
Type 1: The grain particles are regarded as dust agglomerates made by dust 
growth. For our particular model, we consider agglomerates associated with the BCCA model. 
The individual agglomerates are also assigned to different grain charges with 
$j = Z_{\rm min}, \dots, Z_{\rm max}$. Details are given in Sec. \ref{sec3:sub2}.\\[.3em]
Type 2:  The grain particles are now characterised as spherical dust grains of a radius $a_0$. 
The individual grain particles are assigned to different grain charges with 
$j = Z_{\rm min}, \dots, Z_{\rm max}$. These models are discussed in Sec. \ref{sec3:sub3}.\\[.3em]
\noindent
For the chemical reaction network, we applied the generalised version of a simplified 
chemical network introduced in Ilgner \& Nelson (2006a) who named this the ''modified 
Oppenheimer-Dalgarno model''.  It is a modification of the reaction scheme proposed 
by Oppenheimer \& Dalgarno (1974) and was obtained by adding the grain and mantle 
chemistry. We now dropped the restriction $|Z_{\rm min}| =  |Z_{\rm max}| = 2$ we made in Ilgner \& 
Nelson (2006a) and allow $Z_{\rm min}$ and $Z_{\rm max}$ to become free of choice. 
\begin{table}[t]
\caption{List of reactions considered for the modified Oppenheimer-Dalgarno network. M and 
m denote the neutral metal and molecule component, represented by 
magnesium and molecular hydrogen, while X = $\{ \rm Mg, H_2\}$. $k_5$ and $k_6$ represent 
the chemical processes that are not covered by the Umebayashi-Nakano network.}
\label{ref:tab1}
\begin{center}
\begin{tabular}{llclcll}\hline\hline\\
  $k_1$: &   $\mathrm{m^+}$    & + & M & $\rightarrow$ &
                m         +$ \mathrm{M^+} $&  \\
  $k_2$: &   $\mathrm{m^+}$    & + & $\mathrm{e^-}$ & $\rightarrow$  &
                m       &  \\
  $k_3$: &   $\mathrm{M^+}$     & + & $\mathrm{ e^-}$ & $\rightarrow$ &
                M          +   $h\nu$  & \\
  $k_4$: &   m        &     &                         & $\rightarrow$  &
                $\mathrm{m^+}$    +  $\mathrm{e^-}$  & \\
  $k_5$:  &  X & + & grains  & $\rightarrow$ & 
                gX    &     \\
 $k_6$:  &  gX &  &  &$\rightarrow$  &
               X          +  grains  & \\
 $k_7$: &           $\mathrm{X^+}$         & + & $\mathrm{gr^Z}$ & $\rightarrow$ &
\(
 \left\{
 \begin{array}{lcl}
               \mathrm{gX}              & + & \mathrm{gr^{Z + 1}} \quad \text{for $Z \ge 0$}\\
                \mathrm{X}              & + & \mathrm{gr^{Z + 1}}  \quad \text{else}
 \end{array} 
 \right.
 \)
 & \\[1em] \hline\hline
\end{tabular}
\end{center}
\end{table}
A full list of the chemical reaction schemes applied is given in Table \ref{ref:tab1}. 
To keep the terminology simple, we still refer to this model as the 
''modified Oppenheimer-Dalgarno model''.\\
\indent
We benefit from the fact that for our particular chemical model the numerical solution of 
the kinetic equations associated with the equilibrium state is accompanied by an 
analytical solution. The key idea needed to construct the analytical solution was taken 
from Okuzumi (2009), who succeeded to obtain an analytical solution for the network of 
Umebayshi-Nakano (1990). Our network, i.e. the modified Oppenheimer-Dalgarno model, 
and the more complex Umebayashi-Nakano network use of the same chemical key 
processes. However, they differ significantly in the way neutral gas-phase species may stick onto 
grains. To facilitate the comparison with the Umebayashi-Nakano network, we precede this section by 
comparing the modified Oppenheimer-Dalgarno with the Umebayashi-Nakano network. 
Both networks and the chemical models presented in this section are evaluated at disc midplane 
$z/h_{\rm g} = 0$ under minimum mass solar nebula conditions applying Eqs. (\ref{eq02:sec2}), 
(\ref{eq19:sec2}) - (\ref{eq21:sec2}), and (\ref{eq25:sec2}). Except for the sticking 
coefficient $s_{\rm e} = 0.3$, the ionisation rate and the elemental abundance of metals 
$x_{\rm M}$ were taken from Sano et al. (2000) to facilitate the comparison with previous studies. In 
particular, $x_{\rm M} = 7.97 \cdot 10^{-5} \delta$ where $\delta$ denotes the assumed heavy 
metal gas depletion with respect to solar abundances. $\delta = 0.02$ is regarded as the most 
favourable condition for retaining metals in the gas-phase. In  Sec. \ref{sec6} we follow a 
complementary approach by making a conservative estimate with $x_{\rm M} \ll 7.97 \cdot 10^{-5} \delta$.

\subsection{Comparison: Modified Oppenheimer-Dalgarno vs Umebayashi-Nakano network}
\label{sec3:sub1}
Both chemical networks are frequently applied, e.g., in studies on the dead zone structure of 
protoplanetary discs. In particular,  under conditions typically applied for protoplanetary discs 
the ionisation degree and grain charging processes are closely related, cf  Ilgner \& Nelson 
(2006a). Tracing the dead zone structure of protoplanetary discs, the chemical 
network associated with the modified Oppenheimer-Dalgarno model has been 
compared with more complex chemical networks, cf Ilgner \& Nelson (2006a) and Bai \& Goodman 
(2009). While Ilgner \& Nelson were concerned with conventional $\alpha$-disc 
models, Bai and Goodman applied minimum mass solar nebula conditions. Considering 
a huge range of  grain depletion, Bai and Goodman showed that the complex network 
generates more free electrons, which are less  than $ \lesssim 2$ times the values obtained 
for the modified Oppenheimer-Dalgarno model.\\
\indent
Similar to Oppenheimer \& Dalgarno (1974), the network of Umebayashi \& Nakano (1990) 
was originally designed under dense interstellar cloud conditions. Sano et al. (2000) adopted 
the scheme of Umebayashi and Nakano (with some modifications) to estimate the size of the 
dead zone under minimum mass solar nebula conditions. Okuzumi (2009) modified the 
Umebayshi-Nakano network to account for grain charging processes on irregular porous 
dust agglomerates. Again, to keep the terminology simple, the chemical networks presented 
in Sano et al. (2000) and Okuzumi (2009) are referred to as the Umebayashi-Nakano model.\\
\indent
We recall that the modified Oppenheimer-Dalgarno model and Umebayashi-Nakano model 
are quite similar in the way they are constructed. Grain charges up to $|Z| \le 3$ are considered. 
The networks 
\begin{figure*}[ht]
\includegraphics[width = 9cm]{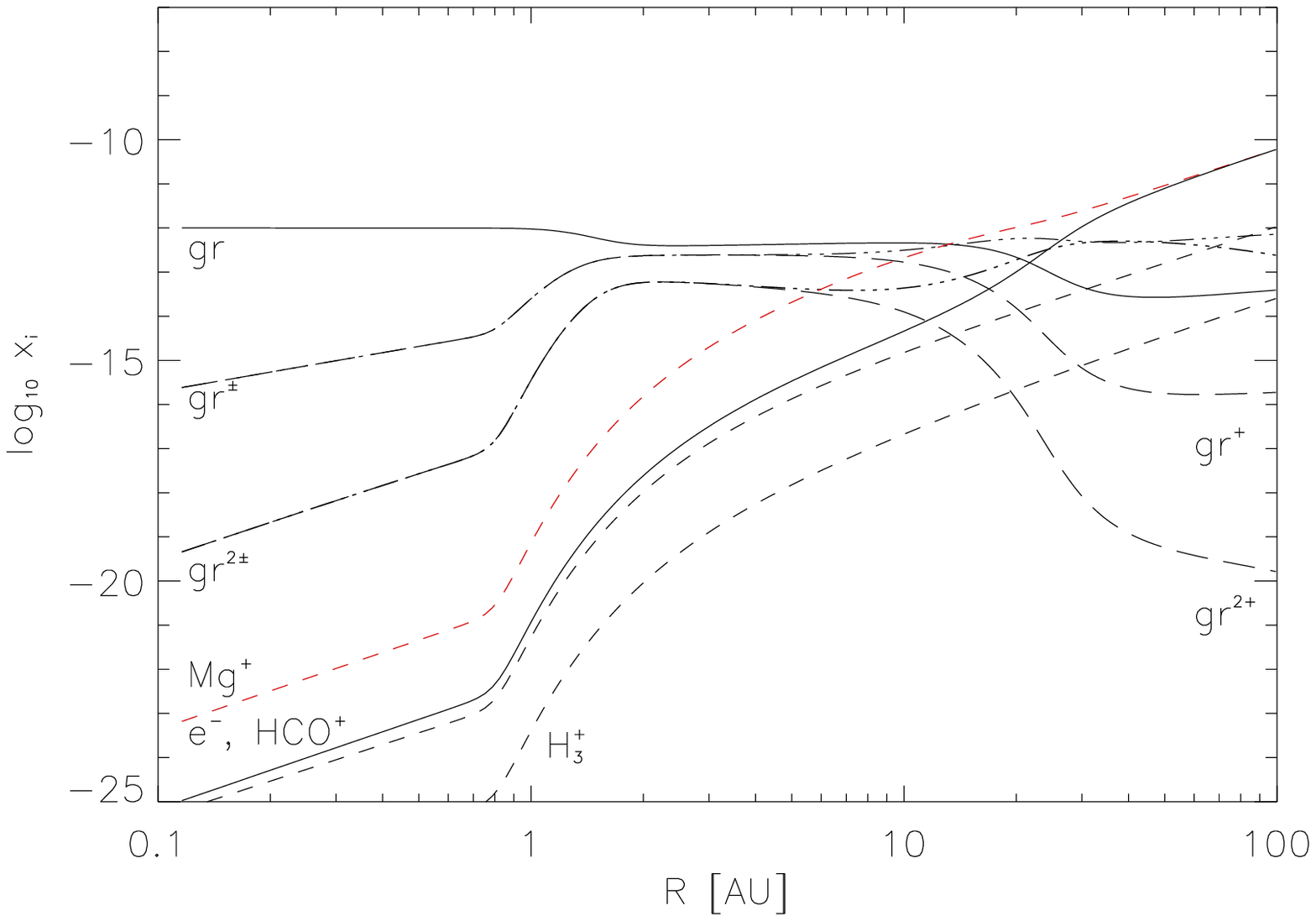}
\includegraphics[width = 9cm]{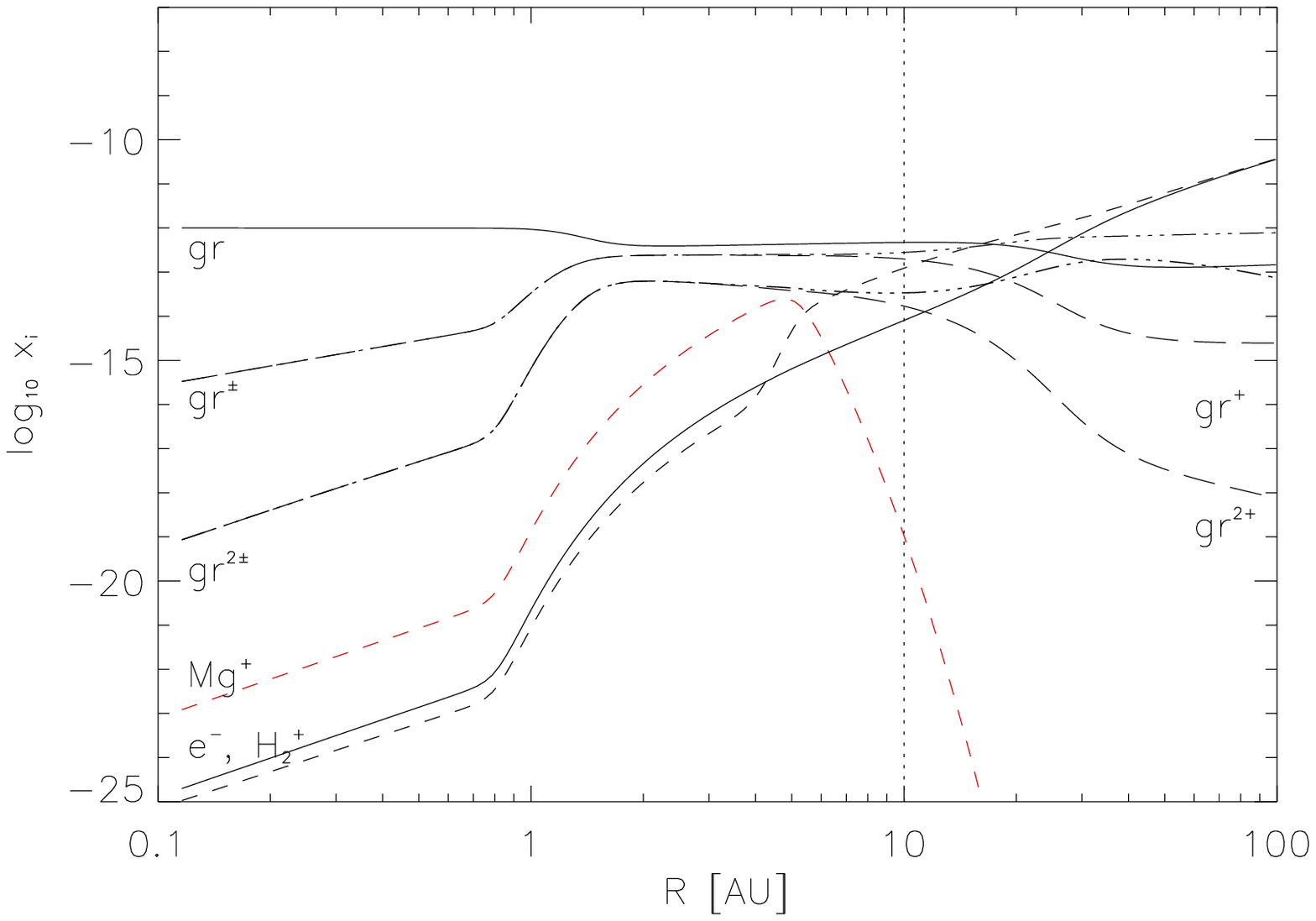}
\caption{Equilibrium concentrations at disc midplane shown for some representative components 
of the Umebayashi-Nakano (left panel) and the modified Oppenheimer-Dalgarno network (right 
panel), respectively. Apart from $s_{\rm e} = 0.3$ the parameter setup is taken from Sano et al. (2000). 
The elements m and M of the modified Oppenheimer-Dalgarno network are specified by molecular 
hydrogen and magnesium. The line $x_i = \infty$ at $R = 10 \rm \ AU$ is marked to aid comparison with 
the modified Oppenheimer-Dalgarno network applied to dust agglomerates, cf Figs. \ref{fig2:sec3} 
and \ref{fig3:sec3}.}
\label{fig1:sec3}
\end{figure*}
use the schematic components m, m$^+$, M, and M$^+$. The components M and  M$^+$ 
represent a heavy metal atom and its ionized counterpart. The networks differ when specifying 
the molecular ion m$^+$. In the modified Oppenheimer-Dalgarno model, m$^+$ simply denotes 
the ion of the dominant molecule, while the ions 
$\rm O_{}^+, O_2^+, OH^+, O_2^{}H_{}^+, CO_{}^+, CH_2^+, \ and \ HCO_{}^+$ 
contribute to m$^+$ in the scheme of the Umebayashi-Nakano model. Umebayashi \& Nakano introduced 
this extension because molecular ions react differentially. Only a few of the ions do react 
via charge transfer, others do not. In addition to the steady-state assumption for the neutral 
gas components $\rm CO, Mg, O_2^{}, \ and \ O$, two chemical processes make the 
chemical networks different. The reaction scheme of the modified Oppenheimer-Dalgarno model includes the 
thermal adsorption of neutral gas-phase particles on grain surfaces and its corresponding reverse desorption 
process, the scheme of Umebayashi \& Nakano does not. As Bai \& Goodman (2009) pointed out, one can 
estimate the temperature range beyond the adsorption that dominates the desorption process, and vice versa. For 
heavy metal atoms M the corresponding temperature range is much more narrow. For gas temperatures below 
a critical value,  the grains are very effective at sweeping up metal atoms. This was 
demonstrated by Ilgner \& Nelson (2006a) when studying which effect this has on the size of 
the dead zone in conventional $\alpha$-discs.\\
\indent
In a previous publication (Ilgner \& Nelson 2006a) we have already examined the 
Umebayashi-Nakano network under minimum mass solar nebula conditions that reproduces 
the results of Sano et al. (2000). Varying $s_{\rm e}$ to $0.3$, we have calculated the 
equilibrium abundances obtained for the modified Oppenheimer-Dalgarno and the 
Umebayashi-Nakano network, respectively. The profiles obtained for  $z/h_{\rm g} = 0$ are shown 
in the left panel (Umebayashi-Nakano network) and in the right panel (modified Oppenheimer-Dalgarno network) 
of Fig. \ref{fig1:sec3}. The components m and M of the modified Oppenheimer-Dalgarno model are specified 
by molecular hydrogen and magnesium with evaporation energies of $450 \ \rm K$ and $5300 \ \rm K$, 
respectively. We moreover assumed a standard value of $1.5 \times 10^{15} \ \rm cm^{-2}$  for the surface density 
of sites available on the grain surface. The comparison of the corresponding profiles revealed major changes in the metal 
ion $\rm M^+$ abundance, which are remarkable, but not unexpected. According to Umebayashi 
\& Nakano,  M is steadily exchanging charges with the ionized molecules 
$\rm H_3^+, C_{}^+, \ and \ HCO_{}^+$ via charge transfer while keeping the reservoir of 
neutral metal atoms $x_\infty[\rm M]$ constant and equal to the total fractional abundance 
of metals.\footnote{Problems related to the steady-state assumption of neutral 
metal atoms are discussed for the original Oppenheimer-Dalgarno model in Ilgner \& Nelson 
(2006a), where it was labelled \texttt{model1}.} In the modified Oppenheimer-Dalgarno 
model at $135 \ \rm K$ ($R \sim 4.2 \ \rm AU)$, the amount of metals adsorbed onto grain particles 
balances the gas-phase heavy metals. For lower temperatures essentially no metals are left 
in the gas phase and hence no metal atoms are available to be processed to metal ions via 
charge transfer reactions. Instead, the molecular ion $\rm m^+$ is becoming the dominant 
gas-phase ion for  ($R > 5 \ \rm AU)$. However, the profiles for the ionisation fraction and the 
grain charges for the modified Oppenheimer-Dalgarno model and the Umebayashi-Nakano 
model agree very well.\\
\indent
We note that metals serve as the key ingredients in governing the chemistry in the 
original Oppenheimer-Dalgarno model as well as in the Umebayashi-Nakano model.\footnote{Under 
certain conditions, this may have important consequences for ongoing planet formation 
in discs: In the ion-electron plasma limit (see Subsec. \ref{sec6:sub1}) the structure of the 
dead zone may be significantly altered by the combined action of recombination of metal ions and 
turbulent transport (Ilgner \& Nelson 2006b, 2008).} Comparing with the corresponding profiles obtained 
for the Umebayashi-Nakano network, we find that the amount of metal ions available is tremendously 
reduced in the modified Oppenheimer-Dalgarno network. Whether or not this may have significant 
consequences on the grain charging through reaction $k_7$ listed in Tab. \ref{ref:tab1} will be 
analysed in the following subsection.

\subsection{Grain chemistry with irregular porous agglomerates}\label{sec3:sub2}
The charging of irregular porous agglomerates was introduced by Okuzumi (2009), who 
applied the BCCA model to characterise the agglomerates. Okuzumi presented an analytical 
solution to obtain the equilibrium state of the ionisation degree and the excess charges $Z$ 
of the dust agglomerates. The latter is well described by a Gaussian with a mean value 
of $<\!\!Z\!\!>$ and a variance $<\!\!\Delta Z^2\!\!>$, which is nicely detailed in Okuzumi (2009). 
However, at first glance the analytical solution might serve to verify the 
numerical solution rather than providing an a priori estimate. To recap: solving the 
corresponding equations (these are Eqs. (27) - (31) in Okuzumi, 2009) requires knowing 
the final composition of the ions in order to calculate the mean ion velocity $\overline{u}_i$ 
and the mean gas-phase recombination rate $\overline{\beta}$. For a metal-rich environment, we 
have seen that the metal ion $\rm Mg^+$ is by far the dominant ion in the 
Umebayashi-Nakano network (see previous subsection), which compensates its lower mean 
thermal velocity. We therefore simply 
\begin{figure*}[tbh]
\includegraphics[width = 9cm]{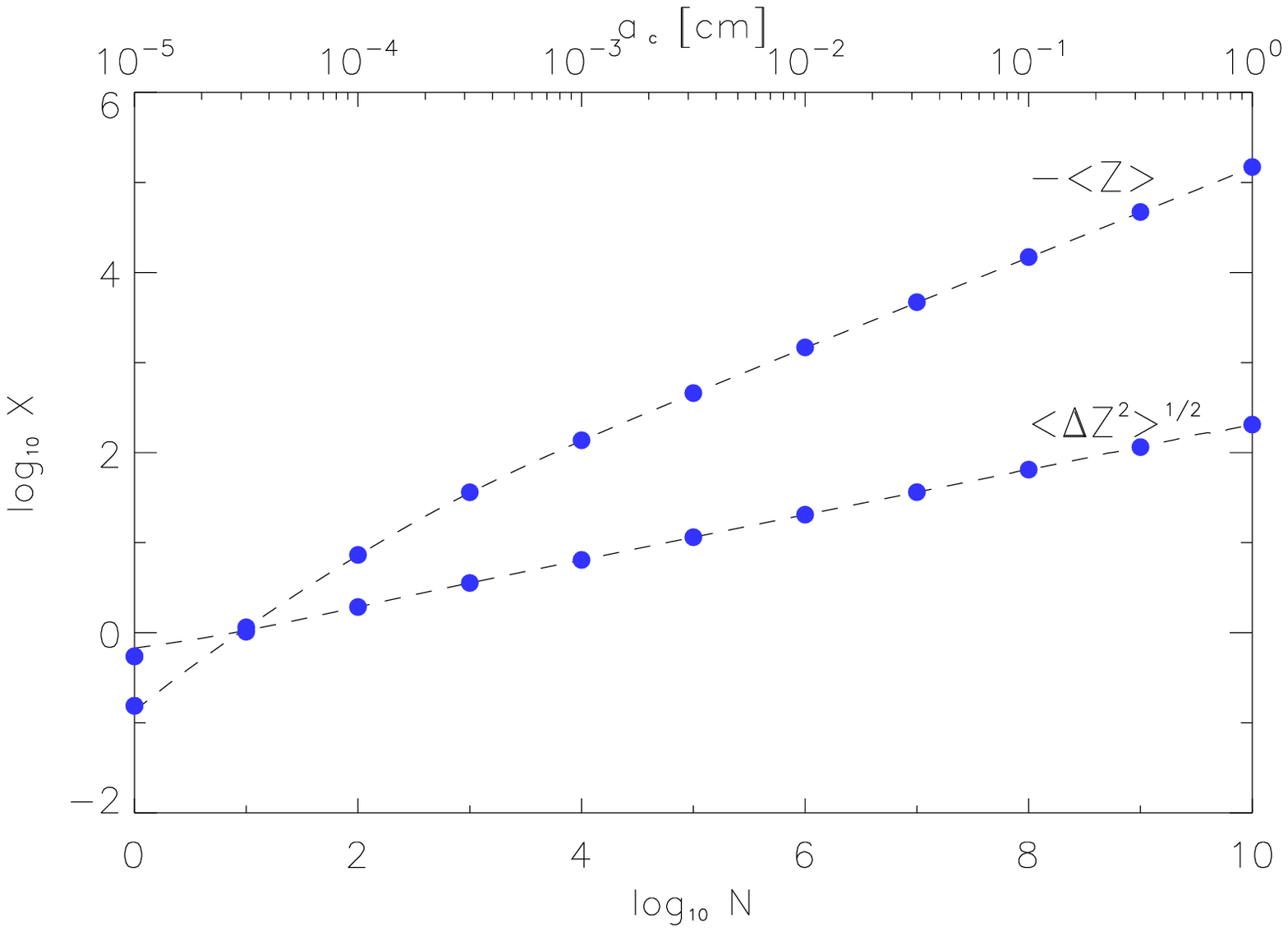}
\includegraphics[width = 9cm]{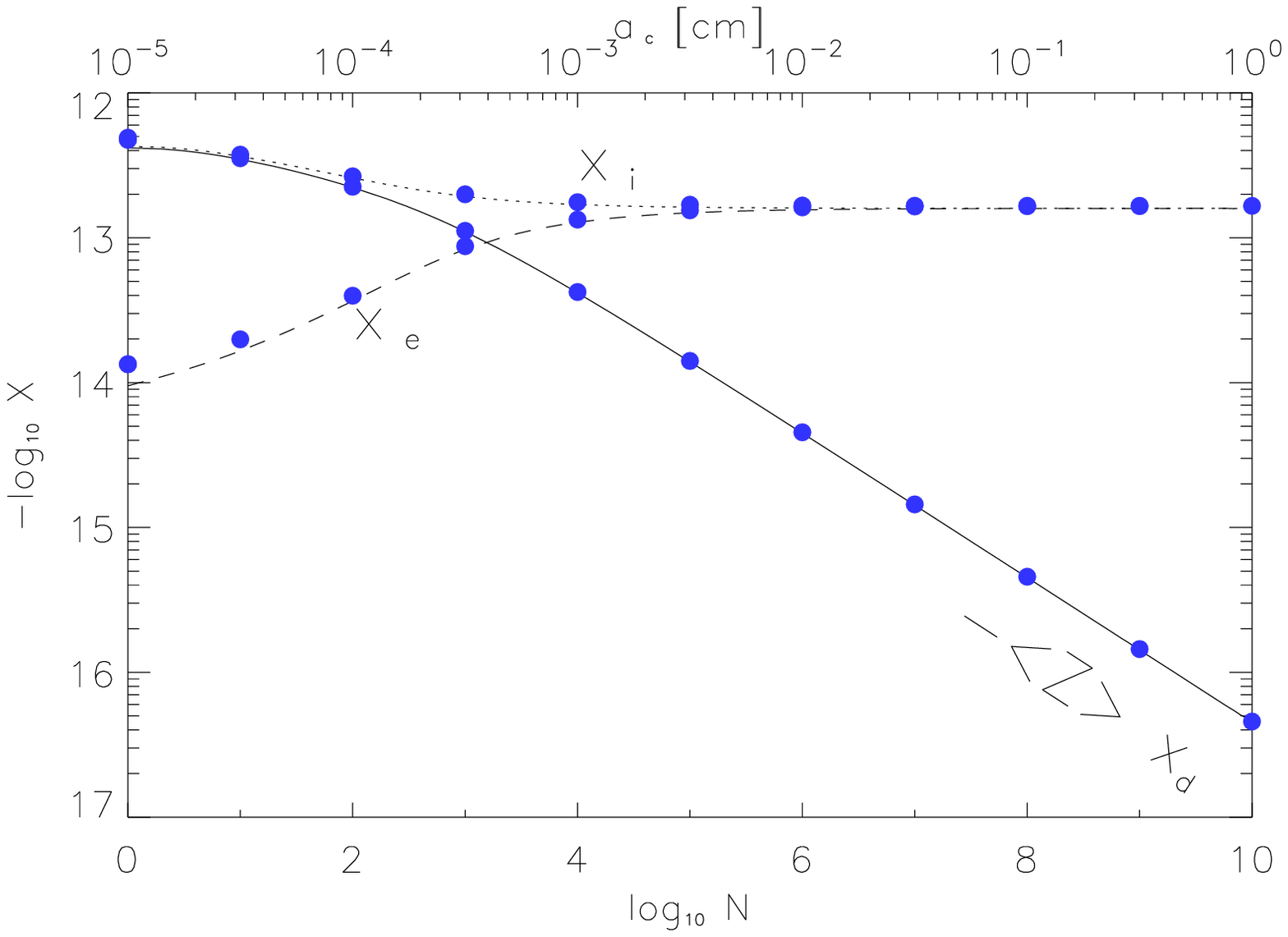}
\caption{Mean charge $\!<\!Z\!>$ of  dust agglomerates and the standard 
deviation $\sqrt{<\!\Delta Z^2\!>}$ as a function of the number of constituent monomers $N$ 
(left panel). The upper abscissa relates $N$ to the characteristic radius $a_c$ of the BCCA 
agglomerate. The electron fraction $x_e$, the ion concentration $x_i$ (with M$^+$ as the 
dominant ion) and the mean charge of the dust $<\!\!Z\!\!>$ per dust agglomerate are shown at 
the right panel using solid, dashed, and dotted lines. The profiles shown are obtained at 
$R = 10 \ \rm AU$ and $z/h_{\rm g}= 0$ applying the ''reduced''  network of the modified 
Oppenheimer-Dalgarno model with adsorption and desorption process switched off. 
$D_{BCCA} = 2$. The lines correspond to the values obtained by numerical integration while 
the values marked by (blue) filled circles are obtained using the semi-analytical solution.}
\label{fig2:sec3}
\end{figure*}
set $\overline{u}_i \approx v_{\rm Mg^+}$. As already mentioned in Okuzumi (2009), the 
contributions of $g(\overline{\beta})$ are negligible for low values of the gas-phase 
recombination rate coeffcients applied.\\
\indent
With respect to the electrostatic polarisation of dust material caused by the electric field of an 
approaching charged particle, agglomerates and spherical particles differ significantly. Polarisation 
contributes to the mean collision rate coefficient $< \! \! \sigma v  \! \! > \ \propto \tilde{J}$ with $\tilde{J}$ 
denoting the so-called ''reduced rate'' (Draine \& Sutin, 1987):
\begin{eqnarray}
\tilde{J}   & = & 
\begin{cases}
 1 + \Big( \frac{\pi}{2\tau} \Big)^{1/2}  & \text{if } \nu = 0  \\
 \Big( 1 - \frac{\nu}{\tau}  \Big) 
 \Big( 1 + \Big( \frac{2}{\tau - 2\nu}\Big)^{1/2} \Big) & \text{if } \nu < 0 \\ 
 \exp\Big\{ - \frac{\Theta_{\nu}}{\tau} \Big\}
 \Big(1 + \Big(4\tau + 3\nu \Big)^{-1/2} \Big)^2 & \text{else } 
\end{cases}  
\label{eq01:sec3} 
\end{eqnarray}
with $\Theta_{\nu} \approx \nu / (1 + \nu^{-1/2})$. $\nu = Ze/q_i$ relates the grain charge $Ze$ to the 
charge of the colliding particle ($q = e$ for singly charged ions and $q = -e$ for electrons) while 
$\tau = akTq^{-2}$ refers to the so-called ''reduced temperature''.
In contrast to spherical dust particles, agglomerates are considered to be hardly affected 
by electrostatic polarisation, see discussion in Okuzumi (2009). For $\tau \to \infty$ and $|\nu| \gg 1$ 
the contributions to $\tilde{J}$ through polarisations become negligible and the expression 
for $\tilde{J}$ reduces to (Draine \& Sutin, 1987)
\begin{eqnarray}
\tilde{J}   & = & 
\begin{cases}
 1 - \frac{\nu}{\tau}  & \text{if } \nu < 0 \\ 
 \exp\Big\{ - \frac{\nu}{\tau} \Big\} & \text{else \ .}
\end{cases}
\label{eq02:sec3} 
\end{eqnarray}
The polarisation also effects the sticking coefficient $s_e(a, Z, T)$ which is defined as 
\begin{equation}
s_e = \frac{\int_0^{\infty} P_e(E) \sigma(E) \exp\{ - \frac{E}{kT}\} dE}{\int_0^{\infty} \sigma(E) \exp\{ - \frac{E}{kT}\} dE} \ ,
\label{eq03:sec3} 
\end{equation}
assuming that the electron obeys a Maxwellian distribution at infinity. We also have to specify 
the grain topology since $P_e$ depends on grain properties such as the grain surface. For example, 
Umebayashi \& Nakano (1980)  assumed planar surfaces to derive $P_e$. We did not investigate this 
question for dust agglomerates and simply assumed that $s_e = 0.3$ is independent of the grain charge. 
In particular, we regard $s_e$ as an adjustable parameter of the simulations presented, which agrees 
with Okuzumi (2009).\\
\indent
When applying the formalism proposed by Okuzumi (2009) to the modified Oppenheimer-Dalgarno 
model, we followed a two step approach. To be able to compare our results with Fig. \ref{fig1:sec3}, 
we applied the same parameter setup. We began with switching off the adsorption and desorption 
process of the modified Oppenheimer-Dalgarno network, i.e., $k_5 = k_6 = 0$ (referring to  Tab. \ref{ref:tab1}), 
and ended in the analytical solution presented in Okuzumi (2009). The results obtained, for example at disc 
midplane at $R = 10 \ \rm AU$, are shown in Fig. \ref{fig2:sec3}. The mean value $<\!\!Z\!\!>$ and standard deviation 
$\sqrt{<\!\!\Delta Z^2\!\!>}$ of the excess charge $Z$ of the dust agglomerate are shown in the left panel while 
the right panel reveals the outcome of the comparison between analytical and numerical values for the electron 
fraction $x[\rm e^-]$, the molecule and metal ion concentration $x[\rm m^+]$,  $x[\rm M^+]$, and the grain charge 
density  approximated by $<\!\!Z\!\!> x_d$. The analytical and numerical values agree excellently, which proves that 
our implementation of Okuzumi's method is correct.\\
\indent
In a final step we constructed an analytical solution that covers $k_5 k_6 \ne 0$. The rates 
associated with thermal adsorption and desorption contribute exclusively to the rate equations 
\begin{figure*}[ht]
\includegraphics[width = 9cm]{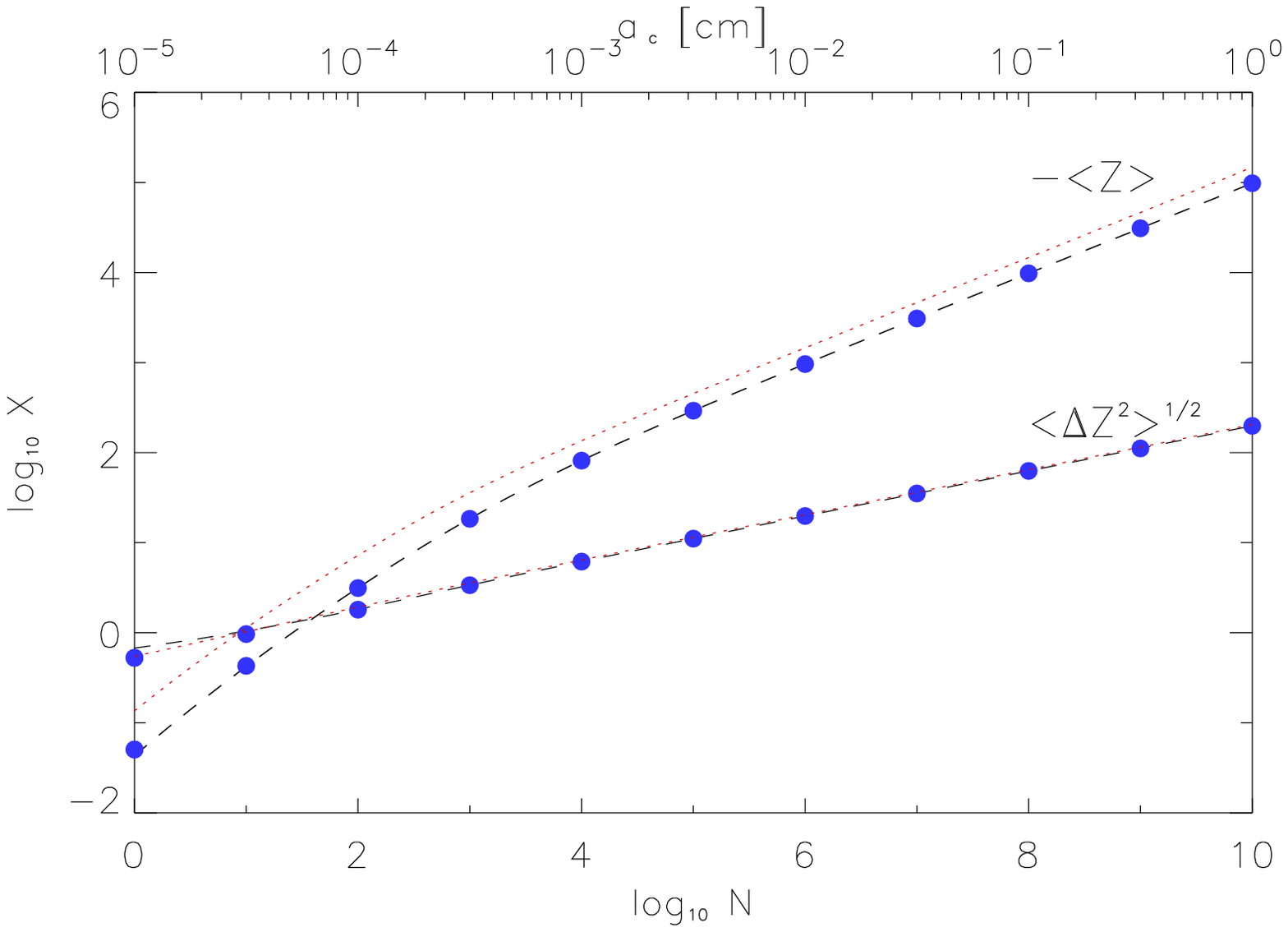}
\includegraphics[width = 9cm]{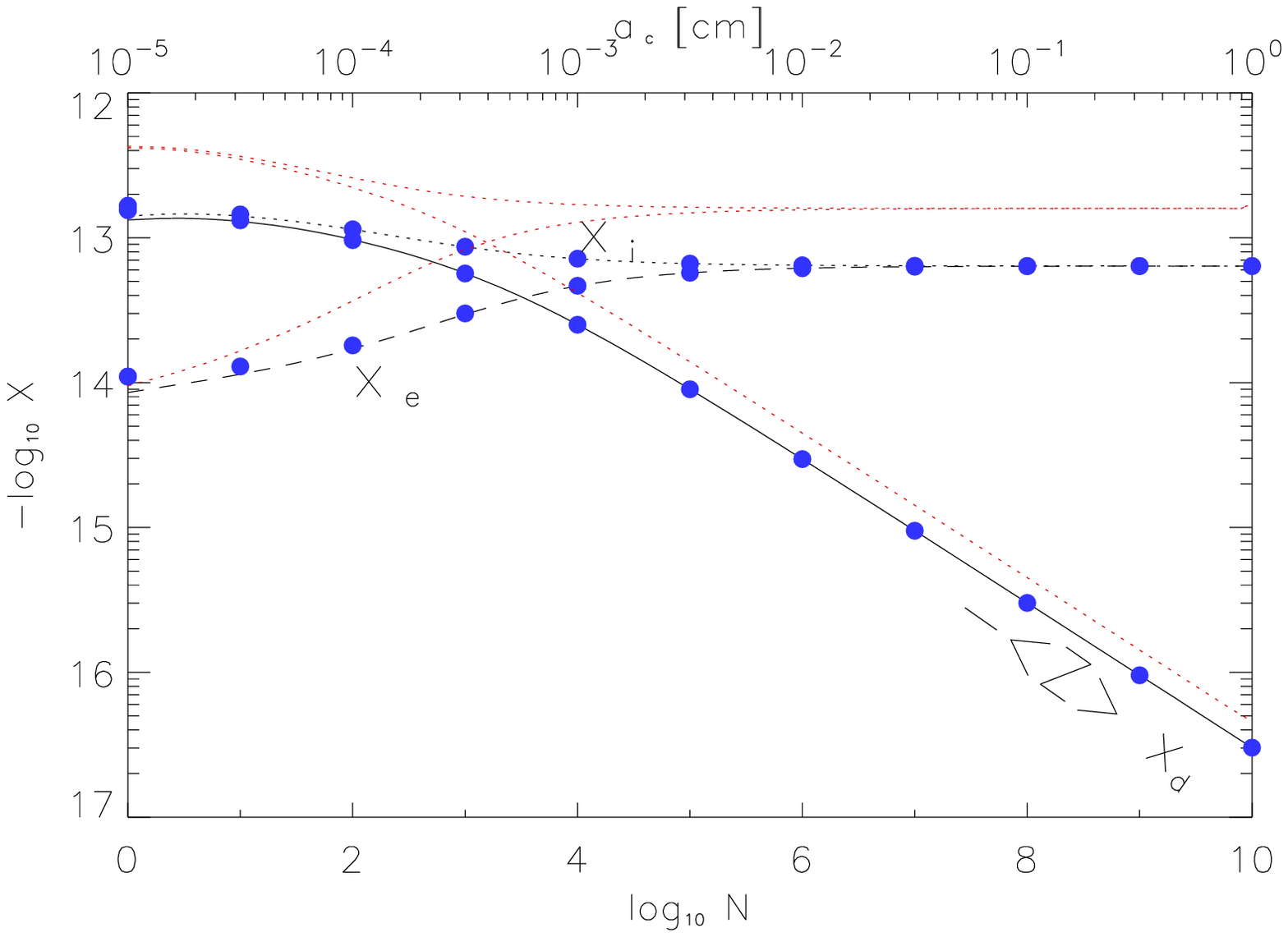}
\caption{Mean charge $\!<\!Z\!>$ of  dust agglomerates and the standard 
deviation $\sqrt{<\!\Delta Z^2\!>}$ as a function of the number of constituent monomers $N$ 
(left panel). The upper abscissa relates $N$ to the characteristic radius $a_c$ of the BCCA 
agglomerate. The electron fraction $x_e$, the ion concentration $x_i$ (with M$^+$ as the 
dominant ion) and the mean charge of the dust $<\!\!Z\!\!>$ per dust agglomerate are shown in 
the right panel using solid, dashed, and dotted lines. The profiles shown are obtained at 
$R = 10 \ \rm AU$ and $z/h_{\rm g} = 0$ applying the modified Oppenheimer-Dalgarno model  
with adsorption and desorption process switched on. $D_{BCCA} = 2$. The lines correspond to the 
values obtained by numerical integration, while the values marked by (blue) filled circles are obtained 
using the semi-analytical solution. The unlabelled (red-coloured) dotted lines are the corresponding 
profiles obtained for the ''reduced'' network of the modified Oppenheimer-Dalgarno model with adsorption 
and desorption process switched off, which we already showed in Fig. \ref{fig2:sec3}.}
\label{fig3:sec3}
\end{figure*}
of the neutral gas-phase components and the mantle components. The kinetic equations 
for the non-neutral gas-components remain unchanged and are (for $t \to \infty$)
\begin{eqnarray}
N[\mathrm{m^+}]  & = & 
k_4  N[\mathrm{m}]  /  \label{eq04:sec3} \\ 
& &  
\left(
k_1 N[\mathrm{M}]  + 
k_2 N_e + 
v_{\mathrm{m^+}} n_d <\! \sigma[\mathrm{m^+}, N_d(Z)]\!>
\right) \nonumber \\
N[\mathrm{M^+}]  & = & 
k_1  N[\mathrm{M}] N[\mathrm{m^+}]  /  \label{eq05:sec3} \\ 
& & \left(
k_3 N_e +
v_{\mathrm{M^+}} n_d <\! \sigma[\mathrm{M^+}, N_d(Z)]\!> 
\right) \nonumber \\
N_e & = & 
k_4 N[\mathrm{m}]  /  \label{eq06:sec3} \\ 
& & \left(
\beta N_i +  
v_{\mathrm{e^-}} n_d <\! \sigma[\mathrm{e^-}, N_d(Z)]\!> 
\right) \nonumber \\
\rm with & & \nonumber \\
n_d & = &  \sum_Z N_d(Z)  \nonumber \\
\beta N_i & = & k_2 N[\mathrm{m^+}]  + k_3 N[\mathrm{M^+}] \nonumber ,
\end{eqnarray}
where $N[\rm x]$ denotes the number density of the component x while $N_{\rm d}(Z)$ 
is the number density of grains with charge $Z$.\\
\indent
However, the (thermal) adsorption/desorption of the neutral gas-phase components 
may have an indirect effect on both the ionized gas-phase species and the charge 
state of the dust since this process may control the amount of gas-phase species 
available. We will demonstrate that the set of Eqs. (\ref{eq04:sec3}) - (\ref{eq06:sec3})
and the charge balance equation
\begin{equation}
N_i - N_e + \sum_Z Z N_d(Z) = 0
\label{eq07:sec3}
\end{equation}
can be solved for a single parameter $<\!\!\!Z\!\!>$ if $N[\mathrm{m}]$ and $N[\mathrm{M}]$ serve 
as constants. Indeed,  $N[\mathrm{m}] = \rm const $ and $N[\mathrm{M}] = \rm const$ 
can be extracted from the modified Oppenheimer-Dalgarno model. Molecular hydrogen serves as 
the neutral molecule component, which is assumed to relax towards hydrostatic equilibrium on 
a dynamical time scale $t_z = \Omega_{\rm K}^{-1}$ (see subsection 2.2). From the kinetic equation 
for the mantle component $\rm gM$ we obtained $N[\mathrm{M}] = k_6 N_{\mathrm{M}} / (k_5 n_d + k_6)$ 
assuming $N_{\mathrm{M}} \approx N[\mathrm{M}]  +  N[\mathrm{gM}]$; reaction $k_7$ 
contributes to $<\!\!Z\!\!> \ > 0$ only.  Solving the set of  Eqs.  (\ref{eq04:sec3}) - (\ref{eq07:sec3}), we 
obtained the semi-analytical solution for  
$N_\infty[\mathrm{m^+}]$, $N_\infty[\mathrm{M^+}]$, $N_\infty[\mathrm{e^-}]$, and 
$<\!\!Z\!\!>_\infty$ (with $N_\infty$ denoting the asymptotic limit for $t \to \infty$). The results 
obtained by numerical integration as well as by applying the semi-analytical approach are 
presented in Fig. \ref{fig3:sec3}. The agreement is excellent. We notice that 
the values associated with the ''reduced'' network of the modified Oppenheimer-Dalgarno model 
(i.e., with adsorption and desorption switched off $k_5 = k_6 = 0$) differ significantly from those of 
the ''full network'' with adsorption and desorption switched on ($k_5k_6 \ne 0$). For $N = \rm const$, 
the ''reduced'' network produces higher mean excess charges (associated with unlabelled 
red-coloured dotted lines in Fig. \ref{fig3:sec3}) than the same network does for $k_5k_6 \ne 0$. 
The same applies for the electron concentration $x_e$, $<\!\! Z\!\!> x_{\rm d}$ and the ion concentration 
$x_i$ with $\rm m^+$ becoming the dominant ion in the modified Oppenheimer-Dalgarno model. 
\begin{figure*}[ht]
\includegraphics[width = 9cm]{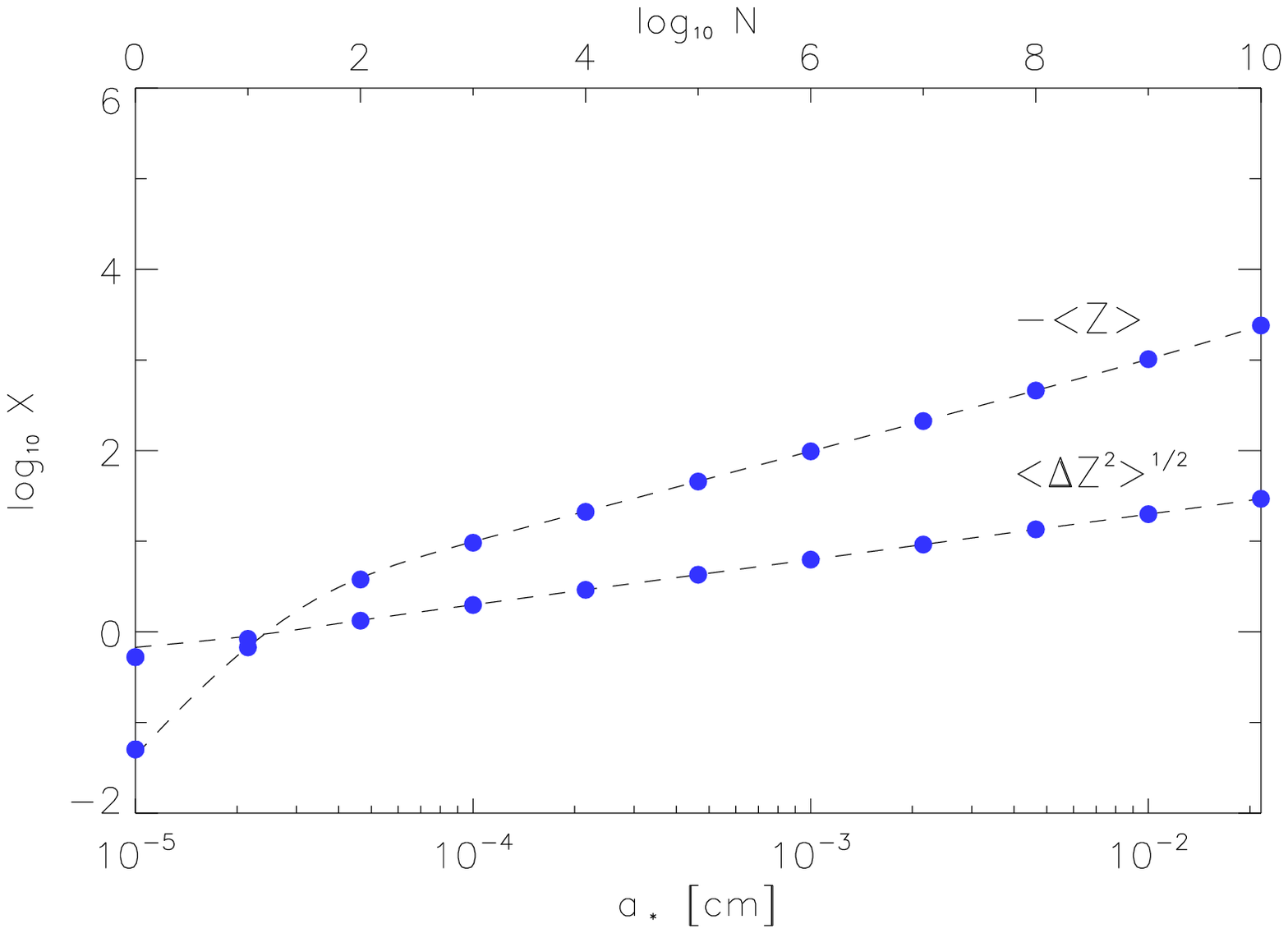}
\includegraphics[width = 9cm]{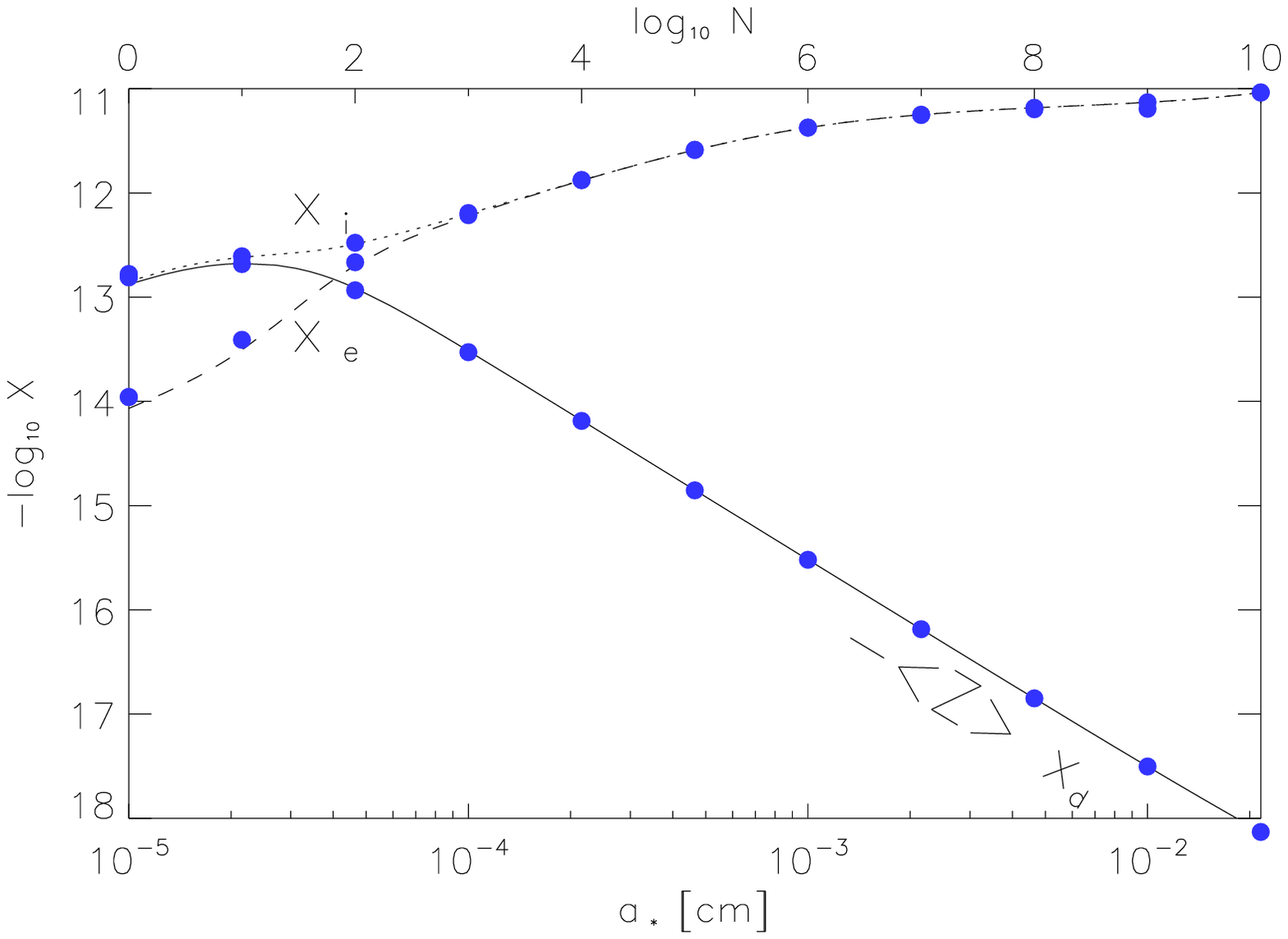}
\caption{Mean charge $\!<\!Z\!>\!$ of spherical dust grains and the standard 
deviation $\sqrt{<\!\Delta Z^2\!>}$ as a function of the grain radius $a_{\ast}$ (left panel). 
The upper abscissa relates $a_{\ast}$ to a BCCA agglomerate of $N$ monomers 
assuming $m_{\rm d} = \rm const$. The electron fraction $x_e$, the ion concentration 
$x_i$ (with m$^+$ as the dominant ion) and the mean charge of the dust $<\!\!Z\!\!>$ per 
dust grain are shown in the right panel using solid, dashed, and dotted lines. The profiles 
shown are obtained at $R = 10 \ \rm AU$ and $z/h_{\rm g} = 0$ applying the 
modified Oppenheimer-Dalgarno network for spherical grains. The lines correspond to the values 
obtained by numerical integration while the values marked by (blue) filled circles are 
obtained using the semi-analytical solution. To be compared with Fig. \ref{fig3:sec3}, which 
is associated with the corresponding modified Oppenheimer-Dalgarno model for BCCA agglomerates.}
\label{fig4:sec3}
\end{figure*}
Before we draw our conclusions, we investigate this problem for the Umebayashi-Nakano 
network, too. In particular, we analysed the effect of the dominant molecular ion $\rm HCO^+$ 
on the grain charging under the assumption that the gas-phase metal is significantly depleted 
through the action of the thermal adsorption. We therefore have modified the Umebayashi-Nakano 
network by adding the thermal adsorption/desorption process for $\rm Mg$.\footnote{Technically speaking, 
we included two additional rate equations for the new components $\rm Mg$ and $\rm gMg$.} The 
results obtained for calculations with rate coefficients $k_5 = k_6 = 0$ and $k_5 k_6 \ne 0$, respectively, 
are quite different compared with the results discussed above. Switching on the adsorption/desorption 
process for metals (i.e., $k_5 k_6 \ne 0$) leads to higher values for $x_e$ and $x_i$ and to higher negative 
mean charges on dust. However, the changes are just in the range of a few percent and therefore less severe 
than the changes observed in the modified Oppenheimer-Dalgarno network. 
These different features of the modified Oppenheimer-Dalgarno and the Umebayashi-Nakano 
network can be traced down to differences in the thermal velocity of dominant gas-phase ions. Under metal 
rich conditions, $\rm Mg^+$ is the dominant ion in the modified Oppenheimer-Dalgarno network with 
$k_5 = k_6 = 0$, while $\rm H_2^+$ for $k_5 k_6 \ne 0$. Since $\rm H_2^+$ has a much higher thermal velocity 
than $\rm Mg^+$ it is quickly captured by grains. Applying the same conditions,  $\rm HCO^+$ becomes the 
dominant ion in the modified Umebayashi-Nakano network with a thermal velocity lower 
than that of $\rm Mg^+$, which is the dominant ion for $k_5 = k_6 = 0$. Because of the slightly reduced contributions 
owing to collisions between grains and $\rm HCO^+$, the values for $x_e$ and $x_i$ are increasing. The 
drop/rise in the mean grain charge observed in the modified Oppenheimer-Dalgarno and Umebayshi-Nakano 
model also originates in differences between the thermal velocities of the dominant gas-phase ions:  the higher 
the thermal velocity of the dominant ion, the less the negative mean charge on grains. Considering all this, we 
find indeed that the adsorption/desorption of metals effects the mean value for grain charges, the electron fraction, 
and the ion abundances. Whether or not the value increases/decreases depends on the ratio of the thermal 
velocities for the metal ion and the dominant molecular ion.

\subsection{Grain chemistry with spherical grains}\label{sec3:sub3}
Instead of agglomerates we now consider spherical dust grains of a given radius $a_0$. 
Applying the modified Oppenheimer-Dalgarno model, we aim to determine their mean excess 
charge $<\!\!Z\!\!>$. If we neglect the effect of polarisation, $<\!\!Z\!\!>$ can be calculated semi-analytically 
following the same procedure detailed in the previous subsection. We simply repeated 
the procedure and obtained an identical set of equations to Eqs. (\ref{eq04:sec3}) - (\ref{eq07:sec3}) 
by replacing 
\begin{eqnarray*}
 N_{\rm d}(Z) & \to & N_{\rm d0}(Z) \\
 n_{\rm d}      & \to & n_{\rm d0} \ ,
\end{eqnarray*}
where $N_{\rm d0}(Z)$ and $n_{\rm d0}$ denote the number density of spherical grain particles 
with excess charge $Z$ and the total number density of spherical grains, respectively. The results obtained 
are shown in Fig. \ref{fig4:sec3}.  In the left panel the mean grain charge and its the standard deviation 
are plotted as a function of the grain radius, while the fractional abundances obtained are 
shown in the right panel. We first note the excellent agreement between the numerical 
and semi-analytical values. We can compare the results with those obtained for agglomerates 
when identifying the grain radius with $a_{\ast}$, see Eq. (\ref{eq18:sec2}). Assuming 
monomers of size $a_0 = 10^{-5} \ \rm cm$, $a_{\ast}$ scales with the number $N$ of constituent 
monomers as shown in the upper x-axis of Fig. \ref{fig4:sec3}. A compact spherical dust grain of, 
for example, $a_{\ast} = 10^{-3} \ \rm cm$ corresponds to an agglomerate made by $N = 10^6$ monomers 
of size $a_0 = 10^{-5} \ \rm cm$. Comparing Fig. \ref{fig3:sec3} with Fig. \ref{fig4:sec3}, we find that the 
mean charge carried by compact spheres is much lower than the value obtained for 
agglomerates. For $N = 10^6$, it is about one order of magnitude. 
This is not caused by the changes in the projected surface area associated with the transition 
between fractal and compact grain topology. It is because of the governing ion-electron regime 
in which the mean grain charge is proportional to the radius, i.e.  $a_c/a_{\ast} = N^{1/6}$. 
For the same reason, the modified Oppenheimer-Dalgarno network applied to compact spherical 
grains produces a much larger amount of  gas-phase ions m$^+$ and e$^-$ and a much lower 
dust charge density $<\!\!Z\!\!> x_{\rm d}$ than the values obtained for agglomerates. We conclude 
that dust agglomerates have a higher charge-to-mass ratio carrying more charges than the 
corresponding compact spheres.\\
\indent
We completed the cycle by taking the polarisation of compact spherical grains into account. 
We cannot apply the equations presented in Okuzumi (2009) though, which were derived under the 
assumption of  Eq. (\ref{eq02:sec3}). Substituting Eq. (\ref{eq02:sec3}) with Eq. (\ref{eq01:sec3}) 
leads to different expressions for $<\!\!\sigma_{\rm di}\!\!>$ and $<\!\!\sigma_{\rm de}\!\!>$. This may also 
affect the Gaussian distribution for $N_{\rm d0}(Z)$ since the coefficient $W(Z)$ of the corresponding 
first-order equation changes (cf appendix in Okuzumi, 2009). However, we did not consider the 
semi-analytical approach and performed a numerical integration of the rate equations taking Eq. 
(\ref{eq01:sec3}) into account. The values for $<\!\!Z\!\!>$ and $<\!\!\Delta Z^2\!\!>$ obtained under the 
assumption of Eq. (\ref{eq02:sec3}) were used as an initial guess for the range of grain charges 
$[Z_{\rm min}, Z_{\rm max}]$ considered. If necessary, the charge range was modified and the 
integration was repeated again until the calculated mean value $<\!\!Z\!\!>$ fitted the peak value of 
the Gaussian profile for $N_{\rm d0}(Z)$. The values obtained by numerical integration match the 
analytical values associated with the non-polarised case very well. The corresponding profiles show 
no distinguishing features when applying the logarithmic scaling of Fig. \ref{fig4:sec3}. Changing to 
linear scales, we observe for sizes $a < 10^{-4} \ \rm cm$ that the analytical values (i.e. the model 
neglecting the polarisation) obtained for $<\!Z\!>$ and $<\!\Delta Z^2\!>^{1/2}$ are slightly lower than 
those obtained by numerical integration (model with polarisation). We conclude that at least for our 
particular disc model polarisation has only a minor effect on grain charging.

\section{Ionisation rates}\label{sec4}
Because of the simplicity of the chemical network applied, all processes that might result 
in ionising disc material could not be treated individually. Instead, the contributions of the 
dominant processes were combined into an effective ionisation rate $\zeta$ with 
$k_4 = \zeta$ (see Tab. \ref{ref:tab1}).\\
\indent
We  included the ionisation of the dominant molecule by incident X-rays that originate in the 
corona of the central T Tauri star. The corresponding X-ray ionisation rates $\zeta_{\rm xray}$  
can be obtained from radiative transfer calculations. The results of these calculations are 
presented in Igea \& Glassgold (1999), who modelled the transport of stellar, low-energy X-rays 
($< 20 \ \rm keV$). The ionisation rates  are published primarily for the minimum solar nebula 
model of Hayashi et al. (1985), which corresponds to the static disc model discussed above. 
Taking the photoelectric absorption and Compton scattering into account, Igea \& Glassgold 
specified absorption-dominated and scattering-dominated regimes, respectively. Bai \& Goodman 
(2009) presented a best fit to the X-ray ionisation rates of Igea \& Glassgold (1999), which 
facilitates employing the results of Igea \& Glassgold (1999).\\
\indent
Igea \& Glassgold (1999) presented ionisation rates obtained for power indices $p_s$ of the 
gas surface density $\Sigma_{\rm g}$ except for $p_s = -3/2$.  For a specified orbital radius 
at $R = 1 \ \rm AU$ they report that the deviation in the X-ray ionisation rate is generally less 
than a factor of two. Igea \& Glassgold stated that the ionisation rates plotted vs the vertical 
column density ''manifest a universal form'' that is independent of the surface density profiles 
considered. Glassgold et al. (2000) note that this ''scaling is also found at other 
disk radii and is insensitive to changes in disk parameters such as disk mass, surface density, 
and slope $q.$''\footnote{In Glassgold et al. (2000) $q$ refers to the power index associated 
with the radial variation of the gas density.} However,  X-ray ionisation rates obtained from 
X-ray transfer calculations have not been made publicly available yet for profiles of the gas surface 
density different from the static model of  Hayashi et al. (1985).\\
\indent
Instead of a complex X-ray radiative transfer model, we attempted to apply a simple ray 
tracing model neglecting scattering effects. The techniques employed are described 
in detail in, e.g., Fromang et al. (2002) and Ilgner \& Nelson (2006a). For static and 
stationary disc models of minimum mass solar nebula type, these calculations can be performed 
in straight line because of the non-discrete nature of the local gas density $\varrho_{\rm g}$. We find 
significant differences when comparing the X-ray ionisation rates we obtained with the rates 
of  Igea \& Glassgold (1999). Similar findings are reported in Bai \& Goodman (2009). The effect 
of scattering may explain these differences, but additional radiative transfer calculations would 
make valuable contributions.\\
\indent
Acknowledging the potential X-ray scattering may have on the ionisation rate, we applied 
X-ray ionisation rates based on the simulations of Igea \& Glassgold (1999). In particular, 
we applied the fitting formula associated with Eq. (21) of Bai \& Goodman (2009), which is 
valid for  plasma temperature $k_{\rm B}T = 3 \ \rm keV$. However, because of the problem 
discussed above we regard the X-ray ionisation rates  $\zeta_{\rm Xray}$ as an order of magnitude estimate.\\
\indent
We also considered cosmic ray particles as a source for ionisation with ionisation rates $\zeta_{\rm CR}$ 
given by eq. (19) of Sano et al. (2000). We adopted the nominal value $\zeta_0 = 10^{-17} \ \rm s^{-1}$ 
of galactic cosmic rays associated with unattenuated penetration of cosmic ray particles. We neglected 
the minor contributions from the decay of radioactive elements. Taking the X-ray and the cosmic ray ionisation 
rates at face value, X-rays dominate the cosmic ray ionisation for regions close to the disc surface $z/h_{\rm g} = 3$. 
At higher optical depths, ionisation through cosmic rays is the dominant source of ionisation. We regard 
X-rays as our standard ionisation source while ionisation through cosmic rays is restricted to specific disc regions.  
For inner disc regions $R < R_1$ we excluded cosmic rays and set $\zeta = \zeta_{\rm Xray}$. We furthermore 
assumed that the shielding effect of the T Tauri winds peters out between $R_1 < R < R_2$. We assumed a 
simple exponential decay
\begin{equation}
\zeta = \zeta_2 \exp \left\{ 
- \left( \frac{R - R_2}{R_1 - R_2}  \right)^2 \ln \frac{\zeta_2}{\zeta_1}\right\}
\label{eq01:sec4}
\end{equation}
to mimic this effect for that particular transition region from X-ray dominated to 
cosmic ray dominated regions with $\zeta_1 = \zeta_{\rm Xray}(R_1, z)$ 
and $\zeta_2 = \zeta_{\rm CR}(R_2, z)$.

\section{Numerical method}\label{sec5}
Although we applied a two-dimensional geometry $(R, z)$, the dynamics of the gas-dust disc 
refers to a one-dimensional description. Therefore, no serious attempt was made to couple 
the dynamical and the chemical evolution of the gas-dust disc by means of conventional 
operator splitting techniques. The coupling is actually not required since the grain 
charging operates on a much shorter time scale than the disc dynamics. However, we 
aimed to simulate the disc dynamics to halt the simulation at different 
evolutionary stages. Assuming that the local disc structure is set up instantaneously, the kinetic 
equations are integrated numerically for $t = 10^4 \ \rm yr$ at each point of the two-dimensional 
domain. In parallel we determined the limiting behaviour (i.e., for $t \to \infty$) following the 
semi-analytical approach.\\ 
\indent
In order to simulate the disc dynamics we continued by applying the so-called ''method of line'' (MOL) 
approach as we did, e.g., in Ilgner \& Nelson (2006b), to solve the set of coupled advection-diffusion 
equations  (\ref{eq04:sec2}) and (\ref{eq05:sec2}). We applied a radial grid 
$\{ R_i: i = 1, \cdots, n_{\rm R} \}$ with an equidistant mesh size near the inner boundary and 
non-equidistant mesh sizes elsewhere. The radial domain $R = [0.1, 10^3] \ \rm AU$ was discretised 
using $n_{\rm R} = 222$ grid cells. The transport terms were discretised in the flux form to secure 
the conservation of mass. We applied the concept of staggered mesh on which the scalars 
and vectors representing the discrete values of the independent variables are centred at 
different locations. Scalars are located at zone centers while vectors are defined at zone 
edges. We limited the maximum absolute step size corresponding to a Courant-Friedrichs-Lewy 
(CFL) number of 0.5.\\
\indent
We applied the Crank-Nicholsen scheme when discretising the diffusion operator 
in Eq. (\ref{eq05:sec2}), as we did in Ilgner \& Nelson (2006b). Regarding the advection operator, 
we employed a nonlinear advection discretisation scheme that prevents negative values 
for the individual mass densities. In particular, we employed an upwind biased discretisation 
scheme with flux limiting. We opted for the flux limiter function proposed by Koren (1993), 
which - depending on the smoothness of the profile for the quantity to be advected  - is associated 
with the accuracy of a third-order scheme.\\
\indent
The kinetic equations associated with the chemical model are given by a set of ordinary 
differential equations (ODE). We employed stiff ODE integrators to obtain stable numerical 
solutions. Stiffness was introduced for both physical and numerical reasons because of, e.g., the 
huge range of time scales for the chemistry and because of the semi-discretisation associated 
with the MOL approach. In particular, we used standard stiff ODE solvers based on linear 
multistep methods, namely the backward differentiation formulas.\\
\indent
Concerning the semi-analytic approach, we employed the procedure proposed by Muller (1956) 
to find the roots of the polynomial associated with Eqs. (\ref{eq04:sec3}) -  (\ref{eq07:sec3}).

\section{Results}\label{sec6}
In this section we present the results of simulations that examined the modified Oppenheimer-Dalgarno network 
under non-static disc conditions. We examined the modified Oppenheimer-Dalgarno network under stationary 
disc conditions which we later used for a comparison with the results obtained for non-stationary discs. 
We evolved the disc chemistry for $t = 10^4 \ \rm yr$. However, by applying the semi-analytic approach 
discussed in Sec. \ref{sec3}, we are in the favourable position to calculate the long-term behaviour of the 
chemical models a priori. That is why we can prove the time interval needed to establish an equilibrium 
solution, which is (under the conditions applied) shorter than the dynamical time scale $t_{\rm K}$.\\
\indent
Apart from the radius $a_{\ast}$ and $a_{\rm c}$, respectively, we kept all other parameters fixed:  
$M_{\ast} = M_{\odot}$,  $\mu = 2.33$, $\alpha = 10^{-3}$, $\varrho_p = 1.0 \ \rm gcm^{-3}$, 
$\rm Sc = 1$, $h_{0, \rm g} = 3.33 \times 10^{-2} \ \rm AU$, and $|\dot{M}_{\rm d}| = 10^{-10} \ \rm M_{\odot}yr^{-1}$. 
We adopted values $L_{\rm X} = 10^{29} \rm \ erg \ s^{-1}$ and $k_{\rm B} T_{\rm X} = 3 \ \rm keV$ for the 
total X-ray luminosity and the plasma temperature while the transition between X-ray dominated and cosmic 
ray dominated ionisation region was fixed to $R_1 = 25 \ \rm AU$ and $R_2 = 30 \ \rm AU$ (see Sec. \ref{sec4}). 
The values for $R_1$ and $R_2$ were motivated by Perez-Becker \& Chiang (2011), who discussed the problem 
of so-called ''sideways cosmic rays''. Regarding the parameter setup associated with the chemical model we fixed 
the sticking coefficients $s_{\rm e} = 0.3$ and $s_{\rm i} = 1.0$, which is in line with the values Okuzumi (2009) 
applied. As we did in Ilgner \& Nelson (2006a), we  assumed the standard value of $1.5 \times 10^{15} \ \rm cm^2$ 
for the surface density of sites, while the value of energy $\Delta E_{\rm S}$ transferred to the grain particle caused 
by lattice vibration was approximated by $2.0 \times 10^{-3} \ \rm eV$. We approximated 
the dissociation energy $D$ for neutral gas-phase components by its binding energy for 
physical adsorption. Complementary to the results discussed in Sec. \ref{sec3}, we applied 
a conservative estimate of $x_{\rm M} = 10^{-11}$ in this section.\\
\indent
We are aware that the disc model we applied serves as a toy model and does not cover the complexity of the 
dust-gas dynamics. $\alpha = 10^{-3}$ is indeed a very naive approximation 
for the turbulence structure in discs and, in particular, in dead zones. However, recent simulations 
have shown that sound waves propagating into dead zones promote the diffusion of dust 
particles (Okuzumi \& Hirose 2011). Lacking reliable models, we excluded grain size 
distributions. Instead we considered mono-sized particles and agglomerates and set limits on $a_{\rm c}$ 
and $a_{\ast}$. We chose $a_{\rm \ast} = [4.64 \cdot 10^{-5}, 10^{-3}] \ \rm cm$, which corresponds to agglomerates 
made of $N = [10^2,10^6]$ monomers of size $a_0 = 10^{-5} \ \rm cm$. 
This limitation is motivated by recent studies of Okuzumi et al. (2011ab). Evaluating a set of growth criteria 
for BCCA agglomerates, they made predictions on the location of so-called ''frozen zones''. In frozen zones the 
electro-static barrier inhibits the agglomerates to growth further. The authors showed that planet forming regions 
are associated with frozen zones. The inferred sizes of the agglomerates are 
$[4 \cdot 10^{-5}, 3\cdot 10^{-2}] \ \rm cm$, depending on the local position.

\subsection{Stationary disc}\label{sec6:sub1}
\begin{figure*}
\includegraphics[width = 8.5cm]{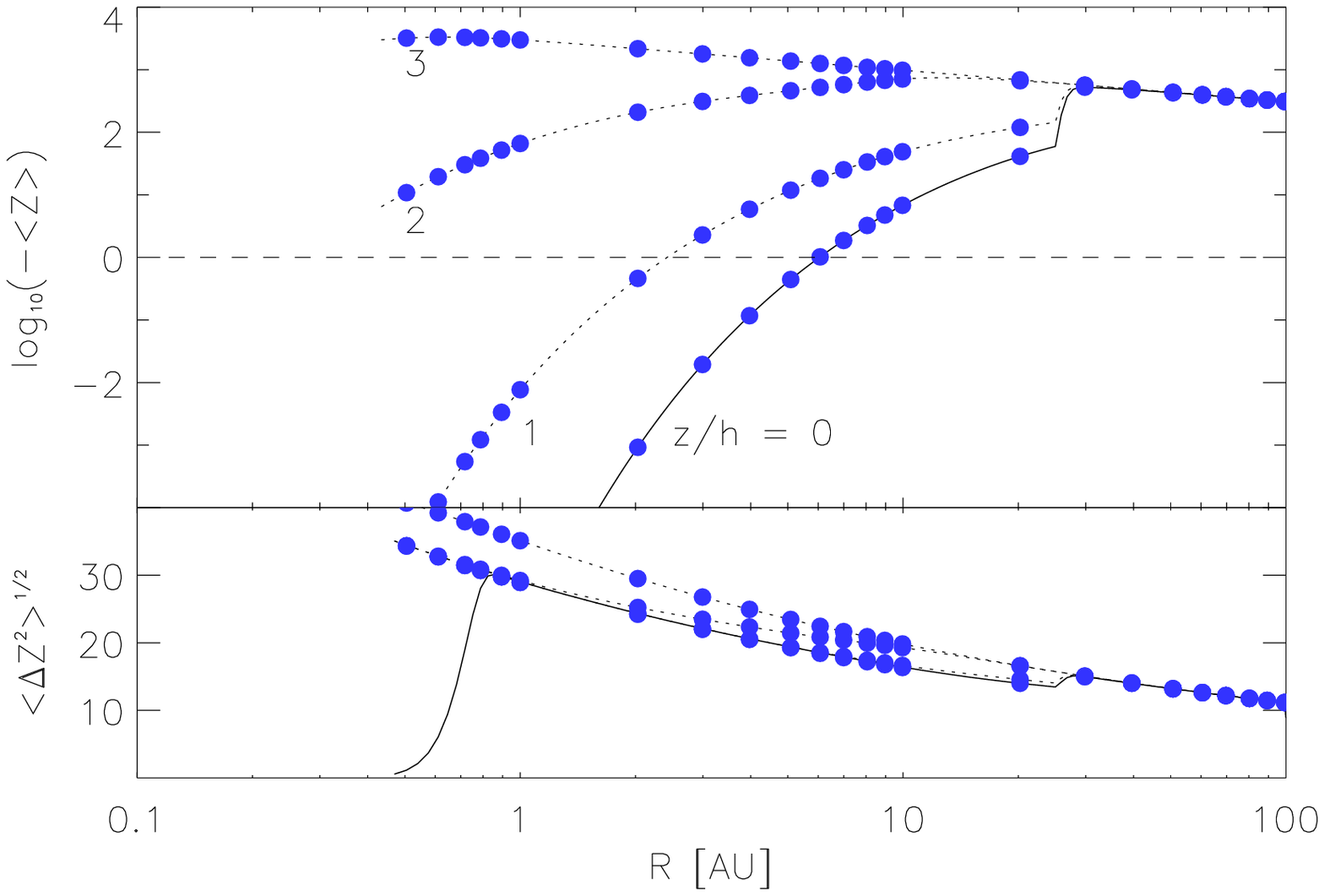}
\includegraphics[width = 8.5cm]{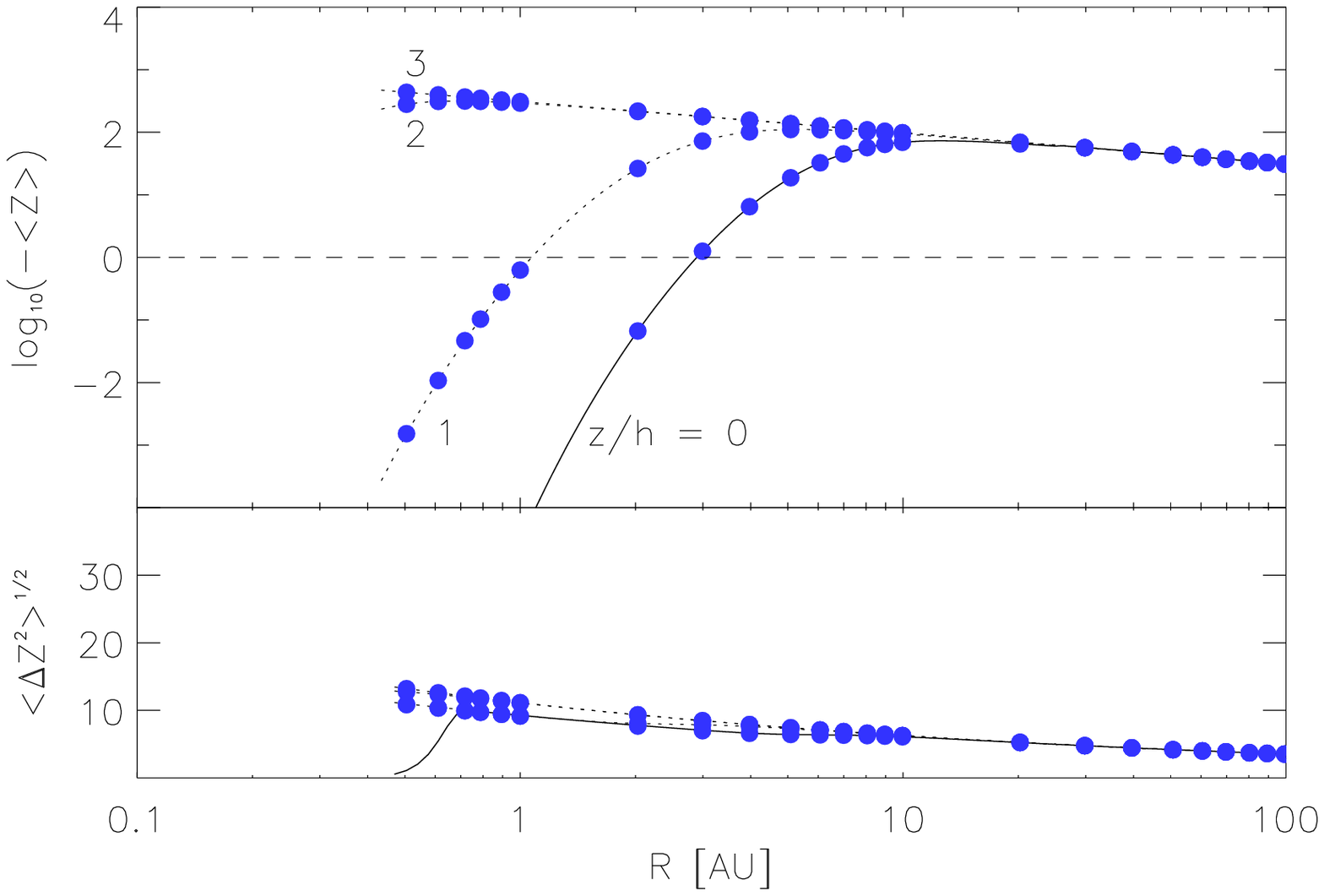}\\
\includegraphics[width = 8.5cm]{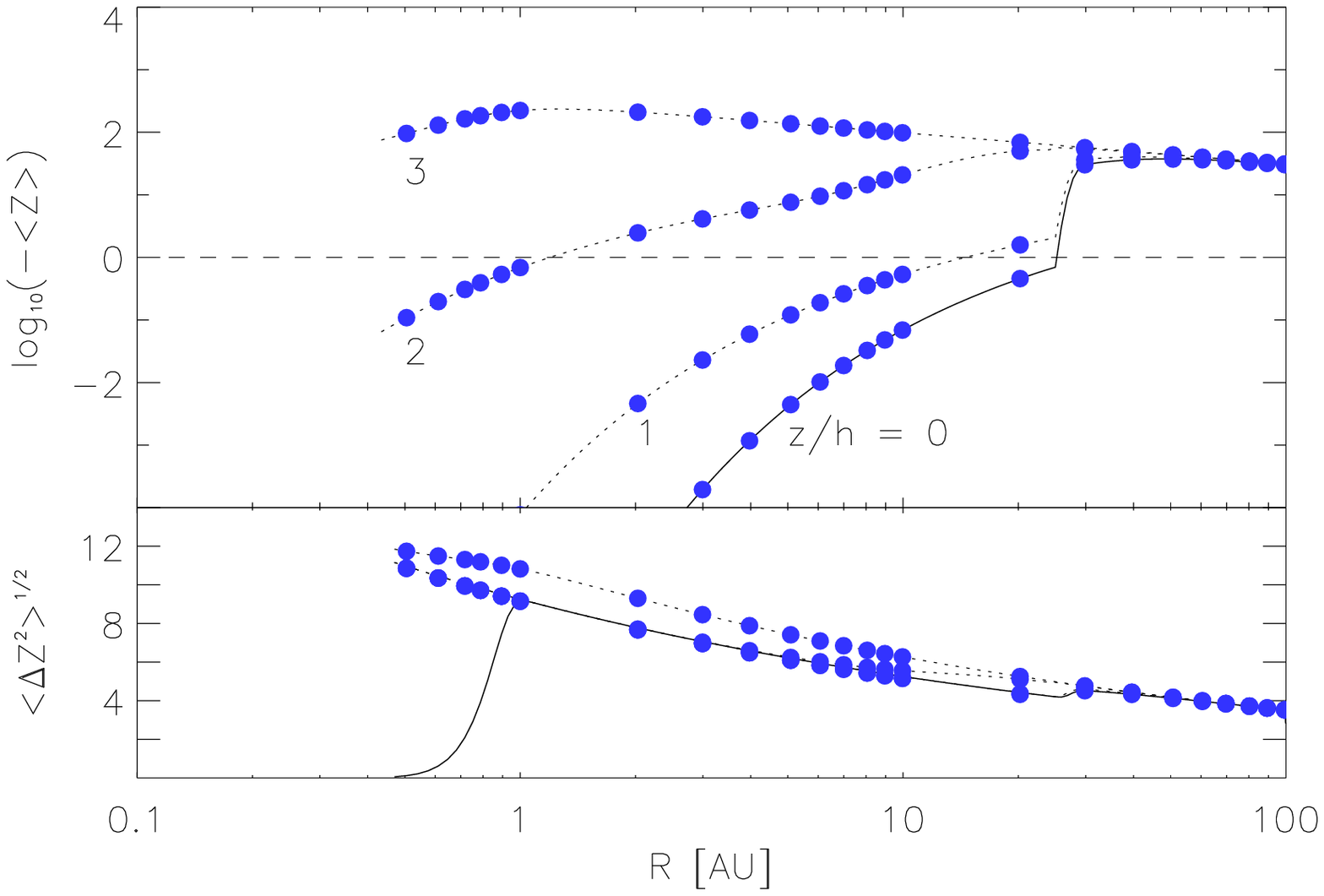}
\includegraphics[width = 8.5cm]{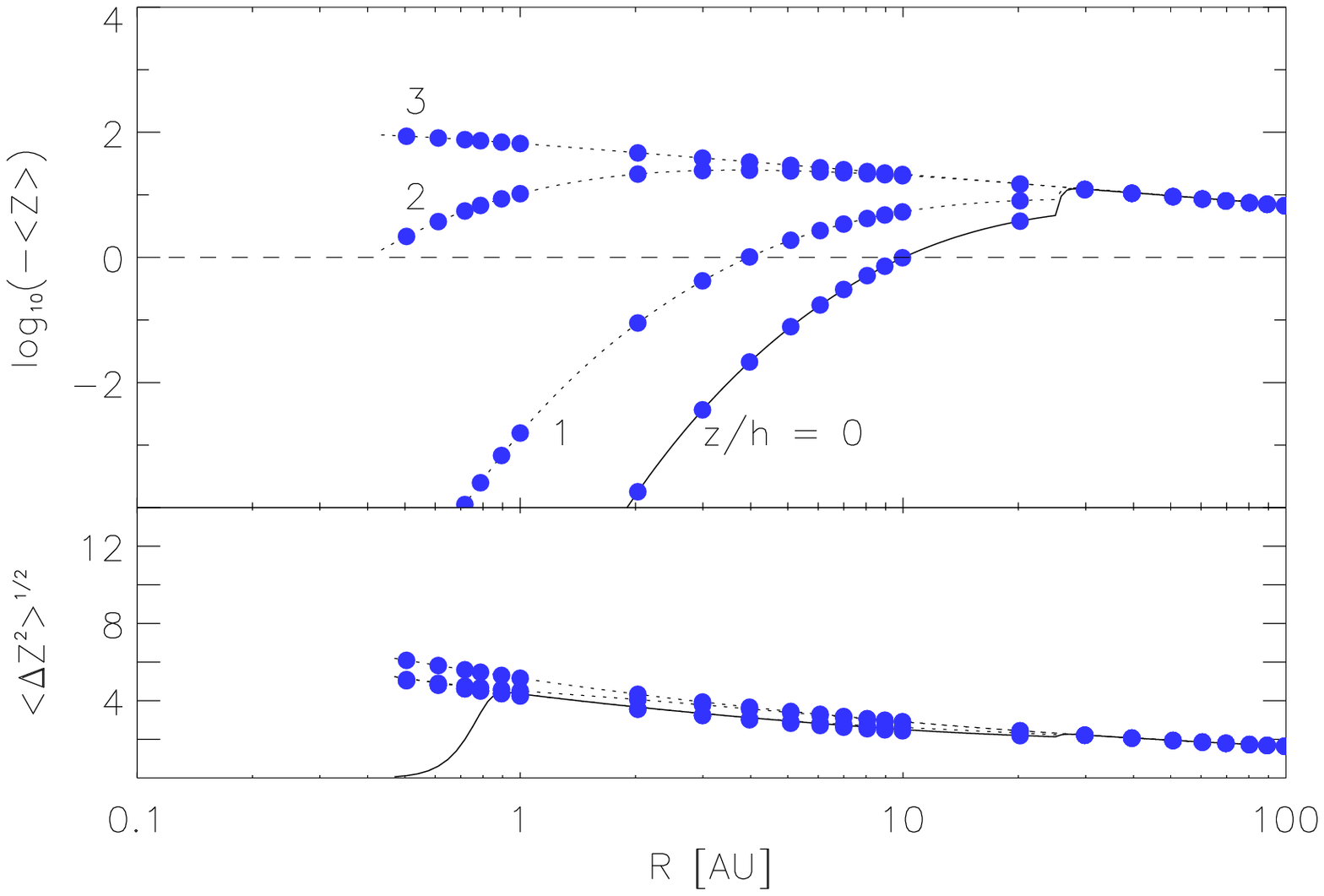}\\
\includegraphics[width = 8.5cm]{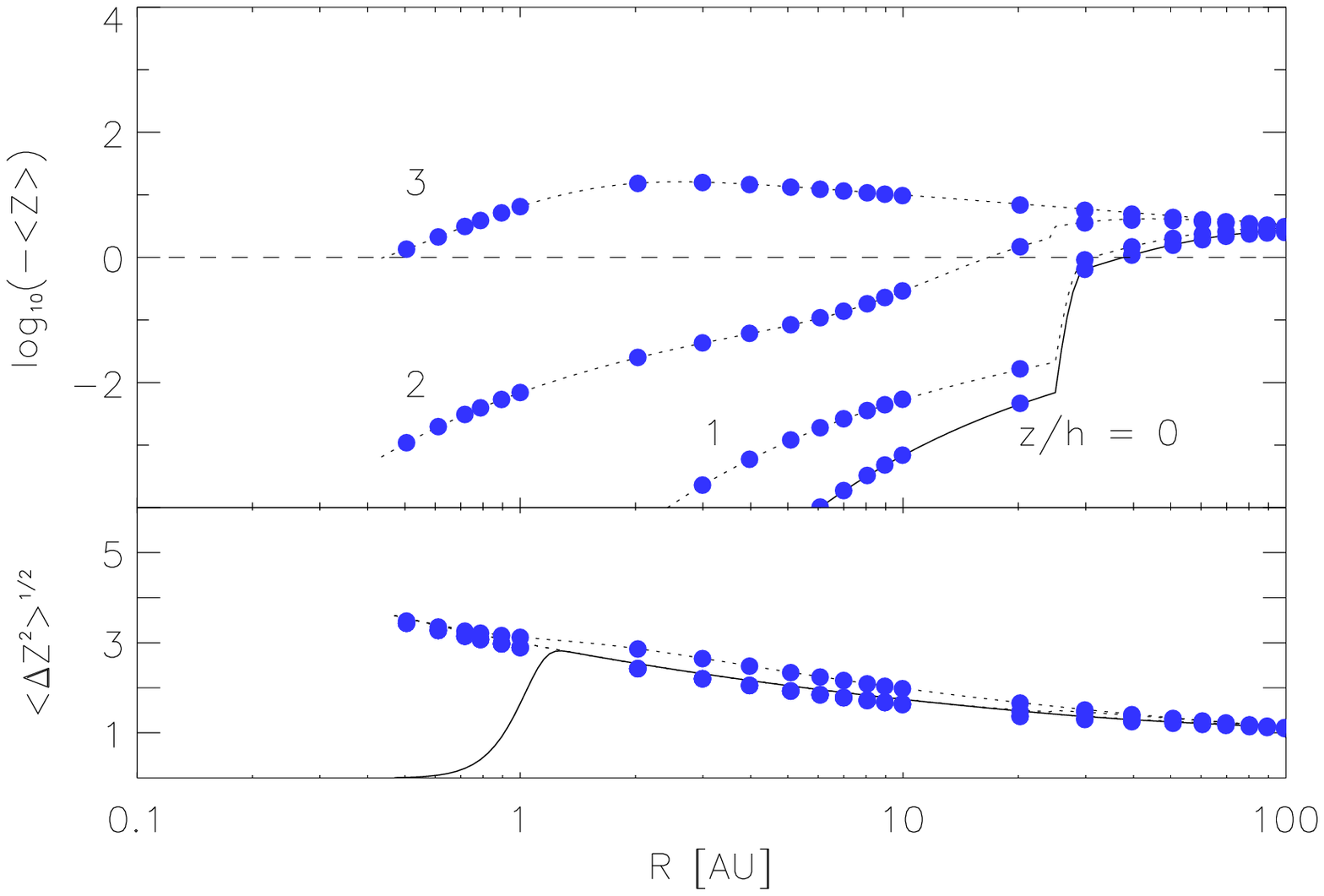}
\includegraphics[width = 8.5cm]{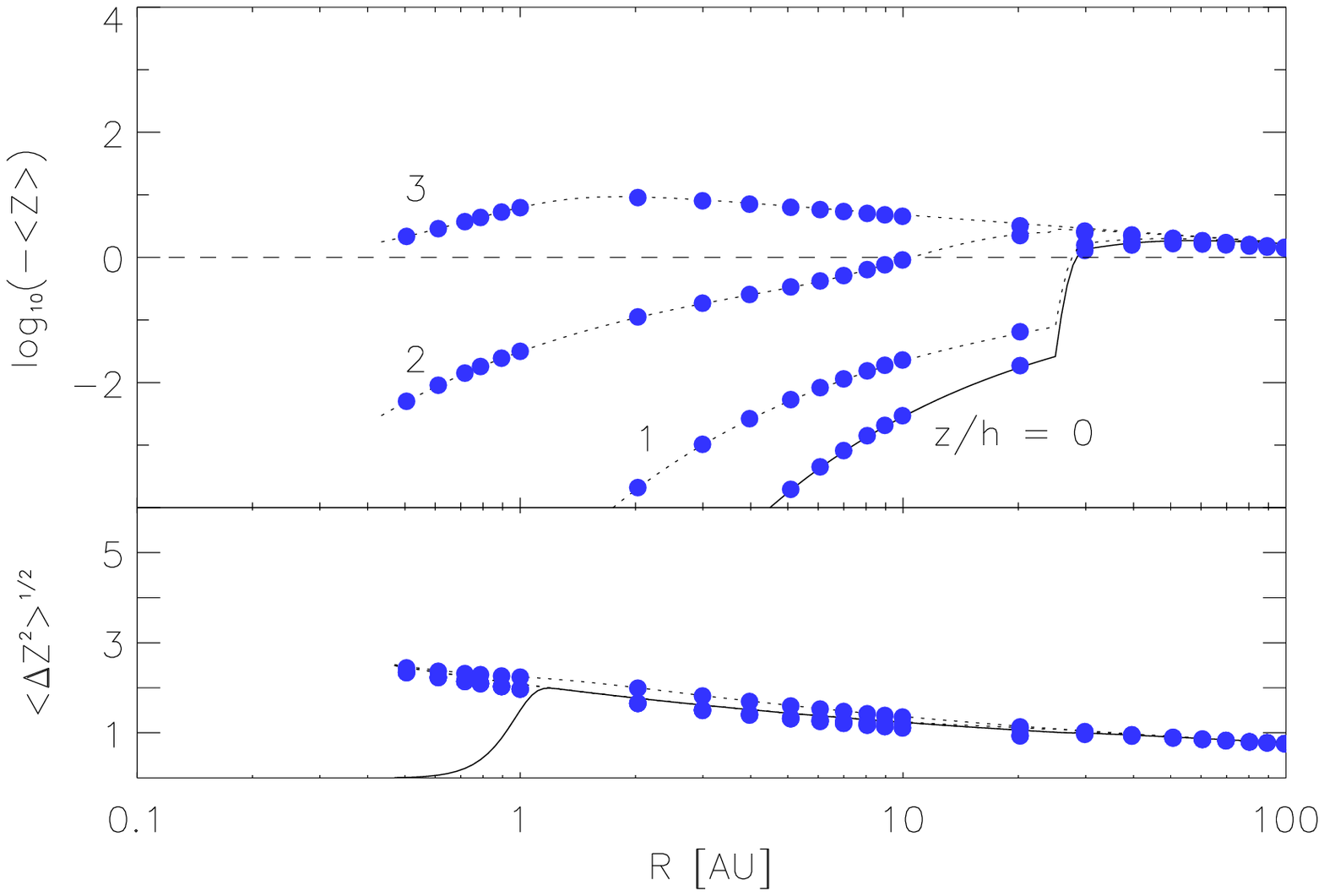}
\caption{Mean charge $\!<\!Z\!>\!$ and the standard deviation $\sqrt{<\!\Delta Z^2\!>}$ 
associated with two different dust topologies as a function of the orbital radius $R$ with 
$|\dot{M}_{\rm d}| = 10^{-10} \ \rm M_{\ast} \ yr^{-1}$. The profiles shown in the left panel refer 
to BCCA agglomerates with $N=10^6$ (top),  $N=10^4$ (middle), and  $N=10^2$ (bottom) 
monomers of $a_0 = 10^{-5} \ \rm cm$. The results obtained for the corresponding 
compact spheres are shown in the right panel. Here, the radius $a_{\ast}$ of the compact 
sphere varies from $a_{\ast} = 10^{-3} \ \rm cm$ (top) to $a_{\ast} = 4.64 \cdot 10^{-5} \ \rm cm$ 
(bottom) passing $a_{\ast}  = 2.15 \cdot 10^{-4} \ \rm cm$. The lines correspond to the values 
obtained by numerical integration while the values marked by (blue) filled circles are 
obtained using the semi-analytical solution.}
\label{fig1:sec6}
\end{figure*}

The assumption of stationary discs, i.e., if the mass flow is constant with the orbital radius, corresponds to 
a gas surface density profile given in Eq.  (\ref{eq08:sec2}) with $\Sigma_0 = 350 \ \rm gcm^{-2}$. 
One primary aim of this work is to determine to what extend compact spherical grains and dust agglomerates 
may allow for different grain charges $Z$. We already answered this question for a fixed dust-to-gas ratio 
$\Sigma_{\rm d}/\Sigma_{\rm g} = 10^{-2}$ (see the preceding section \ref{sec3}); in this section we conduct 
a similar exercise for a fixed mass flow of dust material $\dot{M}_{\rm d}$.\\
\indent
We already know that the population $N_{\rm d}(Z)$ (and $N_{\rm d0}(Z)$) follows a Gaussian and therefore is 
completely characterised by the mean value  $<\!\!Z\!\!>$  and the variance $<\!\! \Delta Z^2 \!\!>$. In particular we 
compare compact spherical grains and dust agglomerates of the same mass, which constrains the radius $a_{\ast}$ 
of the spherical grain. For BCCA agglomerates (with $D_{\rm BCCA} = 2$) relation Eq. (\ref{eq18:sec2}) holds.\\
\noindent
Figure \ref{fig1:sec6} shows the result of the one-to-one comparison. In the left panel of 
Fig. \ref{fig1:sec6} the mean value $<\!\!Z\!\!>$ and the standard deviation $\sqrt{<\! \Delta Z^2 \!>}$ 
obtained for dust agglomerates at different altitudes are plotted against the orbital radius $R$. 
The agglomerates consist of $N$ monomers of size $a_0 = 10^{-5} \ \rm cm$;  the size of the 
agglomerate varies from $N = 10^6$ (top) to $N = 10^2$ (bottom) passing 
$N = 10^4$. The profiles of $<\!\!Z\!\!>$ and $\sqrt{<\! \Delta Z^2 \!>}$ obtained for the corresponding 
compact spherical grains are shown in the right panel with $a_{\ast} = 10^{-3} \ \rm cm$ (top), 
$a_{\ast} = 2.15 \cdot 10^{-4} \ \rm cm$ (middle), and  $a_{\ast} = 4.64 \cdot 10^{-5} \ \rm cm$ 
(bottom). While the solid and dotted lines refer to the values obtained by numerical integration, 
\begin{figure}
\includegraphics[width = 9cm]{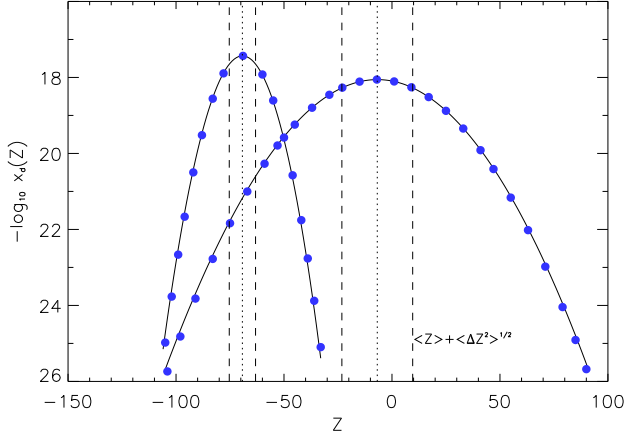}
\caption{Population $N_{\rm d}(Z)$ and $N_{\rm d0}(Z)$ (measured in particle concentration), respectively, vs charge 
excess $Z$ evaluated at disc midplane at $R = 10 \ \rm AU$. The Gaussian distribution of $N_{\rm d}(Z)/n_{\rm g}$ with 
a peak around $Z = - 8$ refers to agglomerates made of $N=10^6$ monomers of size $a_0 = 10^{-5} \ \rm cm$ while 
the profile with a peak at $Z = -70$ is obtained for compact spheres with $a_{\ast} = 10^{-3} \ \rm cm$. The dotted and 
the dashes lines mark $<\!\!Z\!\!>$ and $<\!\!Z\!\!> \pm \sqrt{<\!\! \Delta Z^2 \!\! >}$, respectively. The solid lines correspond 
to the values obtained by numerical integration while the values marked by (blue) filled circles are obtained using the 
semi-analytical solution.}
\label{fig2:sec6}
\end{figure}
the values marked with (blue) filled circles correspond to the semi-analytical solution.\\
\indent
The first thing we comment on is the significant deviation of the analytical value of 
$\sqrt{<\! \Delta Z^2 \!>}$ from the numerical value at disc midplane $z/h_{\rm g} = 0$ 
for $R \le 1 \ \rm AU$. Each numerical value shown in Fig. \ref{fig1:sec6} corresponds 
to the numerical solution of the ODE system for $\Delta t = 10^4 \ \rm yr$. We confirmed by 
numerical integration that  $<\! \Delta Z^2 \!>$ approaches its equilibrium state beyond 
$\Delta t = 10^4 \ \rm yr$ because of the low ionisation rates in this particular region. 
For agglomerates with $N = 10^6$ we obtained, for example, at disc midplane at 
$R = 0.7 \ \rm AU$ $\sqrt{<\! \Delta Z^2 \!>_{\infty}} = 31.54$ instead of $\sqrt{<\! \Delta Z^2 \!>} = 18.97$. 
However, we also remark that because of $<\!\!Z\!\!> \ \to 0^- $ with decreasing $R$, the semi-analytical 
approach may fail to comply with the assumption $Z < 0$, which was made to construct the 
semi-analytical solution.\\
\indent
We found for the population of BCCA agglomerates that the grains follow the same population 
$N_{\rm d}(Z)$ for regions $R > 30 \ \rm AU$ independent of height $z/h_{\rm g}$. 
That is partly because cosmic ray particles become the dominant source of ionisation beyond 
$R = 25 \ \rm AU$ and therefore can penetrate easily towards disc midplane because of the low 
gas surface densities. However, we stress the more important matter of the ion-electron regime 
associated with these regions. Here $N_{\rm d}(Z)$  is determined by the grain size and the gas 
temperature, which is assumed to be independent of $z$.\\
\indent
Because $T_{\rm S} \propto a_0$, the three simulations presented in the left panel of Fig. \ref{fig1:sec6} 
correspond to the same radial profile of $\Sigma_{\rm d}$. We find that the mean value and the standard 
deviation decrease by reducing $N$ and $<\!\!Z\!\!> \ \propto N$ for the ion-dust regime in particular (Okuzumi 2009). 
The profiles in the right panel of  Fig. \ref{fig1:sec6} show similar features.\\[.5em]
\noindent
In a next step we examined the population of BCCA agglomerates and compact spherical grains with respect to 
their excess charge. To recap: The populations $N_{\rm d}(Z) $ and $N_{\rm d0}(Z) $ are assumed to obey a 
Gaussian 
\begin{equation}
N_{\rm d}(Z) = \frac{n_{\rm d}}{\sqrt{2\pi<\! \Delta Z^2\!>}} \exp\left\{ - \frac{1}{2} \frac{(Z - <\!Z\!>)^2}{<\!\Delta Z^2\!>}\right\} \ .
\label{eq01:sec6}
\end{equation}
Recalling the substantial changes in $<\!\!Z\!\!>$ and $\sqrt{<\! \Delta Z^2 \!>}$ (cf Fig.  \ref{fig1:sec6}), we 
expect significant changes with respect to their population. This is indeed what we observe in Fig. \ref{fig2:sec6}. 
There we plot the population $N_{\rm d}(Z) $ and $N_{\rm d0}(Z)$ (in terms of particle concentration) at disc midplane 
at $R = 10 \ \rm AU$ for BCCA agglomerates with $N = 10^6$ and compact spheres of $a_{\ast} = 10^{-3} \ \rm cm$. Again, 
the solid lines refer to the values obtained by numerical integration while the values marked with (blue) filled circles 
correspond to the semi-analytical solution. Comparing with the corresponding profile for BCCA agglomerate, we find that 
for a given $\dot{M}_{\rm d}$ 
i) the mean value obtained for spherical grains is shifted to lower values while 
ii) the distribution becomes much narrower. Differences regarding $\max [x_{\rm d}(Z)]$ are 
caused by small changes in $\Sigma_{\rm d}$ because of slightly different values for $T_{\rm S}$.\\[.5em]
\noindent
We compared the results obtained for BCCA agglomerates and the corresponding compact spheres 
\begin{figure*}[ht]
\includegraphics[width = 8.5cm]{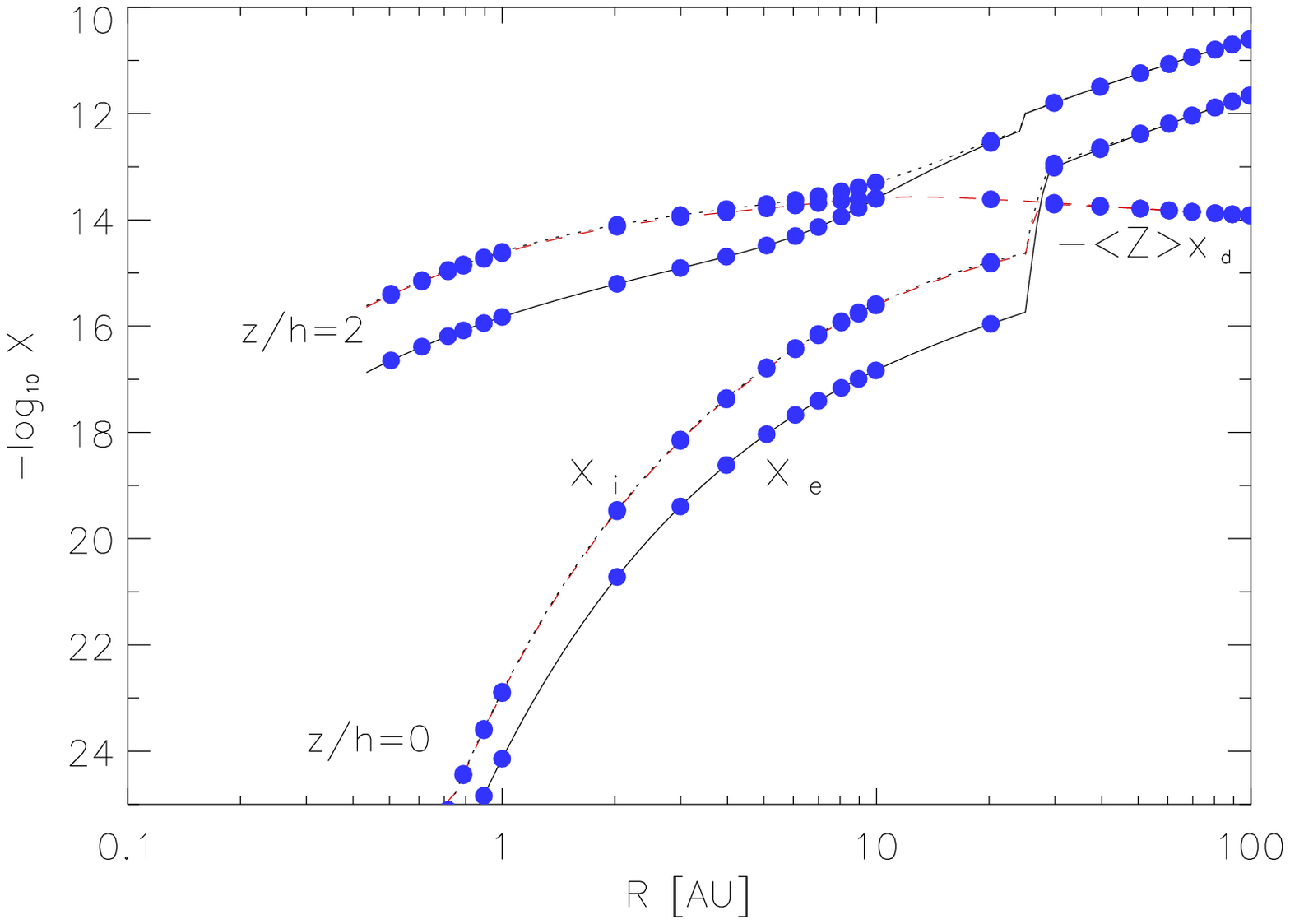}
\includegraphics[width = 8.5cm]{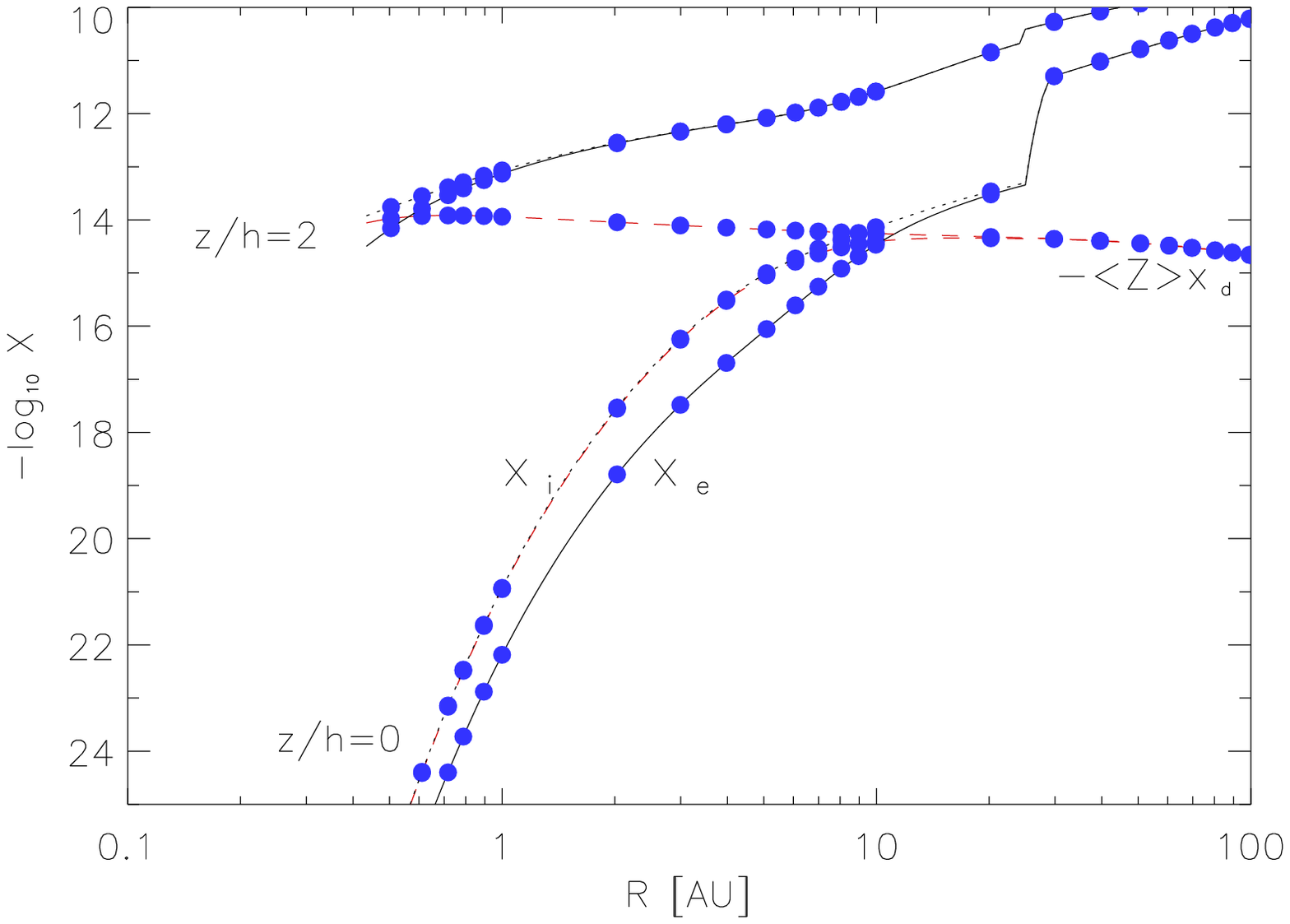}\\
\includegraphics[width = 8.5cm]{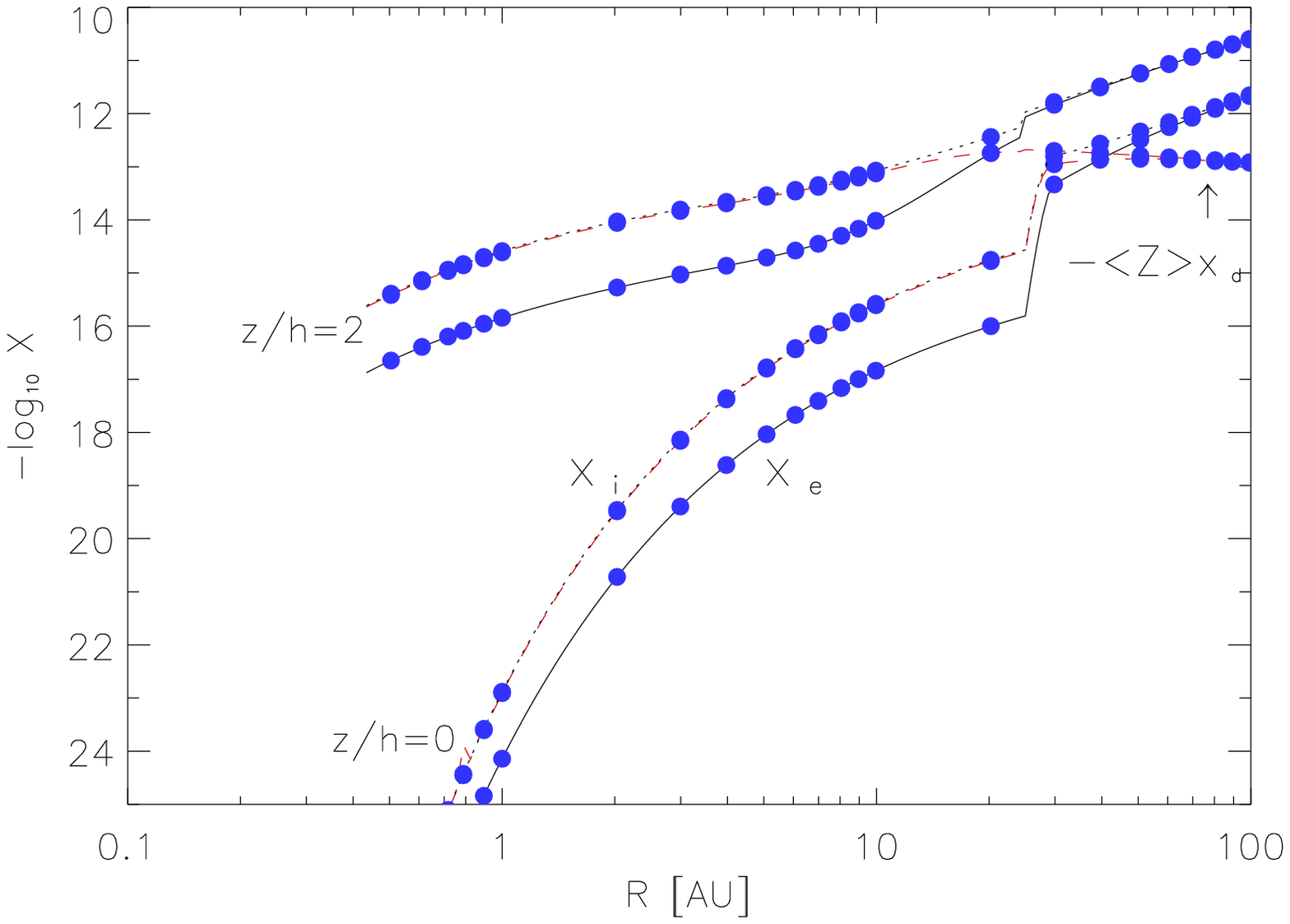}
\includegraphics[width = 8.5cm]{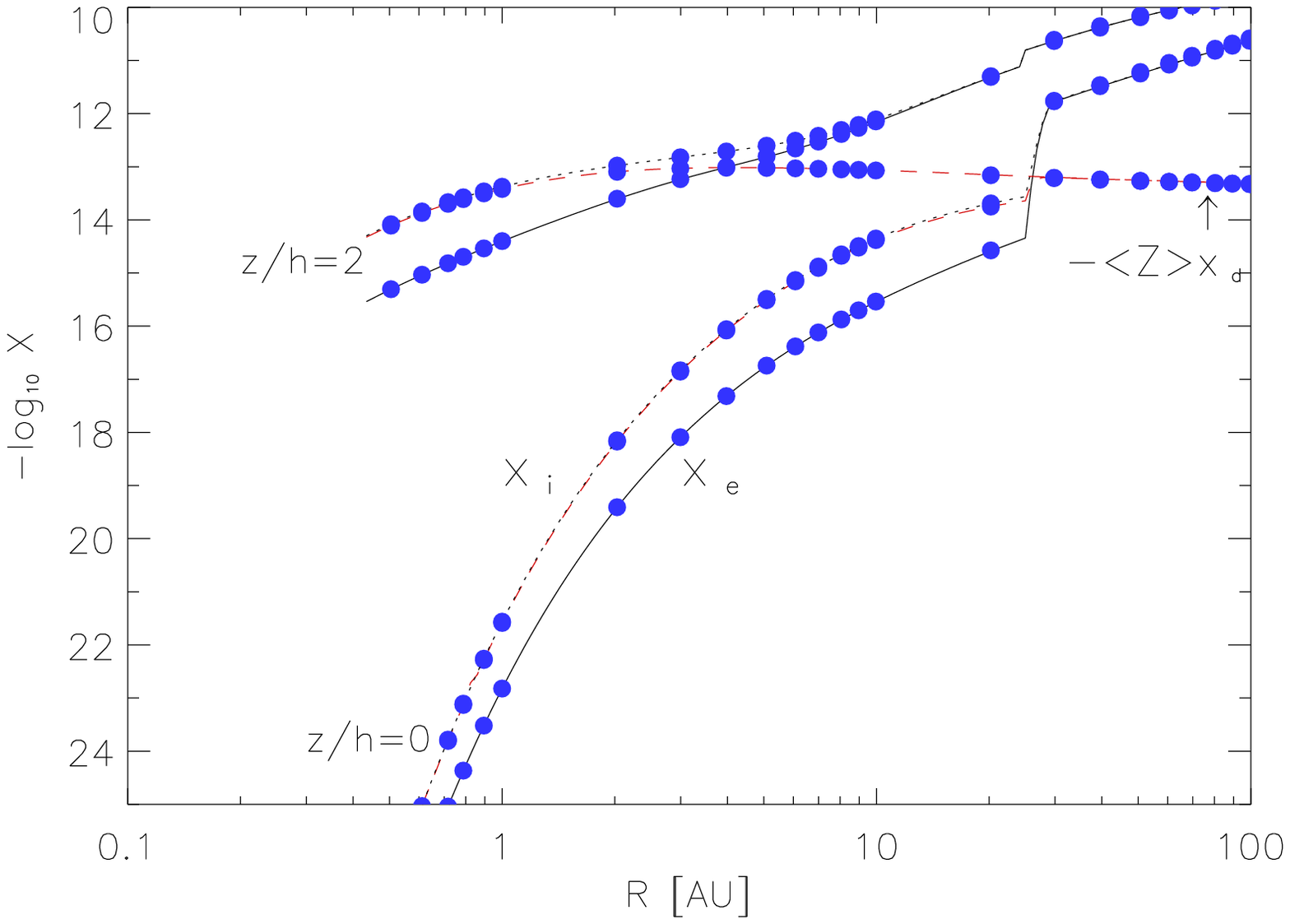}\\
\includegraphics[width = 8.5cm]{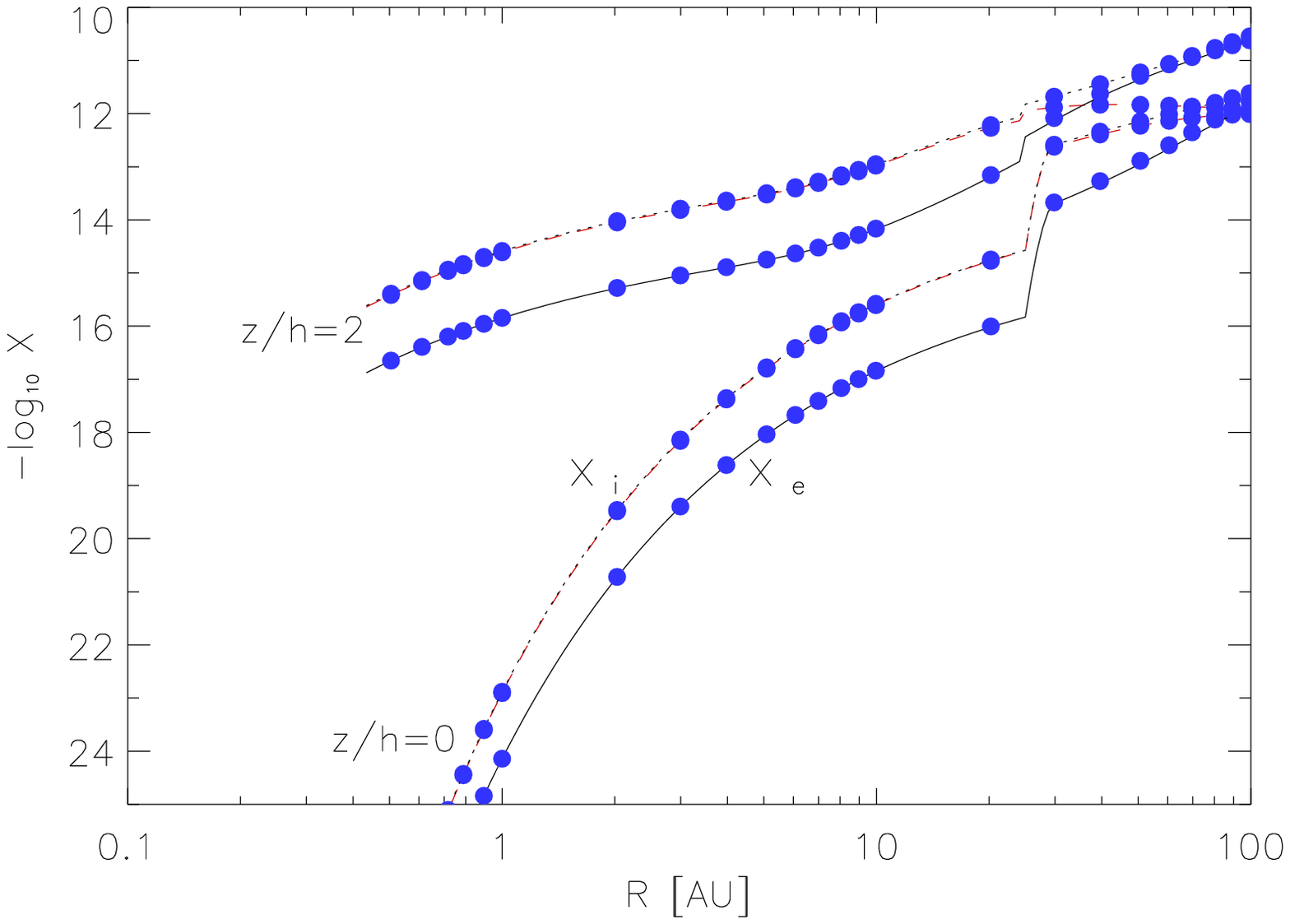}
\includegraphics[width = 8.5cm]{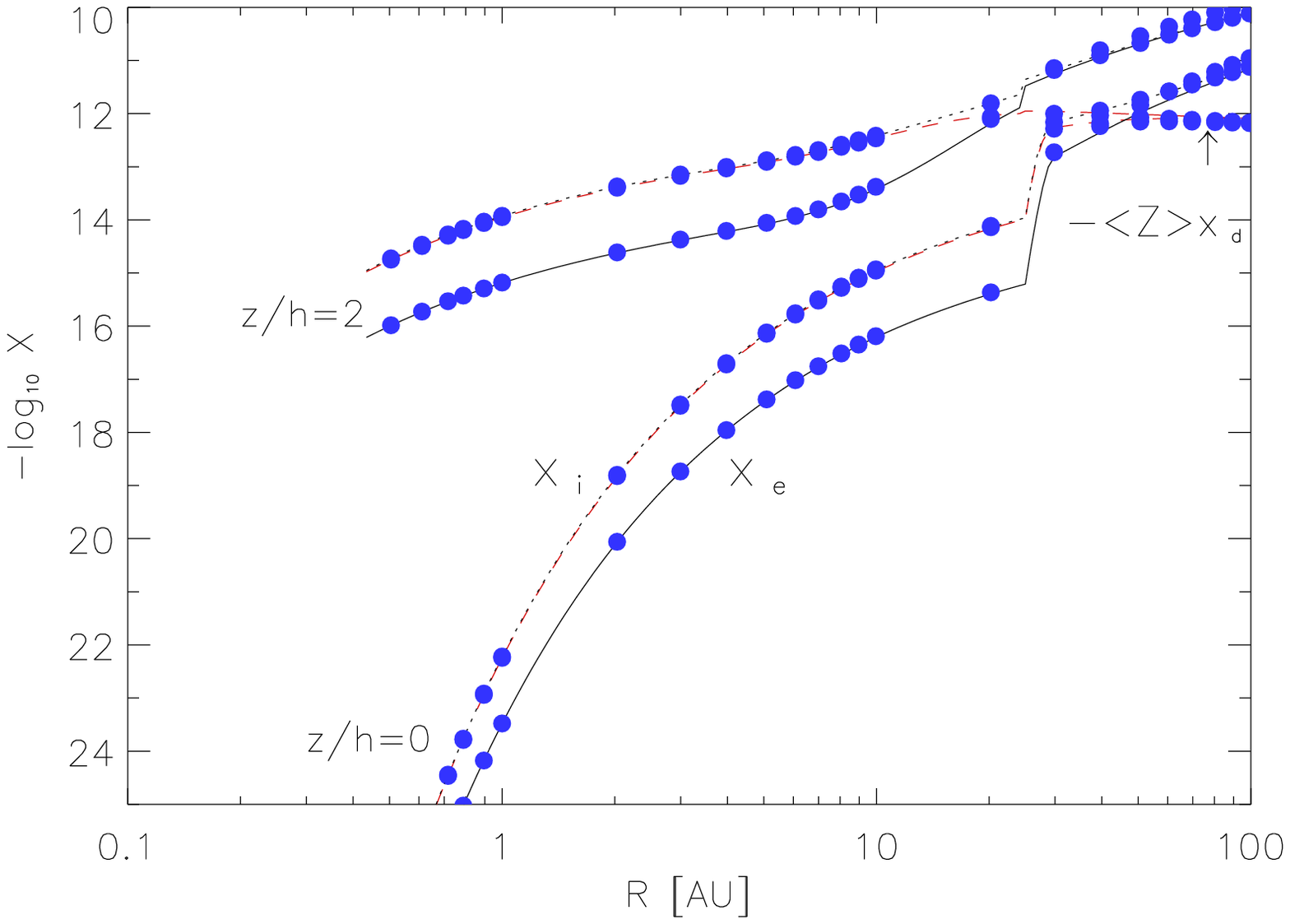}
\caption{Fractional abundances of the charge carrier associated with two different dust topologies 
as a function of the orbital radius $R$ with $|\dot{M}_{\rm d}| = 10^{-10} \ \rm M_{\ast} \ yr^{-1}$. The profiles 
shown in the left panel refer to BCCA agglomerates with $N=10^6$ (top),  $N=10^4$ (middle), and 
$N=10^2$ (bottom) monomers of $a_0 = 10^{-5} \ \rm cm$. The results obtained for the corresponding 
compact spheres are shown in the right panel. Here, the radius $a_{\ast}$ of the compact 
sphere varies from $a_{\ast} = 10^{-3} \ \rm cm$ (top) to $a_{\ast} = 4.64 \cdot 10^{-5} \ \rm cm$ 
(bottom) passing $a_{\ast}  = 2.15 \cdot 10^{-4} \ \rm cm$. The lines correspond to the values 
obtained by numerical integration while the values marked by (blue) filled circles are 
obtained using the semi-analytical solution, see the discussion in the text. The (red-coloured) dashed 
lines refer to $<\!\!Z\!\!> x_{\rm d}$.}
\label{fig3:sec6}
\end{figure*}
with respect to the fractional abundances of the charge carrier $x_{\rm i}$, $x_{\rm e}$, and $<\!\!Z\!\!> x_{\rm d}$.
The fractional abundances are shown in the left and right panels of Fig. \ref{fig3:sec6}, which correspond to those of 
Fig. \ref{fig1:sec6}. The values obtained by numerical integration and the semi-analytical solution (marked with (blue) 
filled circles) agree excellently. An important discriminant is given by $x_{\rm i} \approx x_{\rm e}$ separating the 
ion-electron plasma limit from the ion-dust plasma limit. The ion-dust plasma limit corresponds to $x_{\rm i} \gg x_{\rm e}$. 
The results obtained for BCCA agglomerates indicate that the planet forming regions are always associated with the 
ion-dust regime. The profiles for $x_{\rm e}$, $x_{\rm i}$, and $<\!\!Z\!\!> x_{\rm d}$ at disc midplane are also interesting. 
The profiles for $N=10^2$ exactly fit those obtained for $N=10^4$ and $N=10^6$, and so does $N=10^4$ and $N=10^6$. 
The reason for this is simple since the cumulative projected surface area of all BCCA agglomerates is independent of $N$. 
The situation changes slightly when comparing the profiles obtained for compact spheres, as shown in the right panel of Fig. 
\ref{fig3:sec6}. Here, the transition from the ion-dust to the ion-electron regime becomes apparent in the profiles at 
$z/h_{\rm g} = 2$ in particular. The corresponding profiles at disc midplane also change systematically with increasing 
$a_{\ast}$. 
\begin{figure*}[ht]
\includegraphics[width = 9cm]{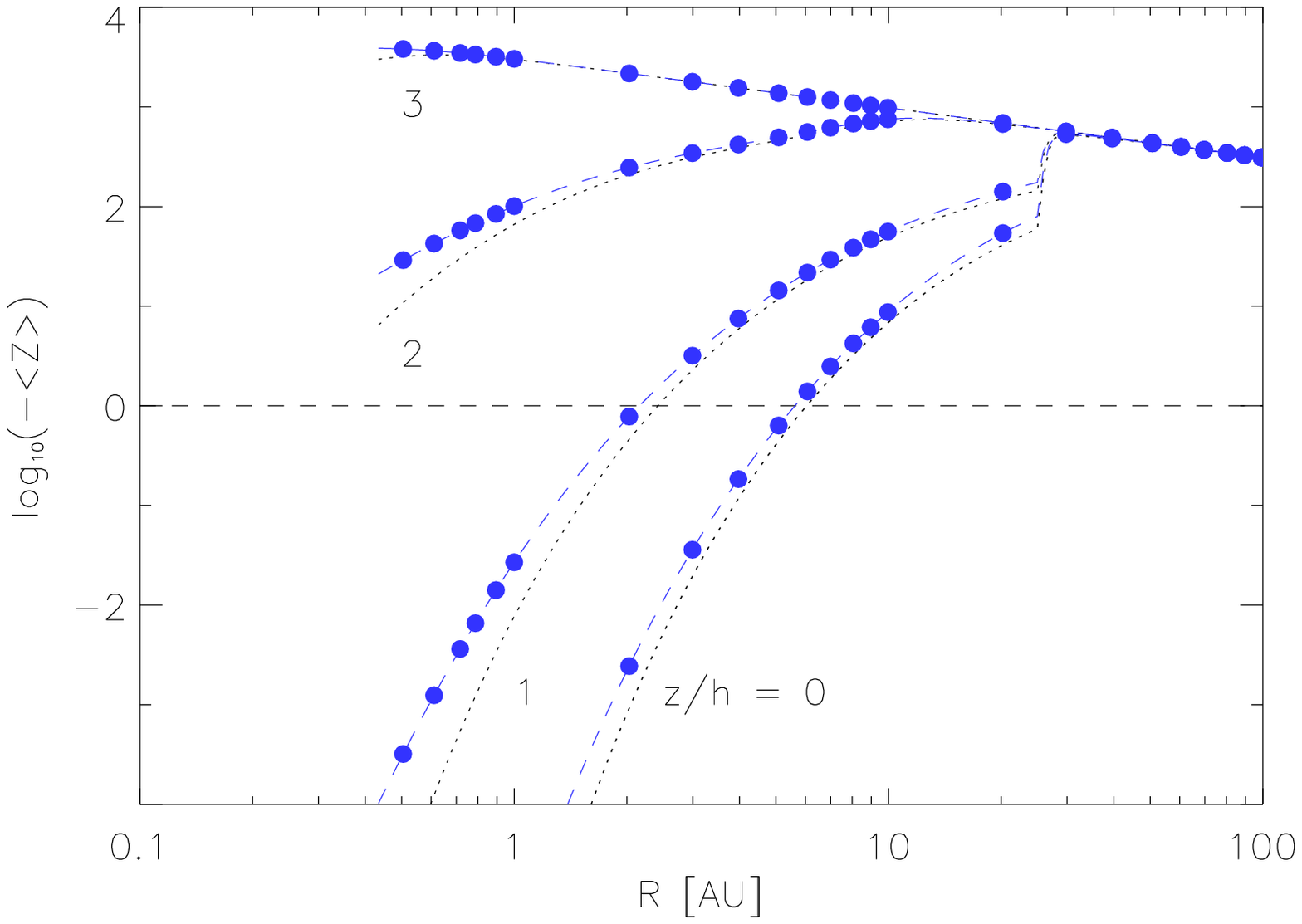}
\includegraphics[width = 9cm]{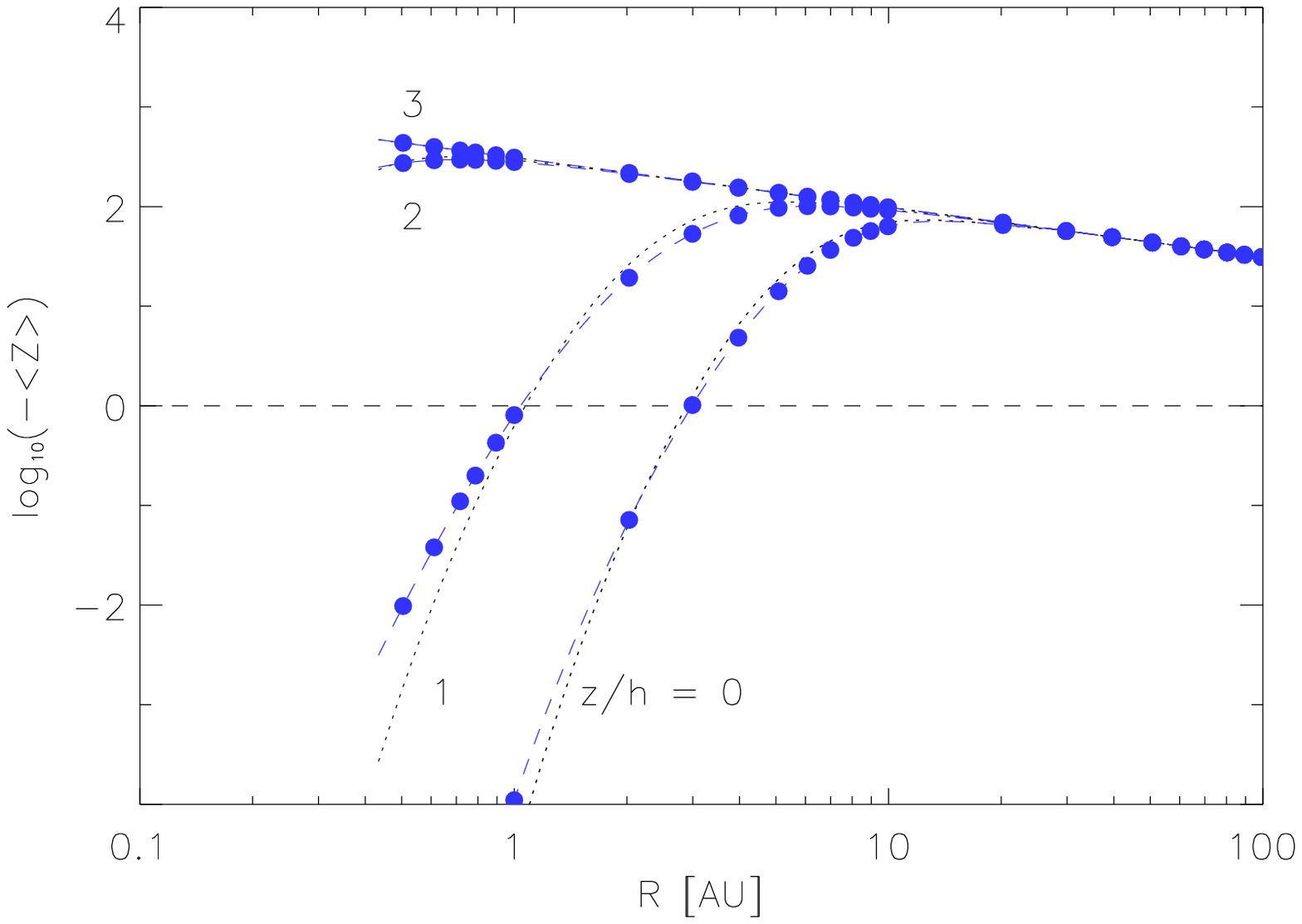}\\
\includegraphics[width = 9cm]{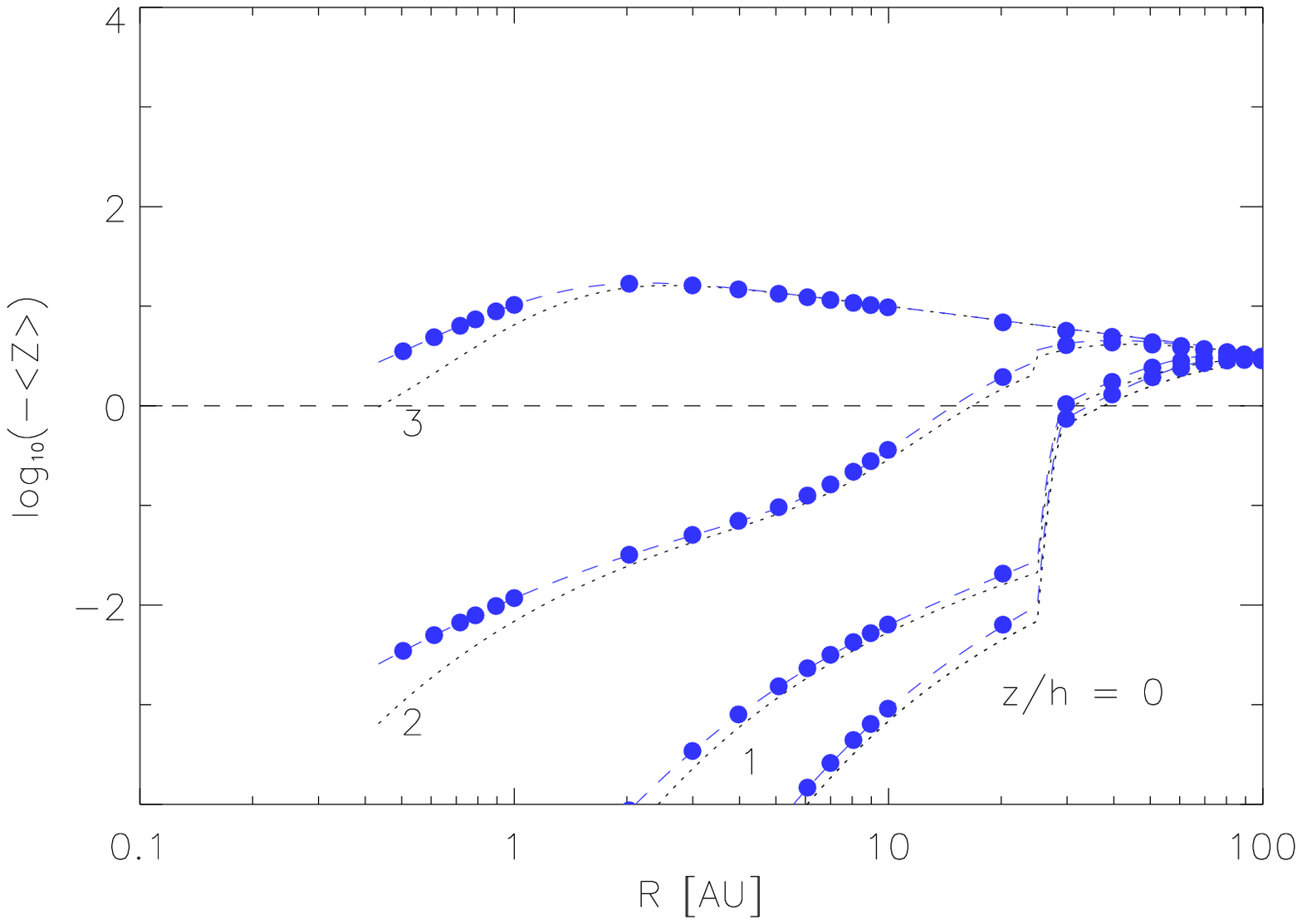}
\includegraphics[width = 9cm]{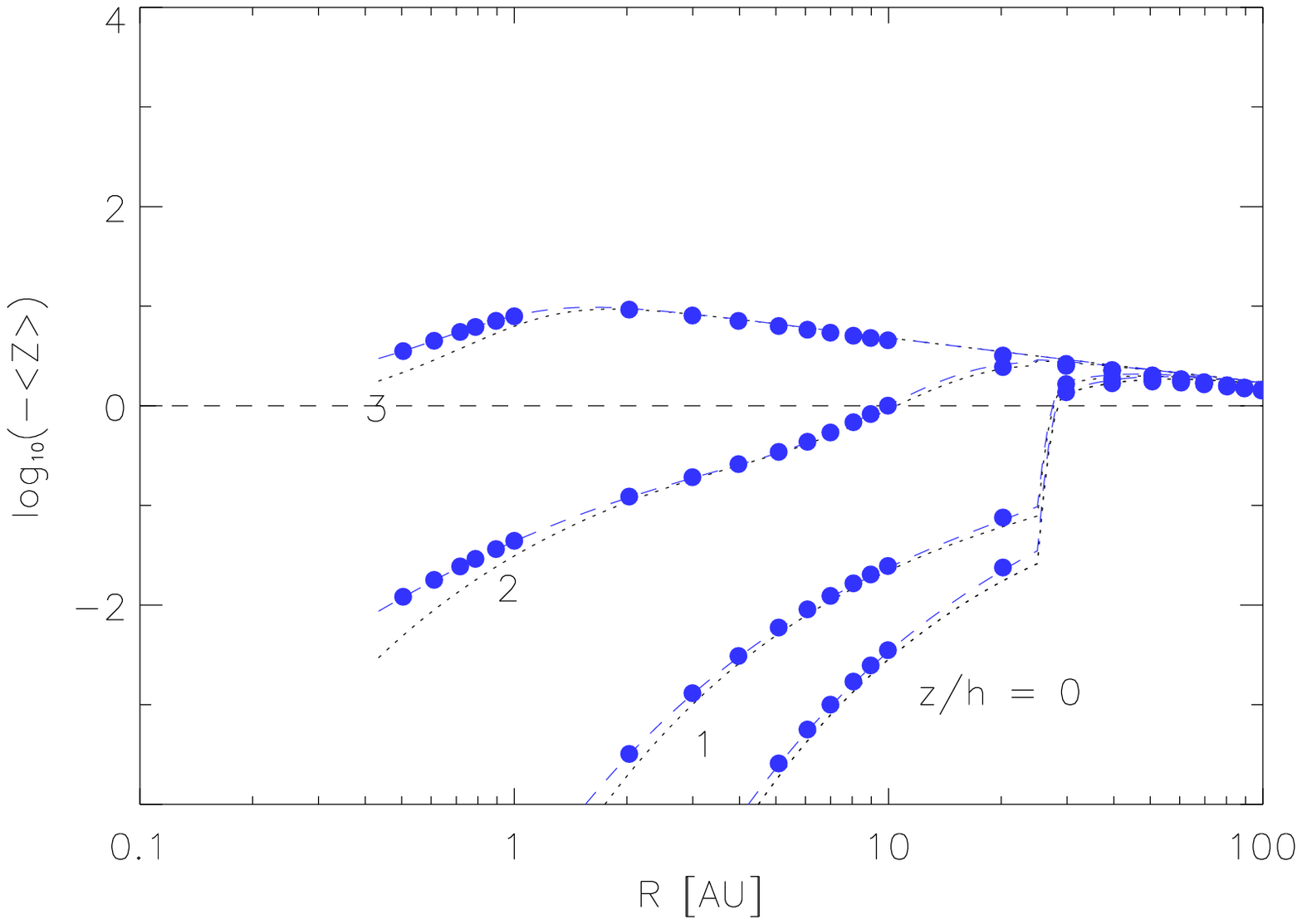}
\caption{Snapshots at $t=0 \ \rm yr$ and $t=10^6 \ \rm yr$ of dynamical disc evolution showing 
the mean charge $<\!\!Z\!\! >$ associated with two different dust topologies as a function of the orbital 
radius $R$. The profiles shown in the left panel refer to BCCA agglomerates with $N = 10^6 $ 
(top) and $N = 10^2$ (bottom) monomers of $a_0 = 10^{-5} \ \rm cm$. The results obtained for the 
corresponding compact spheres are shown in the right panel. Here, the radius $a_{\ast}$ of the 
compact sphere varies from $a_{\ast} = 10^{-3} \ \rm cm$ (top) to $a_{\ast} = 4.64 \cdot 10^{-5} \ \rm cm$ 
(bottom). The dashed lines are associated with the profiles evaluated after $t=10^6 \ \rm yr$ disc 
evolution; to aid comparison with the initial profiles we superimposed the corresponding profiles 
of Fig. \ref{fig1:sec6} using dotted lines. As we did in the preceding figures, we mark the semi-analytical 
solution with (blue) filled circles.}
\label{fig5:sec6}
\end{figure*}

\subsection{Non-stationary disc}\label{sec6:sub3}
We already analysed the  grain charging i) under static disc conditions (see section \ref{sec3}) 
and ii) under stationary disc conditions, see the preceding subsection. Comparing these with our results 
obtained under stationary disc conditions, we now explore the impact of the gas-dust dynamics 
on the grain charging. We recall the gas-dust dynamics that we analysed in section \ref{sec2}: 
Compact spherical grains with $a_{\ast} < 10^{-2} \ \rm cm$ are tightly coupled to the gas because 
of $T_{\rm S} \ll 1$. The turbulent mixing therefore dominates the advection of grains towards the central 
object. For dust particles treated as compact spheres we therefore expect that the evolving gas-dust disc 
dynamics causes minor changes in $<\!\!Z\!\!>$, $<\!\!\sqrt{\Delta Z^2}\!\!>$, and the fractional abundances 
compared with the corresponding profiles obtained under stationary disc conditions. Regarding the BCCA 
agglomerates, we already know that the agglomerates with different $N$ obey the same gas-dust disc dynamics. 
In particular, we find that the BCCA agglomerates are well coupled with the gas with no substantial 
inward migration (cf. right panel of Fig. \ref{fig1:sec2}). We conclude that the BCCA agglomerates considered 
(with $N \le 10^6$) and the corresponding compact spheres (with $a_{\ast} \le 10^{-3} \ \rm cm$) basically 
follow the similar dynamical evolution.\\
\indent
We repeated the simulations of section \ref{sec2} and let the gas-dust dynamics evolve. After $t=10^6 \ \rm yr$ 
we stopped the calculation and examined the grain charging under local disc conditions. The profiles for 
$<\!\!Z\!\!>$ are shown in Fig. \ref{fig5:sec6} assuming BCCA agglomerates (left panel) and the corresponding 
compact spheres (right panel). In particular, we focused on grain charging of agglomerates made of $N=10^2$ 
(bottom) and $N=10^6$ (top) constituent monomers and the corresponding compact spheres of 
$a_{\ast} = 10^{-3} \ \rm cm$ (top) and $a_{\ast} = 4.64 \cdot 10^{-5} \ \rm cm$ (bottom). We also compared 
the profiles obtained after  $t=10^6 \ \rm yr$  of dynamical evolution with the initial profiles at $t = 0 \ \rm yr$ 
(marked with dotted lines in Fig. \ref{fig5:sec6}). The results confirm our expectations demonstrating that 
the disc dynamics causes minor changes in the grain charging. However, we are aware that the situation 
may change if a more appropriate X-ray ionisation rate is taken, see discussion in section \ref{sec4}.

\section{Summary}\label{sec7}
We have presented calculations of the grain charging under conditions that mimic different stages 
of protoplanetary disc evolution. Instead of parametrising the dust-to-gas ratio, we inferred the  value for 
$\Sigma_{\rm d}(R)/\Sigma_{\rm g}(R)$ from the evolution of the gas-dust disc. In particular, we 
applied the disc model of Takeuchi \& Lin (2002, 2005) and content ourself with order-of-magnitude 
estimates. We considered collisional charging as the dominant process to determine the charge 
state of grains. For that purpose, we generalised the modified Oppenheimer-Dalgarno model 
introduced in Ilgner \& Nelson (2006a) to account for higher grain charges. Based on that simple chemical 
network, we examined the grain charging for two different types of grain topology: compact spherical grains 
and fractal agglomerates of $D_{\rm f} = 2$. Our main conclusions are:\\
\begin{enumerate}
\item The effect that thermal adsorption/desorption of metals has on grain charging depends on the 
mass of the dominant molecular gas-phase ion. If its mass is heavier than that of metal, the 
inclusion of the thermal adsorption of metals results in high negative excess charges on 
grains on average. Less negative charge excess on average is observed for molecular ions lighter 
than metals. 
\item  We extended the semi-analytical method proposed by Okuzumi (2009), which allowed us to determine 
steady-state solutions for the mean grain charge, electron and ion abundances associated with the 
modified Oppenheimer-Dalgarno model. The semi-analytical solutions were derived for both grain topologies: 
compact grains and fractal aggregates.
\item The results obtained confirm that dust agglomerates have a higher charge-to-mass ratio than the 
corresponding compact spheres.
\item We found that reducing the number $N$ of constituent monomers causes a drop in the mean value 
$<\!\!Z\!\!>$ of the grain charge and the standard deviation $<\!\!\sqrt{\Delta Z^2}\!\!>$. Under conditions valid for 
the ion-dust plasma limit (i.e., $x_{\rm i} \gg x_{\rm e}$) the profiles for $<\!\!Z\!\!>x_{\rm d}$, $x_{\rm e}$, and 
$x_{\rm i}$ remain unchanged with varying $N$. This is because the cumulative projected surface area of all 
BCCA agglomerates is independent of $N$.
\item The results obtained by switching from one type of grain topology (fractal agglomerates) to another (compact 
spheres) reveal that for compact spherical grains $<\!\!Z\!\!>$ is shifted towards more negative values while 
$<\!\!\sqrt{\Delta Z^2}\!\!>$ decreases.
\item The results obtained for agglomerate sizes $a_{\rm c} = [10^{-4}, 10^{-2}] \ \rm cm$ indicate that the grain 
charging of BCCA agglomerates is governed by the ion-dust plasma limit. Transitions from the ion-dust to the 
ion-electron plasma limit are observed for compact spheres depending on altitude $z/h_{\rm g}$.
\item Another conclusion is that grain charging in a non-stationary disc environment is expected to 
lead to similar results as long as the effective ionisation rate and the temperature of the gas match the stationary 
values.
\end{enumerate}
We regard the drastically simplified description of the temperatures of the gas and the dust mentioned above 
as a potentially serious omission of our model. Working out more realistic conditions may result in significant 
changes in the local disc structure and might be a potential problem for future investigations.

\begin{acknowledgement}
\noindent
I appreciate the discussions with Hiroshi Kobayashi, Satoshi Okuzumi, and Taku Takeuchi very much indeed. 
\end{acknowledgement}


\appendix
\section{Initial-boundary-value problems concerning the gas-dust dynamics}
\begin{figure*}[ht]
\includegraphics[width = .5\textwidth]{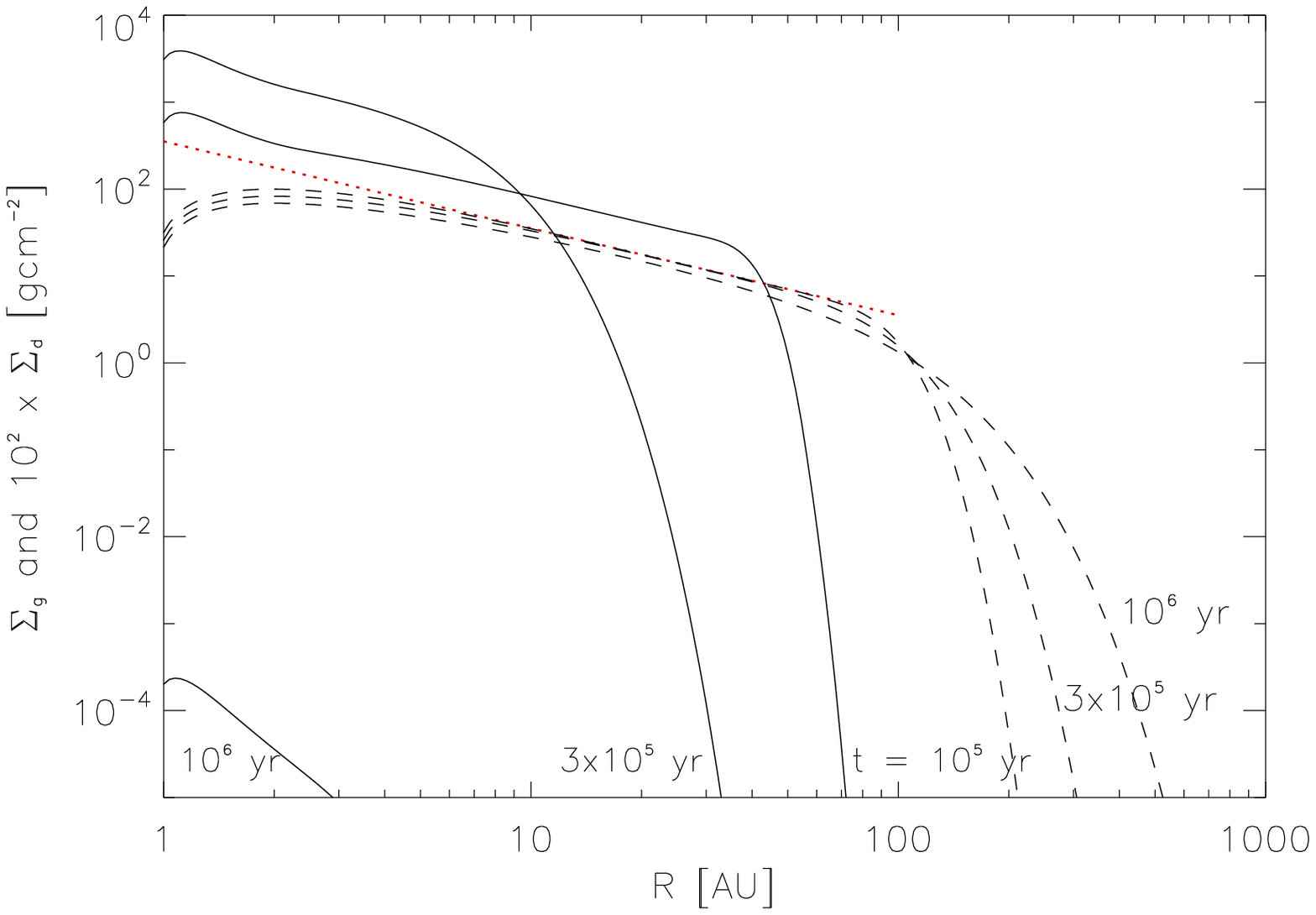}
\includegraphics[width = .5\textwidth]{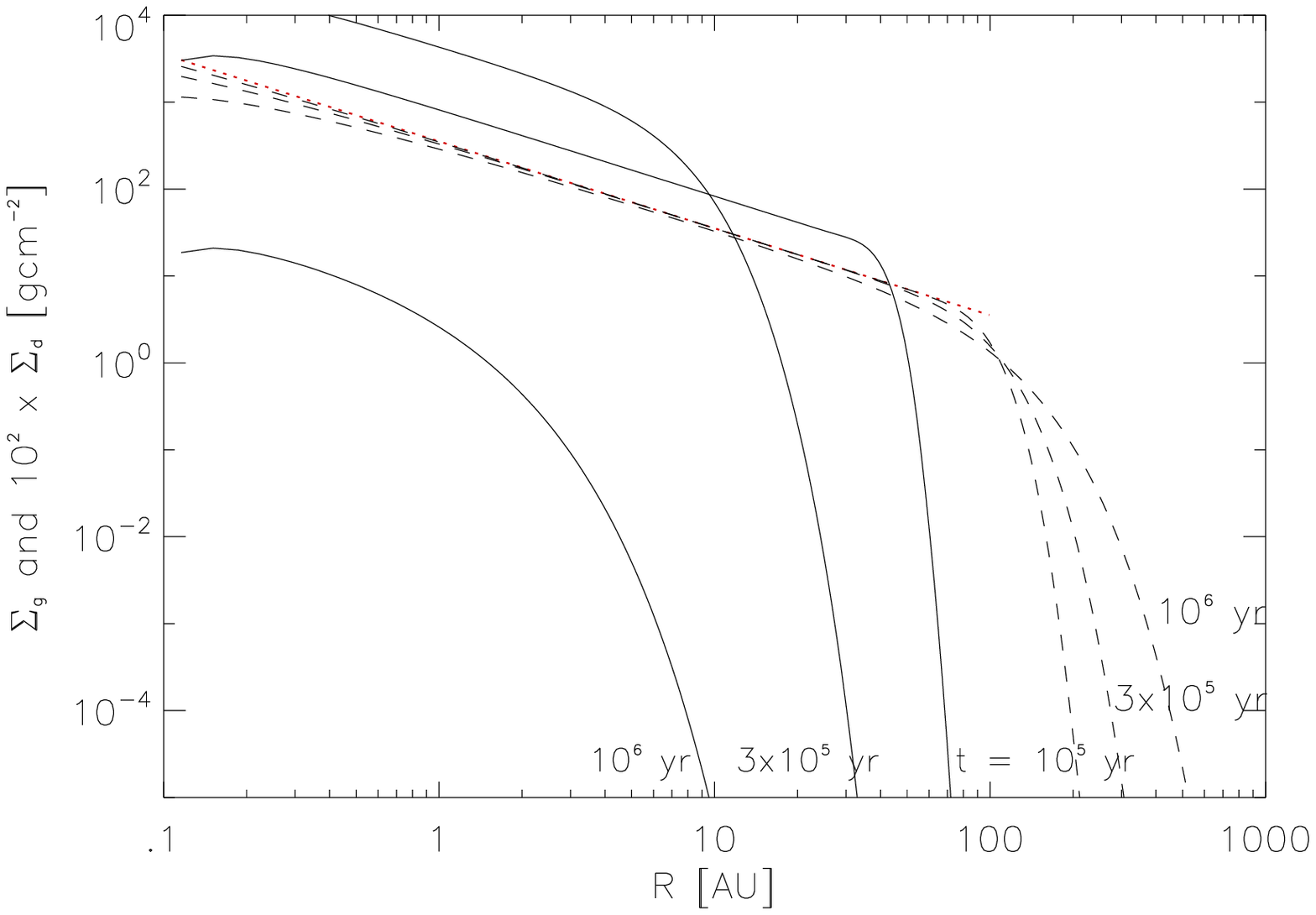}
\caption{Radial profile of the surface densities of the gas and the dust particles at different 
time snaps $t = 10^5, 3 \cdot 10^5$ and $10^6 \ \rm yr$ with $a_0 = 1 \ \rm mm$ and 
$\varrho_{\rm p} = 1.0 \ \rm gcm^{-3}$. The parameter are $\alpha = 10^{-3}$ 
and $\Sigma_{\rm d}/\Sigma_{\rm g} = 10^{-2}$ at $t/t_{\rm K}  = 0$. The solid lines correspond 
to the dust surface density while the gas surface density is shown using the dashed lines. 
The dotted line refers to $\Sigma_{\rm g} \sim R^{-1}$ at $t/t_{\nu} = 0$. The profiles shown in 
the left panel are obtained using the boundary conditions of Takeuchi \& Lin (2005). 
Moving the inner boundary to smaller $R$ and applying an analytical prescription for 
$\Sigma_{\rm g}$ at the inner and outer boundary cause the profiles to change as shown 
in the right panel.}
\label{fig1:app}
\end{figure*}

We adopted the disc model of Takeuchi \& Lin (2005) to simulate the 
dynamics of the gas-dust disc. The corresponding set of coupled partial 
differential equations (\ref{eq04:sec2}) and (\ref{eq05:sec2}) are subjected 
to initial and boundary conditions. We already know that for different 
sets of the dust parameters ($a, \varrho_{\rm p}$) the dynamics of the dust disc 
remains unchanged if the parameter sets correspond to the same stopping 
time $T_{\rm S}$. Takeuchi \& Lin (2005) have introduced a non-dimensional 
parameter $A$ to retain the same dust evolution for different values of the initial 
gas mass. In the appendix we study the effect of the boundary conditions and 
the initial profile for $\Sigma_{\rm d}$ on the dynamics of the gas-dust disc adopting 
the fiducial model of Takeuchi \& Lin (2005).\\
\indent  
We therefore solved Eqs. (\ref{eq04:sec2}) and (\ref{eq05:sec2}) for 
dust sizes of $a_0 = 1 \ \rm mm$ assuming $\Sigma_{\rm g} \propto R^{-1}$ for $t = t_0$ 
while $\Sigma_{\rm d}/\Sigma_{\rm g} = 10^{-2}$ is truncated at $R = 100 \ \rm AU$. 
Concerning the boundary conditions for $\Sigma_{\rm g}$ and $\Sigma_{\rm d}$, Takeuchi 
and Lin applied zero torque boundary conditions at $R_{\rm inner} = 5/6 \ \rm AU$ and 
$R_{\rm outer} = 10^3 \ \rm AU$. The profiles we obtained are shown in the left panel of 
Fig. \ref{fig1:app}, which should be compared with the upper panel of figure 4 of 
Takeuchi \& Lin (2005). They correspond very well. However, we briefly recall the initial 
profile for the gas surface density, which is marked in Fig. \ref{fig1:app} using (red-coloured) 
dotted lines. It represents the steady-state solution $\Sigma_{\rm g} \propto R^{-1}$ discussed 
in Sec. \ref{sec2:sub1}, which was obtained under the assumption $\nu_{\rm t} \propto R$. 
Applying the same power law $\nu_{\rm t} \propto R$, Lynden-Bell \& Pringle (1974) presented 
a similarity solution of Eq. (\ref{eq04:sec2}) with dimensionless variables $X = R/R_0$ and 
$\tau = t/t_0 + 1$. In the limit of  $X \ll 1$ the expression reduces to $\Sigma_{\rm g} \propto X^{-1}$. 
At this point, the temporal change of the gas surface density shown in the left panel of Fig. \ref{fig1:app}  
becomes important. In particular, the positive gradients in the gas surface density at $R \le 2 \ \rm AU$ 
clearly indicate an inner boundary problem associated with the zero torque boundary conditions 
applied at $R = 5/6 \ \rm AU$.\\
\indent
In order to mimic the dynamics of the gas-dust disc at the very inner disc regions more 
\begin{figure*}[ht]
\includegraphics[width = .5\textwidth]{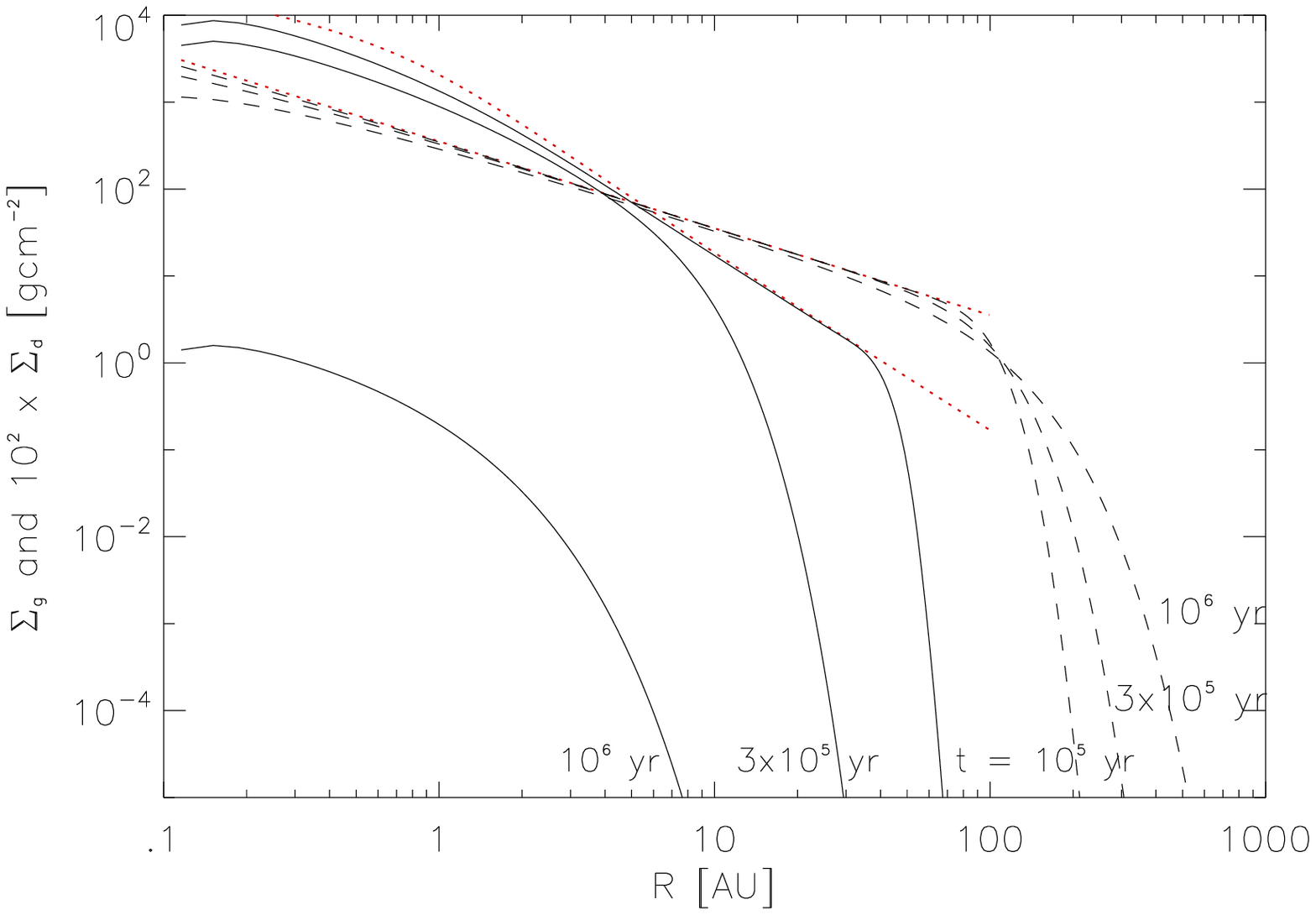}
\includegraphics[width = .5\textwidth]{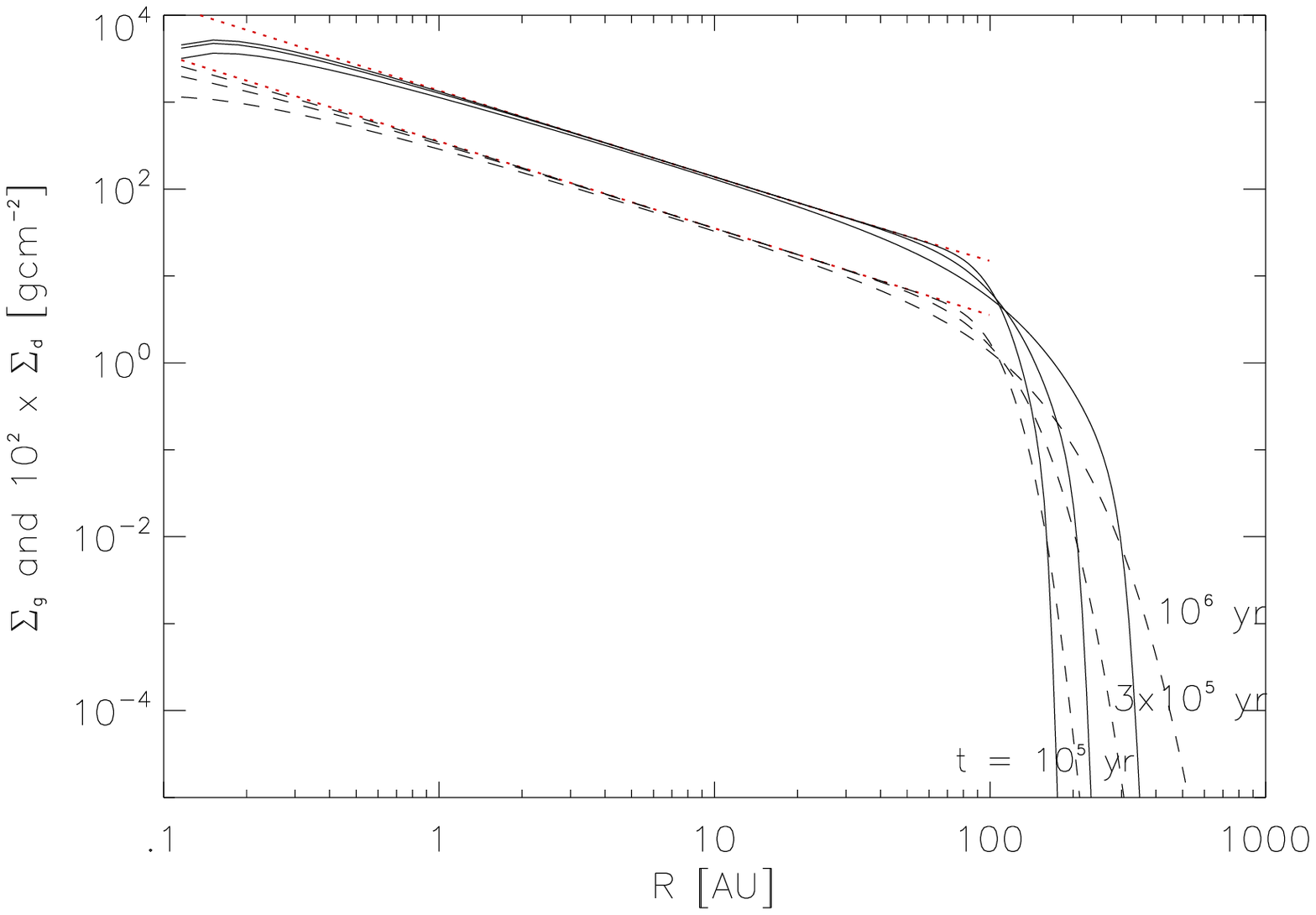}
\caption{Radial profile of the surface densities of the gas and the dust at different 
time snaps $t = 10^5, 3 \cdot 10^5$ and $10^6 \ \rm yr$. The solid lines correspond 
to the dust surface density while the gas surface density is shown using the dashed lines. 
The (red-coloured) dotted line refers to $\Sigma_{\rm g} \sim R^{-1}$ and 
$\Sigma_{\rm d} = \dot{M}_{\rm d} /(2\pi R \! <\!v_{R, \rm d}\!>\!) $ at $t/t_{\rm K} = 0$.   
$|\dot{M}_{\rm d}| = 10^{-10} \rm M_\odot / yr$, $\alpha = 10^{-3}$, and 
$\varrho_{\rm p} = 1.0 \ \rm gcm^{-3}$. The profiles shown in the left panel are obtained 
under the assumption of spherical grains with $a_0 = 1 \ \rm mm$. The corresponding 
profiles obtained for agglomerates with $D_{\rm BCCA} = 2$ and $N = 10^{12}$ monomers 
of $a_0 = 10^{-5} \ \rm cm$ are shown in the right panel.}
\label{fig2:app}
\end{figure*}
precisely, we set the inner boundary at  $R_{\rm inner} = 0.1 \ \rm AU$. We also 
replaced the zero torque boundary conditions with predefined analytical values for 
$\Sigma_{\rm g}(t)$ applying the self-similarity solution given in Eq. (\ref{eq15:sec2}). 
The most appropriate value for the constant $a$ was applied to approximate 
$\Sigma_{\rm g}(t)$ in the outer disc regions. Applying the improved boundary conditions, 
we then repeated the simulation and obtained the profiles shown in the right panel of 
Fig. \ref{fig1:app}. Comparing with the corresponding profiles in the left panel, we find that 
the time evolution of $\Sigma_{\rm g}$ at the outer disc regions remains unchanged while 
we observe $\Sigma_{\rm g} \propto R^{-1}$ in the inner disc regions. The changes in 
$\Sigma_{\rm d}$ at late evolutionary stages are remarkable. While in Takeuchi \& Lin (2005) 
the entire dust disc has vanished for $t = 10^6 \ \rm yr$, the improved boundary conditions cause 
a significant slow-down the rapid accretion of mm-sized compact spherical grain particles onto 
the star.\\
\indent
We proceeded by reassessing the assumption $\Sigma_{\rm d}/ \Sigma_{\rm g} = \rm const$ 
at $t = t_0$ that Takeuchi \& Lin (2005) applied. We already mentioned the steady-state 
solutions (\ref{eq08:sec2}) and (\ref{eq09:sec2}). For small dust sizes $a < 10^{-3} \ \rm cm$, 
$\Sigma_{\rm d}/ \Sigma_{\rm g} = \rm const$ corresponds very well to 
(\ref{eq08:sec2}) and (\ref{eq09:sec2}), for larger grains it does not. In stationary discs larger particles 
migrate inwards much faster, causing a size fractionation reported in Takeuchi \& Lin (2002). 
Hence we repeated the calculation replacing the initial assumption 
$\Sigma_{\rm d}/ \Sigma_{\rm g} = \rm const$ by $|\dot{M}_{\rm d}| = \rm const$. For the sake of 
convenience we applied the value of Takeuchi \& Lin (2002) 
$\dot{M}_{\rm d} = 10^{-10} M_{\odot} \rm yr^{-1}$, which is one order of magnitude lower 
than the corresponding mass flux of the gas $\dot{M}_{\rm g} = 3 \pi \nu \Sigma_{\rm g} 
\sim 2.6 \times 10^{-9} M_{\odot} \rm yr^{-1}$. The results obtained are shown in the left panel 
of Fig. \ref{fig2:app}. The initial profiles of $\Sigma_{\rm d}$ and $\Sigma_{\rm g}$ 
are shown using (red-coloured) dotted lines and confirm the size fractionation and 
$\Sigma_{\rm d}/ \Sigma_{\rm g} \ne  \rm const$ at $t = t_0$.\\
\indent
The grain particles are now regarded as BCCA agglomerates with a characteristic radius 
$r_{\rm c}$ defined by Eq. (\ref{eq16:sec2}). We can relate the single agglomerate made of 
$N$ monomers of size $a_0$ to a compact sphere of the same mass applying Eq. 
(\ref{eq18:sec2}). According to Eq. (\ref{eq18:sec2}), a compact sphere of $a_{\ast} = 10^{-1} \ \rm cm$ 
corresponds to an agglomerate of $N = 10^{12}$ monomers with $a_0 = 10^{-5} \ \rm cm$. The 
surface density profiles obtained for such a dust agglomerate are shown in the right panel of 
Fig. \ref{fig2:app}. We observe that the dust is tightly coupled to the gas dynamics expanding 
because of turbulent mixing towards larger orbital radii $R$.


\begin{thebibliography}{}
\bibitem[]{}
\apj{Bai, X.-N., \& Goodman, J., 2009}{701}{737}
\bibitem[]{}
\apj{Balbus, S., \&Hawley, J., 1991}{376}{214}
\bibitem[]{}
\apj{Chiang, E., \& Goldreich, P., 1997}{490}{368}
\bibitem[]{}
\apj{Desch, S., 2007}{671}{878}
\bibitem[]{}
\apj{Draine, B, \& Sutin, B., 1987}{320}{803}
\bibitem[]{}
\phyrev{Epstein, P., 1923}{22}{710}
\bibitem[]{}
\apj{Fleming, T., \& Stone, J., 2003}{585}{908}
\bibitem[]{}
\mn{Fromang, S., Terquem, C., \& Balbus, S., 2002}{329}{18}
\bibitem[]{}
\pp{Glassgold, A., Feigelson, E., \& Montmerle, T., 2000}{429}
\bibitem[]{}
\mn{Gressel, O., Nelson, R.P., \& Turner, N., 2011}{415}{3291}
\bibitem[]{}
\apj{Hawley, J., \& Balbus, S. 1991}{376}{223}
\bibitem[]{}
\pptwo{Hayashi, C., Nakazawa, K., \& Nakagawa, Y., 1985}{1100}
\bibitem[]{}
\ptps{Hayashi, C., 1981}{70}{35}
\bibitem[]{}
\apj{Hughes, A.M. et al., 2011}{727}{85}
\bibitem[]{}
\apj{Igea, J., \& Glassgold, A., 1999}{518}{848}
\bibitem[]{}
\aa{Ilgner, M., \& Nelson, R.P., 2006a}{445}{205}
\bibitem[]{}
\aa{Ilgner, M., \& Nelson, R.P., 2006b}{445}{223}
\bibitem[]{}
\aa{Ilgner, M., \& Nelson, R.P., 2008}{483}{815}
\bibitem[]{}
\nnfm{Koren, B., 1993}{45}{353}
\bibitem[]{}
\mn{Lynden-Bell, D., \& Pringle, J., 1974}{168}{603}
\bibitem[]{}
\rgp{Meakin, P. 1997}{29}{317}
\bibitem[]{}
\aa{Minato, T., et al., 2006}{452}{701}
\bibitem[]{}
\MTAC{Muller, D.E., 1956}{10}{208}
\bibitem[]{}
\apj{Okuzumi, S., 2009}{698}{1122}
\bibitem[]{}
\apj{Okuzumi, S., Tanaka, H., \& Sakagami, M., 2009}{707}{1247}
\bibitem[]{}
\apj{Okuzumi, S. et al., 2011a}{731}{95}
\bibitem[]{}
\apj{Okuzumi, S. et al., 2011b}{731}{96}
\bibitem[]{}
\apj{Okuzumi, S., \& Hirose., S., 2011}{742}{65}
\bibitem[]{}
\apj{Oppenheimer, M., \& Dalgarno, A., 1974}{192}{29}
\bibitem[]{}
\apj{Perez-Becker, D., \& Chiang, E., 2011}{727}{2P}
\bibitem[]{}
\aarev{Pringle, J., 1981}{19}{137}
\bibitem[]{}
\apj{Sano, T., et al., 2000}{543}{486}
\bibitem[]{}
\aj{Sicilia-Aguilar, A. et al., 2004}{128}{805}
\bibitem[]{}
\apj{Takeuchi, T., \& Lin, D., 2002}{581}{1344}
\bibitem[]{}
\apj{Takeuchi, T., \& Lin, D., 2005}{623}{482}
\bibitem[]{}
\pasj{Umebayashi, T.,  \& Nakano, T. 1980}{32}{405}
\bibitem[]{}
\mn{Umebayashi, T.,  \& Nakano, T. 1990}{243}{103}
\end{thebibliography}
\end{document}